\documentclass[lettersize,journal]{IEEEtran}
\usepackage{amsmath,amsfonts}
\usepackage{algorithmic}
\usepackage{algorithm}
\usepackage{array}
\usepackage[caption=false,font=normalsize,labelfont=sf,textfont=sf]{subfig}
\usepackage{textcomp}
\usepackage{stfloats}
\usepackage{url}
\usepackage{verbatim}
\usepackage{graphicx}
\usepackage{cite}

\usepackage{adjustbox}
\usepackage{amsmath,amsfonts,amssymb,amsthm, bm}
\usepackage{booktabs}
\usepackage{comment}
\usepackage{doi}
\usepackage{float}
\usepackage{gensymb}
\usepackage{multirow}
\usepackage{rotating}   \usepackage{subfig}
\usepackage{xcolor}
\usepackage{soul}
\usepackage{url}

\usepackage{listings}
\newcommand{\etal}{\textit{et al}.~}
\newcommand{\ie}{\textit{i}.\textit{e}.~}

\newcommand{\Loss}{\mathcal{L}}

\begin{document}

\title{Hyperspectral Pixel Unmixing with Latent Dirichlet Variational Autoencoder}

\author{
  Kiran Mantripragada and Faisal Z. Qureshi
  \thanks{
    Visual Computing Lab -- Faculty of Science -- Ontario Tech University.
    2000 Simcoe Street North, Oshawa, ON L1G OC5, Canada.\\
    Corresponding author: kiran.mantripragada@ontariotechu.net.
  }
  \thanks{Manuscript received December DD, 2022; revised MMM dd, 2023.}
}

\markboth{IEEE TRANSACTIONS ON GEOSCIENCE AND REMOTE SENSING, ~Vol.~XX, No.~X, MMM~YYYY}%
{Mantripragada \textit{et} Qureshi: Hyperspectral Pixel Unmixing with Latent Dirichlet Variational Autoencoder}

\maketitle

\begin{abstract}

  We present a method for hyperspectral pixel {\it unmixing}.  The
  proposed method assumes that (1) {\it abundances} can be encoded as
  Dirichlet distributions and (2) spectra of {\it endmembers} can be
  represented as multivariate Normal distributions.  The method solves
  the problem of abundance estimation and endmember extraction within
  a variational autoencoder setting where a Dirichlet bottleneck layer
  models the abundances, and the decoder performs endmember
  extraction.  The proposed method can also leverage transfer learning
  paradigm, where the model is only trained on synthetic data
  containing pixels that are linear combinations of one or more endmembers
  of interest.  In this case, we retrieve endmembers (spectra) from the United
  States Geological Survey Spectral Library.  The model thus trained
  can be subsequently used to perform pixel unmixing on ``real data''
  that contains a subset of the endmembers used to generated the
  synthetic data.  The model achieves state-of-the-art results on
  several benchmarks: Cuprite, Urban Hydice and Samson.  We also
  present new synthetic dataset, OnTech-HSI-Syn-21, that can be used
  to study hyperspectral pixel unmixing methods.  We showcase the
  transfer learning capabilities of the proposed model on Cuprite and
  OnTech-HSI-Syn-21 datasets. In summary, the proposed method can be applied for pixel unmixing a variety of domains, including agriculture, forestry, mineralogy, analysis of materials, healthcare, etc. Additionally, the proposed method eschews the need for labelled data for training by leveraging the transfer learning paradigm, where the model is trained on synthetic data generated using the endmembers present in the ``real'' data.
\end{abstract}


\section{Introduction} \label{sec:introduction}

For Hyperspectral Images (HSI) the pixel intensity values represent the cumulative reflectance of various materials within the Instantaneous Field of View (IFOV). In many cases, a single pixel's intensity results from the combined contributions of multiple materials. Consequently, a pixel can be conceptualized as a blend of these materials, with both the specific materials and their respective proportions remaining unidentified.
This holds particularly true for high-altitude and low-resolution hyperspectral images common in remote sensing scenarios, where individual pixels can span large spatial areas~\cite{heylen:2014}.
\textit{Pixel unmixing}, often known as ``Spectral Unmixing'' in HSI analysis, involves determining the materials in a pixel (endmembers) and their mixing proportions (abundances). It is crucial for understanding the composition and proportions of materials in hyperspectral images and it has received significant research attention.

Existing approaches for pixel unmixing can be broadly classified into two groups: 1) data-driven methods---such as Blind Source Separation
(BSS) models, Analysis of Principal Components (linear and non-linear), Linear Discriminant Analysis---that attempt to ``decompose'' a pixel into spectra of various endmembers, and more recently Deep Learning-based approaches that leverage architectures such as Autoencoders \cite{zhang:2021, garg:2021, khajehrayeni:2021, zhao:2021, plaza:2005}, and 2) the so-called physics-based methods that assume access to phenomenological models for radiance responses of different materials~\cite{heylen:2014, plaza:2012}.
The use of physics-based methods can be challenging in practice due to the laborious task of developing models of radiance response for various imaging scenarios. As a result there is a growing interest in developing data-driven approaches for hyperspectral pixel unmixing.

This paper develops a new data-driven method for hyperspectral pixel unmixing---identifying endmembers and estimating the mixing ratios of these endmembers.
In the following discussion, we will refer to endmembers as the spectral signals of pure materials.
Furthermore, we will refer to the problem of identifying endmembers as endmember extraction, since this terminology is often used within the hyperspectral image analysis community.
Our method casts pixel unmixing as an optimization problem within the variational inference setting.
Specifically, we develop a Latent Dirichlet Variational Autoencoder (LDVAE) whose latent representation encodes endmembers' mixing ratios (solving the abundance estimation problem)~\cite{kingma:2014}.
The decoder is able to reconstruct the endmembers spectra, thus solving the endmember extraction problem.
Our model assumes that endmembers' spectra can be represented by a Multivariate Normal Distribution and that the endmembers' mixing ratios can be represented as a Dirichlet Distribution.
The proposed method solves the two sub-tasks---endmember extraction and abundance estimation---together.

The proposed method leverages a variational autoencoder architecture with a latent dirichlet distribution for two reasons. Firstly, the Dirichlet Distribution is a probability distribution over $n$-simplex vectors that naturally encode the abundances for a given pixel while enforcing the Abundances Sum-to-One Constraint (ASC) and Abundances Non-negative Constriant (ANC).  Secondly, variational autoencoders are powerful deep learning architectures for probabilistic modeling that follow the variational inference paradigm.  These autoencoders are relatively straightforward to train given access to the training data.  Palsson et al.~\cite{palsson:2018} independently reached a similar conclusion that variational autoencoders are appropriate models for abundance representation. In addition to performing both abundance estimation and endmember extraction, the generative nature of our proposed model is also able to synthesize hyperspectral pixels.

We evaluate our approach on four datasets: a synthetic dataset generated using USGS Spectral Library ~\cite{usgs_spectral_library:2017}, Cuprite dataset, HYDICE Urban dataset, and Samson dataset. We demonstrate that the proposed method achieves state-of-the-art performance on these datasets as measured by the commonly-used metrics---SAD and RMSE.  We also show that our method can be applied in situations where training data is absent, e.g., in the case of Cuprite dataset, pixel-level abundance information is missing.  Here the model is trained on synthetic dataset that contains the same materials as those present in the Cuprite dataset.  Subsequently, the model trained on synthetic dataset is applied on the ``original'' Cuprite dataset.
A secondary contribution of this work is a collection of tools that we have developed to (1) manage hyperspectral datasets, (2) parse the USGS spectral library, and (3) generate synthetic hyperspectral cubes using the USGS spectral library.
We plan to release these tools to the community as open-source software.
Finally, we introduce a new synthetic dataset that others may find useful for hyperspectral pixel unmixing.  This dataset is synthesized using the same materials as used by other researchers ~\cite{zhang:2021_cvpr, zhang:2021_iccv, zhang:2020_cvpr, wang:2019_cvpr}.

\section{Related Work}
\label{sec:related_work}

The field of pixel unmixing is primarily divided into two categories: (a) physics-based methods~\cite{heylen:2014} and (b) data-driven methods.
Physics-based schemes use models of light reflection, scattering, transmission and absorption, e.g., Hapke's Bidirectional Reflectance Model (BRDF) \cite{hapke:2012, drumetz:2019_spectral, sun:2021} and the Atmospheric Dispension Model~\cite{janiczek:2020}, for hyperspectral pixel unmixing.
Physics-based models are laborious to use in practice, since these require radiance models that are situation specific.
Conversely, data-driven methods are simpler to apply and to use in practice; therefore, a majority of pixel unmixing methods fall into this category.
However, data-driven approaches require training data.
Other approaches, e.g., the work by Drumetz \etal~\cite{drumetz:2019_spectral} that combines Hapke's BRDF model with linear mixing models, sit at the intersection of physics-based and data-driven approaches.

The method proposed in this paper belongs to the class of data-driven aproaches and the following paragraphs provide a brief overview of data-driven methods for pixel unmixing.
Blind Source Separation (BSS) type methods, such as N-FINDR, PPI, and VCA, divide the problem of unmixing into two steps: 1) endmember extraction and 2) abundance estimation.
Oftentimes abundance estimation (step 2) requires \emph{a priori} knowledge of the endmembers; therefore, it is sensitive to the accuracy of the estimated endmembers from step 1~\cite{winter:1999,drumetz:2019, drumetz:2016}.
N-FINDR, for example, is an iterative algorithm for endmember extraction that seeks to find the vertices, which represent the endmembers, of the $n$-simplex
containing the pixel spectra~\cite{winter:1999}.
Pure Pixel Index (PPI) is another commonly used scheme for endmember extraction that is able to deal with atmospheric, solar, and
instrument-induced artifacts.
PPI achieves the endmember extraction task by compressing (via dimensionality reduction) and denoising (via noise whitening) the input spectra before projecting it to $n$-simplex hyperplane.
The pixels closest to the vertices (of the $n$-simplex) are used to identify endmembers present in the pixel. These methods use ASC and ANC constraints to setup a fully constrained least square optimization problem for abundance estimation~\cite{ibarrola:2019, mahabir:2018, bioucas:2012, nascimento:2005,winter:1999}.

Methods---such as Spectral-Spatial Weighted Sparse Non-Negative Matrix Factorization (SSWNMF)~\cite{zhang:2021}, Spatial Group Sparsity Regularized Nonnegative Matrix Factorization (SGSNMF)~\cite{wang:2017}, Total Variation Regularized Reweighted Sparse Nonnegative Matrix Factorization (TV-RSNMF)~\cite{he:2017}, and Graph-Regularized $L_{1/2}$-NMF (GLNMF)~\cite{lu:2013}---rely upon non-negative matrix factorization to estimate abundances.
The key idea is to express the hyperspectral image as a product of two matrices representing endmembers and abundances.
SSWNMF and SGSNMF use spatial information when performing pixel unmixing.
Whereas SSWNMF define a neighbourhood using a weighted-window around the pixel of interest, SGSNMF define the neighbourhood as a super-pixel.
TV-RSNMF iteratively updates endmembers' matrix and abundance maps.
It can be considered as an abundance map {\it denoising} procedure.
GLNMF extends TV-RSNMF and builds a graph that defines the local neighbourhood around the pixel of interest.
Both TV-RSNMF and GLNMF methods makes sparsity assumptions when solving for hyperspectral pixel unmixing.
Non-negative matrix factorization based methods post compelling results on hyperspectral unmixing benchmarks.
Specifically, SSWNMF achieves the state-of-the-art results on hyperspectral pixel unmixing benchmarks.
Therefore, we have followed the evaluation scheme proposed by SSWNMF, and we use the same benchmark datasets and metrics to evaluate our methods as those used in~\cite{zhang:2021}.

More recently, researchers have been exploring Deep Learning based approaches for HSI pixel unmixing.
DeepGUn~\cite{borsoi:2019} is a deep learning method for pixel unmixing that explores regularization techniques
to learn latent representations that are amenable to Vertex Component Analysis (VCA) for endmember extraction.
The extracted endmembers are subsequently used to train deep learning models to reconstruct pure pixels.

Palsson et al. (2018)~\cite{palsson:2018} employ an autoencoder architecture where the encoder stage learns to represent abundances by enforcing pixel reconstruction at the decoder stage.  This approach assumes a linear mixing of endmembers within a pixel.  Subsequently, Palsson \etal (2022) explores the use of a variational autoencoder to generate synthetic data~\cite{palsson:2022_synthetic}.
DAEN \cite{su:2019:daen} also use an autoencoder architecture for pixel unmixing.  Here, first, a stacked autoencoder uses VCA to identify candidate pixels based upon their purity-index.  Next, a variational autoencoder is used to solve the underlying non-negative matrix factorization problem.  The quality of the candidate pixels identified by the stacked autoencoder influence the overall unmixing results.
Shahid et al. (2022) \cite{shahid:2022} is another method that uses an autoencoder architecture for hyperspectral pixel unmixing.  This method requires an initial abundances estimate, which is provided either via K-means clustering or via Radial Basis Functions.  The decoder stage incorporates one of the following mixing models: Fan, bilinear, or Postpolynomial.

Palsson \etal~\cite{palsson:2020} proposes a Convolutional Neural Network Autoencoder Unmixing (CNNAEU) model for the problem of hyperspectral pixel unmixing.
Similar to the non-negative matrix factorization schemes discussed above, this method uses both spatial and spectral information when performing pixel unmixing.
Unlike our model, CNNAEU assumes a linear mixing model and does not provide generative decoder; therefore, it cannot cover higher spectral variability and is not capable of generating unseen pixels.

Our method differs from the existing schemes in an important way---given an input pixel, our method learns to construct its latent representation that models a Dirichlet Distribution, which perfectly captures the ANS and ANC constraints that arise in abundance estimation.
Due to the generative nature of our architecture, the proposed model is able to synthesize new pixel spectra given known abundances or endmembers.
Unlike CNNAEU~\cite{palsson:2020}, our model currently does not use spatial information.
It is, however, feasible to extend our model to use a CNN-based encoder that will incorporate the  spatial neighbourhood information of a pixel when constructing the latent representation.  We plan to investigate this at another time.
The proposed model requires training data in the form pixel-level abundances.
We show in Section~\ref{sec:results} that it is possible to train the proposed model using only synthetic data when the ``real'' data is missing pixel-level abundances for training purposes.

Others have explored the use of latent Dirichlet VAE in other domains.  Li~\etal~\cite{li:2020_dirichlet} introduced the mathematical framework of Dirichlet Graph Variational Autoencoder (DGVAE).
Their goal was to replace the Gaussian latent space by the Dirichlet latent space and their work deals with cluster membership.
Kim~\etal~\cite{kim:2023} propose a method for anomaly detection in high-dimensional data using a Dirichlet Variational Autoencoder.  Xu~\etal~\cite{xu:2023} proposed a Variational Autoencoder with Dirichlet priors method for feature disentanglement.
This work explores the reparametrization trick using the Laplace approximation.  Our method uses the reparametrization trick as presented in Joo~\etal~\cite{joo:2020}.

In a nutshell, others have explored latent Dirichlet VAEs; however, ours is the first approach that applies this architecture to the problem of hyperspectral unmixing.  Similarly, others have explored autoencoders and variational autoencoders for hyperspectral pixel unmixing, none have used a latent Dirichlet Variational Autoencoder for the task of pixel unmixing.  Consequently, our work represents an important contribution to the field of hyperspectral pixel unmixing.
\section{Method}
\label{sec:method}

The problem of pixel unmixing is similar to the topic modeling problem that aims to discover the topics in a collection of documents and how these topics are related specifically to each individual documents in this collection~\cite{blei:2003}.
We can extend this idea to the problem of pixel unmixing as follows: (1) the hyperspectral image is the collection of document; (2) each pixel is a document: and (3) each endmember is a topic.  The endmembers are unknown \textit{a priori}. Additionally, for any given pixel, the mixing ratios of these endmembers (abundances) are unknown.
Within this setting, we can leverage techniques available in the topic modeling literature for the problem of hyperspectral pixel unmixing.
Latent Dirichlet Allocation (LDA) is a popular technique for topic modeling that, given a corpus, aims to (1) discover these latent topics and (2) estimate to what degree each topic contributes to a particular document.
Inspired by LDA, we represent the abundances as a Dirichlet Distribution. Thus, hyperspectral pixel unmixing becomes the problem of constructing the latent representation that follows a dirichlet distribution. We also seek a method that reconstructs the spectra given a set of endmembers and their mixing ratios.  We propose that both of these tasks can be accomplished with LDVAE, which we describe in the following section.

\subsection{Latent Dirichlet Variational Autoencoder (LDVAE)}
\label{subsec:ldvae}

\begin{figure}
  \centerline{
    \includegraphics[width=0.98\linewidth]{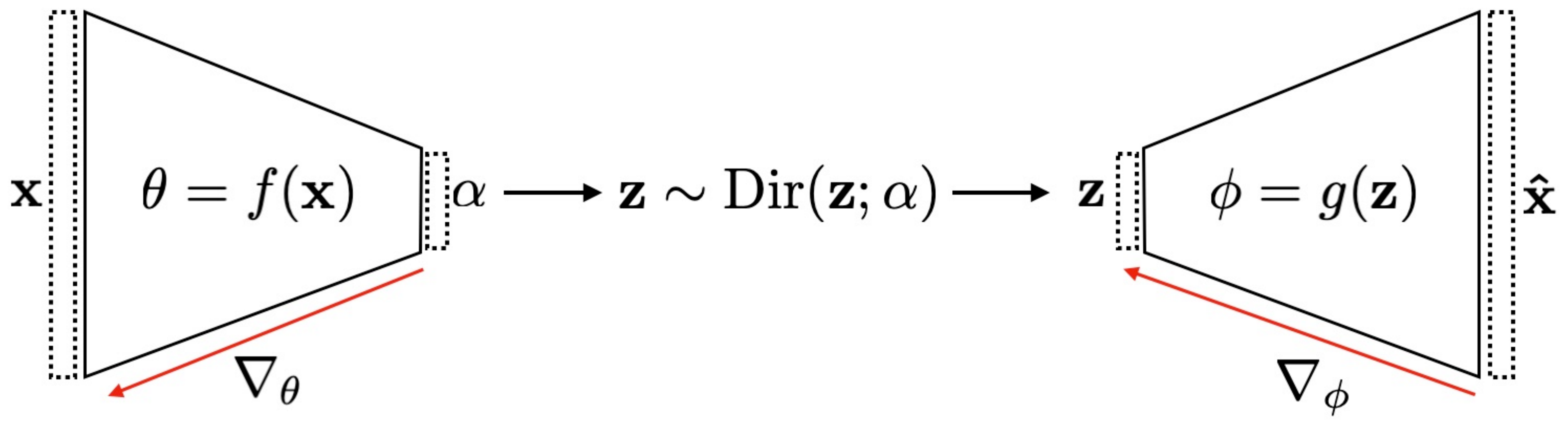}
  }
  \caption{Latent Dirichlet Variational Autoencoder.}
  \label{fig:vae_dirichlet_nobackprop}
\end{figure}

We implemented our model using the VAE architecture as presented in Figure~\ref{fig:vae_dirichlet_nobackprop}.
The encoder function, parameterized by $\theta$, outputs the parameters $\mathbf{\alpha}$ of a dirichlet distribution.
The abundances $\mathbf{z}$ are sampled from the dirichlet distribution and fed to the decoder, which reconstructs the spectral signal $\mathbf{\hat{x}}$.  The decoder is paramterized by $\phi$.
The input $\mathbf{x}$ represents the pixel spectra and $\mathbf{z}$ is a sample from the dirichlet distribution in the n-simplex form.

\begin{figure}
  \centerline{
    \includegraphics[width=0.98\linewidth]{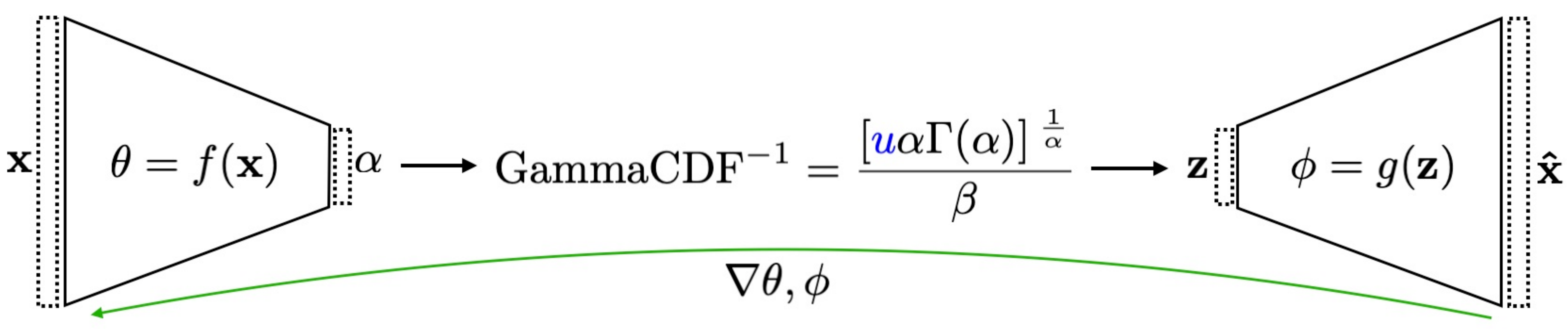}
  }
  \caption{Inverse Gamma Cumulative Distribution Function as a replacement for the sampling function of a Dirichlet probability distribution.}
  \label{fig:vae_dirichlet_backprop}
\end{figure}

The forward pass includes generating a sample from a dirichlet distribution.
However, sampling is not differentiable, which prevents the backpropagation of gradients $\nabla_{\theta}$ and $\nabla_{\phi}$.
Therefore, we need to apply the reparameterization trick, similarly explored by Kingma \etal on Multivariate Normal Distribution~\cite{kingma:2014}.
Specifically, for Dirichlet Distribution, we follow the method proposed in~\cite{joo:2020} and apply a reparametrization as follows:

\begin{equation}
  \label{eq:dir_gamma_cdf}
  \mathbf{z} \sim \operatorname{GammaCDF}^{-1} (u, \alpha, \beta) = \frac {\left[ \color{blue}u\color{black} \alpha \Gamma(\alpha) \right] ^{\frac{1}{\alpha}}} {\beta}.
\end{equation}
\noindent The Dirichlet Probability Density function can be recasted as a Multivariate Gamma, so it becames possible to sample from the dirichlet distribution using the Inverse Gamma Cumulative Distribution Function (Equation~\ref{eq:dir_gamma_cdf}).
Here, $\Gamma(.)$ is the Gamma function, $\mathbf{\alpha}$ is the concentration parameter, $\mathbf{\beta}$ is a normalization factor to ensure that the vector $\mathbf{z}$ is in the $n$-simplex form, and $\color{blue}u\color{black} \sim U(0,1)$.

The variational autoencoder is trained using a reconstruction loss and an Evidence Lower Bound (ELBO) loss.
The decoder reconstructs the input spectra given $\mathbf{z}$, i.e., the abundances.
This serves two purposes: 1) the decoder is able to construct spectra given previously unseen combination of abundances and 2) the decoder is able to perform endmember extraction.
Pragmatically, the decoder generates the endmembers; however, we refer to this process as ``endmember extraction'' to align it with the prevalent terminology in the hyperspectral pixel unmixing community.
The intended purpose (1) further implies that the proposed model is capabble of generating synthetic data that mimics the characteristics of the ``real'' data used to train the model.
The model assumes that spectra follows a multivariate Normal distribution as seen below:

\begin{align}
  \label{eq:mvn}
  \mathbf{x}   & \sim \operatorname{Normal}(\mathbf{x};\mathbf{\mu}, \Sigma)\mathrm{\ where}            \\
  \mathbf{x}   & = \{ x_1, x_2, x_3, \ldots, x_k \}\mathrm{,}                         \nonumber         \\
  \mathbf{\mu} & = \{ \mu_1, \mu_2\, \mu_3, \ldots, \mu_k \} \mathrm{,\ and}                  \nonumber \\
  \Sigma       & = \operatorname{diag} ( \sigma_1, \sigma_2, \sigma_3, \ldots \sigma_k ). \nonumber
\end{align}
\noindent Here $k$ denotes the number of spectral bands. Note that in the current setup, each individual band are not correlated, \ie~ $\Sigma$ is a diagonal matrix.

\subsubsection{ELBO Loss}

For variational autoencoders, in addition to minimizing the reconstruction loss during training, the Kullback-Leibler (KL) divergance between the distribution induced by the latent representation and the desired distribution is also minimized during training.
In our setup the latent representation $\mathbf{\alpha}$ parameterizes a dirichlet distribution. The  leads us to the ELBO loss:
\begin{equation}
  \label{eq:elbo}
  \mathcal{L}(\mathbf{x};\mathbf{\theta},\mathbf{\phi}) = \mathbb{E}_{q_{\theta}}  \left[ \log{p_{\phi}(\mathbf{x}|\mathbf{z})} \right] \color{blue} - KL(q_{\theta}(\mathbf{z} | \mathbf{x}) \| p(\mathbf{z}) ) \color{black}.
\end{equation}
For more details on the derivation of the ELBO loss, please see~\cite{cinelli:2021, blei:2017,kingma:2014,fox:2012}.
In Equation~\ref{eq:elbo} the first term on the right-hand-side represents the reconstruction loss.  The second term on the right-hand side of Equation~\ref{eq:elbo} is a tractable KL-divergence term and it represents the divergence between the prior distributions $p(\mathbf{z})$ and the estimated $q_{\theta}(\mathbf{z}|\mathbf{x})$.  Following~\cite{joo:2020}, we re-write the KL term to account for the dirichlet distribution as follows:
\begin{align}
  \begin{split}
    \operatorname{KL}\left[ q(\mathbf{z}|\mathbf{x};\hat{\alpha}) \| p(\mathbf{z};\alpha) \right] & = \sum{\log{\Gamma(\alpha_k)}} - \sum{\log{\Gamma(\hat{\alpha_k})}} \\
    & + \sum{(\hat{\alpha_k} - \alpha_k )} \frac{d}{dx}\ln{\Gamma(\hat{\alpha_k})},
  \end{split}
\end{align}
where $\Gamma(.)$ is the Gamma function, $\alpha$ is the concentration parameter of the Dirichlet prior, and $\hat{\alpha}$ is concentration parameter of the estimated Dirichlet distribution.

\subsubsection{Final Loss function}
The final loss function is
\begin{align}
  \tiny
  \Loss = & \mathbb{E}_{q_{\theta}}  \left[ \log{p_{\phi}(\mathbf{x}\vert\mathbf{z})} \right]          \nonumber \\
          & - \sum{\log{\Gamma(\alpha_k)}} - \sum{\log{\Gamma(\hat{\alpha}_k)}}                        \nonumber \\
          & + \sum{(\hat{\alpha}_k - \alpha_k )} \ln{\hat{\alpha}}-\dfrac{\hat{\alpha}}{2\hat{\alpha}} \nonumber \\
          & + \omega MSE(\mathbf{z}, \mathbf{\hat{z}}),
\end{align}
where the first terms on right hand side are the reconstruction and the derivation of KL divergence for dirichlet probability.
The last term $\omega MSE(\mathbf{z}, \mathbf{\hat{z}})$ is an ancillary term to ensure the sampled parameters is converging while the underlying dirichlet distribution enforce ASC and ANC constraints.
\subsubsection{Transfer Learning}

The proposed model requires pixel-level abundances for training.  We use transfer learning to apply to the model to these scenarios.  It works as follows.  Say, we are given a hyperspectral image $\mathbf{I}_{\text{no\_abundances}}$ alongwith the list of endmembers $\mathbf{e}$ present in this image.
Pixel-level abundance information is missing, so we cannot use $\mathbf{I}_{\text{no\_abundances}}$ for model training.
Instead, we generate a synthetic datacube $\mathbf{I}_{\text{synthetic}}$  that contains that same endmembers as the original image.  Each pixel $i$ in the generated image is $\sum_{j=1}^n a_j^i \mathbf{e}_j$, where $n$ denotes the number of endmembers, $a_j^i \in [0,1]$ and $\sum_j a_j^i = 1$.
The model is trained on $\mathbf{I}_{\text{synthetic}}$, and the trained model is subsequently used to analyze the origianl image $\mathbf{I}_{\text{no\_abundances}}$.
We show that the proposed model is able to exploit this approach to analyze Cuprite dataset where pixel-level abundances are not available.
\section{Evaluation Metrics and Datasets}

We assess our proposed model using the same evaluation methods employed in \cite{zhang:2021} and \cite{palsson:2020}.
We use two metrics: 1) Spectral Angle Distance (SAD) to evaluate endmembers extraction and 2) Root Means Squared Error (RMSE) to evaluate abundance estimation.
We also evaluate spectral reconstruction using MSE (Means Squared Error) and SAD. We consider spectral reconstruction
an important task for our model, because it underpins endmember extraction. We briefly describe SAD, RMSE, and MSE metrics
below. For a detailed treatment of these metrics, we refer the reader to~\cite{deborah:2015, zhu:2017b, borsoi:2019}.

\subsection{Spectral Angle Distance (SAD)}
\label{sec:sad}
The SAD metric is a distance measurement between two spectral signals:
\begin{equation}
    \operatorname{SAD} = \arccos \left(
    \frac {\mathbf{\hat{x}}_{e}^{T} \mathbf{x}_{e}}
    {\| \mathbf{\hat{x}}_{e}^{T} \|  \| \mathbf{x}_{e} \|}
    \right),
\end{equation}
where $\mathbf{\hat{x}}_{e}$ represents the endmember (spectral signal) generated by the decoder stage of LDVAE
and $\mathbf{x}_{e}$ is the reference endmember (ground truth endmember).  SAD is used to compute the accuracy
of endmember extraction.
In our experiments, we used SAD to evaluate the quality of the endmembers extracted for each material present in
the dataset.
The subscript $e$ denotes that these are spectra corresponding to pure pixels.

\subsection{Root Means Square Error (RMSE)}
\label{sec:rmse}
RMSE is used to evaluate abundance estimation accuracy.  Estimated abundances $\mathbf{\hat{z}}$ are generated by the encoder (stage of the LDVAE).  The difference between estimated abundances and the ground truth abundances capture abundance estimation accuracy.  It is computed as follows:
\begin{equation}
    \operatorname{RMSE}_{} =\sqrt{ \frac{1}{N} \sum_{n=1}^{N} ( \mathbf{z}_{n} - \mathbf{\hat{z}}_{n} ) ^2 }.
\end{equation}
Here, $\mathbf{z}$ denote the ground truth abundances.  $N$ denote the number of pixels used in this computation.

\subsection{Means Square Error (MSE)}
\label{sec:mse}
RMSE is used to capture spectra reconstruction accuracy.  It is defined as follows:
\begin{equation}
    \operatorname{MSE} =\frac{1}{N} \sum_{n=1}^{N} (\mathbf{x}_{n} - \mathbf{\hat{x}}_{n} ) ^2.
\end{equation}
Here $\mathbf{\hat{x}}$ representes the reconstructed spectrum, $\mathbf{x}$ represents the input, and $N$
denotes the number of pixels used in this computation.

\subsection{Datasets}

We evaluate the performance of our model on one synthetic dataset and three real HSI images: 1) OnTech-HSI-Syn-21 Synthetic Dataset 2) Cuprite HSI dataset~\cite{cuprite_dataset}, 3) Urban (HYDICE) dataset~\cite{hydice_urban_dataset}, and 4) Samson dataset~\cite{samson_dataset}.  Cuprite, Urban (HYDICE) and Samson are widely-used HSI benchmarks and these allow us to compare our model with existing schemes.

\subsubsection{OnTech-HSI-Syn-21 Synthetic Dataset}
\begin{figure}
    \centerline{
        \includegraphics[width=.98\linewidth]{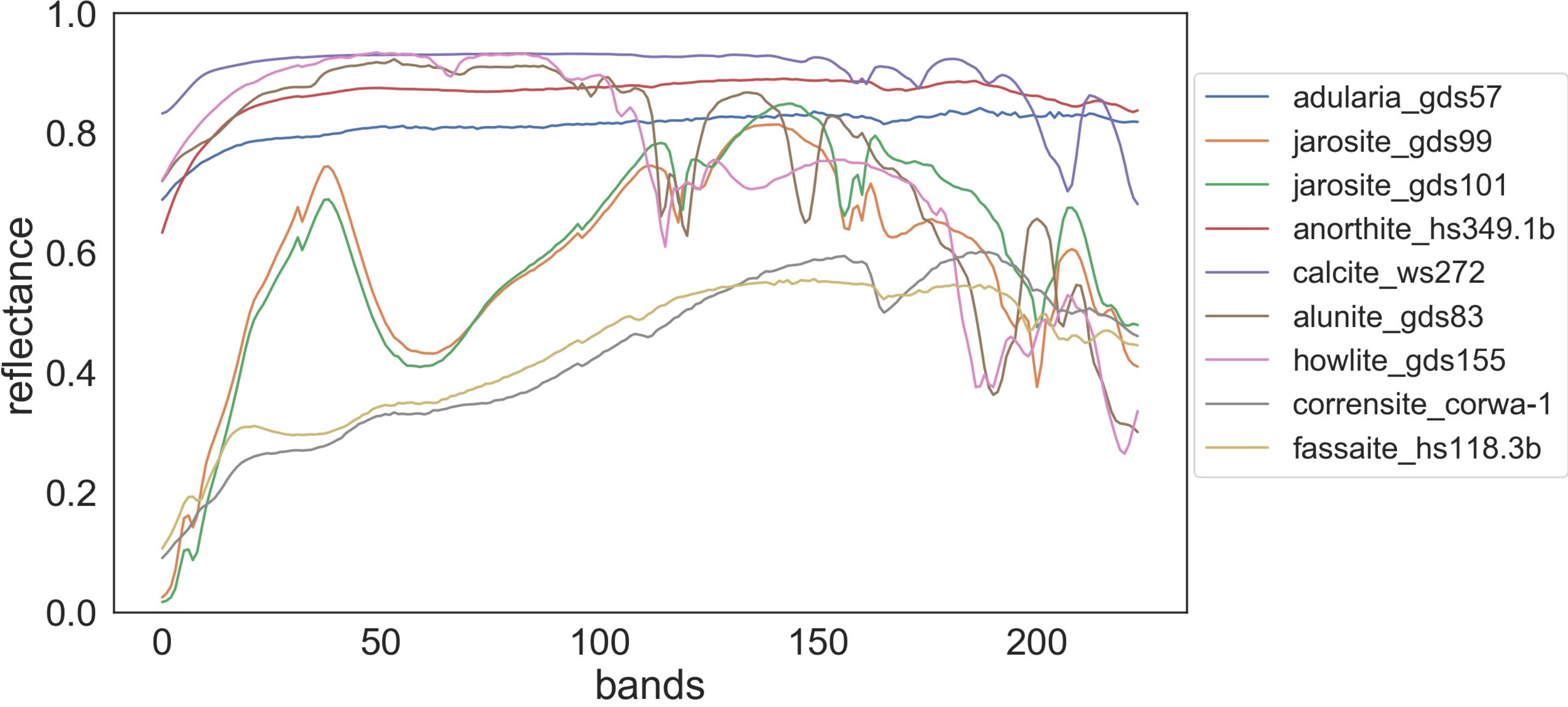}
    }
    \caption{Endmember spectra (taken from USGS spectral library) used to generate OnTech-HSI-Syn-21 dataset.  Each pixel represents a linear combination of these spectra where mixing coefficients are randomly drawn non-negative numbers that sum to one.}
    \label{fig:synthetic_endmembers_test}
\end{figure}

We generated two $128 \times 128$, $224$ channels hyperspectral images.  One image was used for model training and the second image was used solely for model testing and evaluation.  These images contain nine endmembers: {\it Adularia GDS57}, {\it Jarosite GDS99}, {\it Jarosite GDS101}, {\it Anorthite HS349.1B}, {\it Calcite WS272}, {\it Alunite GDS83}, {\it Howlite GDS155}, {\it Corrensite CorWa-1}, and {\it Fassaite HS118.3B}.
The spectra for these endmembers were taken from USGS spectral library.  A similar approach is used by \cite{zhang:2021} and~\cite{li:2021}; however, the datacubes used in those works are not publicly available.
Figure~\ref{fig:synthetic_endmembers_test} show the endmembers spectra used to generate these datacubes.
We refer to this dataset as OnTech-HSI-Syn-21 and we will make these available to the research community.

\subsubsection{Cuprite and Cuprite-Synthetic Datasets}

The Cuprite HSI dataset covers a region around Las Vegas, Nevada, US and comprises a $512 \times 614$, $188$-channel
hyperspectral image.
The area under observation contains twelve minerals (or, for our purposes, endmembers): {\it Alunite},
{\it Andradite}, {\it Buddingtonite}, {\it Dumortierite}, {\it Kaolinite1}, {\it Kaolinite2}, {\it Muscovite},
{\it Montmorillonite}, {\it Nontronite}, {\it Pyrope}, {\it Sphene}, and {\it Chalcedony}.  Cuprite dataset lacks pixel-level abundance information that the proposed model needs for training.  We explored \textbf{transfer learning} to deal with this issue.  We constructed a Cuprite Synthetic dataset that uses the same materials as those found in the Cuprite dataset.  The spectra for these materials were taken from USGS spectral library.  The model was trained on Cuprite Synthetic dataset only, and the trained model was used subsequently used to analyze the original Cuprite dataset.  The results presented in Section~\ref{subsubsec:results_cuprite} showcase the applicability and usefulness of using a model trained on synthetic data to analyze real data.

\subsubsection{Urban (HYDICE) Dataset}
Urban (HYDICE) dataset comprises a $307 \times 307$, $162$-channel hyperspectral image covering a $2 \times 2 \mathrm{m}^2$ region.
This dataset is available in three versions containing four, five, and six endmembers, respectively.
In this work we use the version containing six endmembers.
Further information about this dataset is available in~\cite{zhang:2021}.
We use a $50/50$ training and evaluation split.

\subsubsection{Samson Dataset}

Samson dataset comprises a $95 \times 95$, $156$-channel hyperspectral image.  This dataset contains three endmembers: soil, tree, and water.  We use $80/20$ training and evaluation split.

\section{Experimental Setup and Results}
\label{sec:results}

\begin{table}
  \centering
  \caption{Statistics of the reconstruction errors using SAD and MSE to capture the differences between input pixels and the respective reconstructed signal.}
  \label{table:reconstruction_error_sad_mse}
  \begin{adjustbox}{width=0.98\linewidth}
    \begin{tabular}{l|rr|rr|rr|rr}
      \toprule
           & \multicolumn{2}{c}{OnTech-HSI-Syn-21} & \multicolumn{2}{c}{Cuprite} & \multicolumn{2}{c}{HYDICE Urban} & \multicolumn{2}{c}{Samson}                                     \\
           & SAD                                   & MSE                         & SAD                              & MSE                        & SAD    & MSE    & SAD    & MSE    \\
      \midrule
      mean & 0.0349                                & 0.0031                      & 0.0904                           & 0.0120                     & 0.0833 & 0.0010 & 0.1241 & 0.0107 \\
      std  & 0.0578                                & 0.0023                      & 0.0367                           & 0.0035                     & 0.0799 & 0.0004 & 0.1322 & 0.0044 \\
      min  & 0.0013                                & 0.0007                      & 0.0377                           & 0.0031                     & 0.0125 & 0.0004 & 0.0113 & 0.0002 \\
      25\% & 0.0063                                & 0.0011                      & 0.0671                           & 0.0101                     & 0.0380 & 0.0007 & 0.0429 & 0.0082 \\
      50\% & 0.0100                                & 0.0022                      & 0.0779                           & 0.0120                     & 0.0556 & 0.0009 & 0.0647 & 0.0114 \\
      75\% & 0.0204                                & 0.0049                      & 0.0988                           & 0.0147                     & 0.0916 & 0.0013 & 0.1574 & 0.0135 \\
      max  & 0.2850                                & 0.0101                      & 0.2592                           & 0.0185                     & 0.9507 & 0.0019 & 1.0919 & 0.0197 \\
      \bottomrule
    \end{tabular}
  \end{adjustbox}
\end{table}

\begin{figure}
    \centering
    \includegraphics[width=0.98\linewidth, trim={0cm 0cm 0cm 0cm},clip]{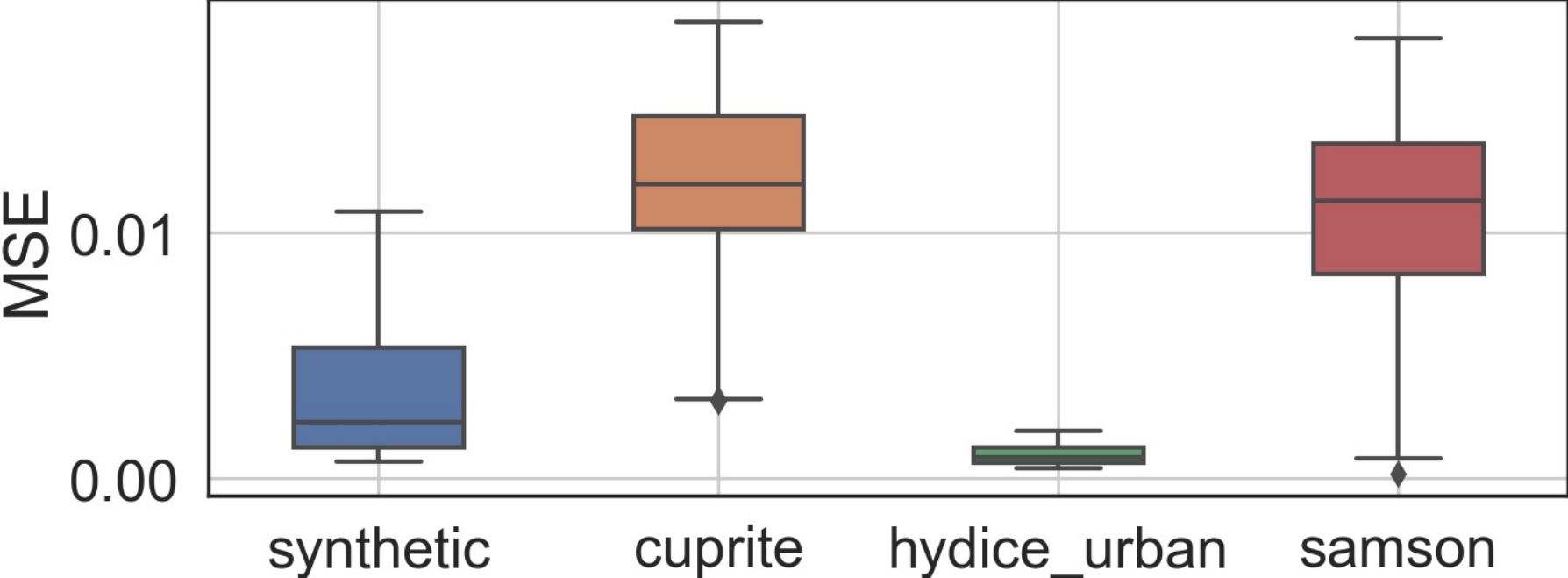}
    \includegraphics[width=0.98\linewidth, trim={0cm 0cm 0cm 0cm},clip]{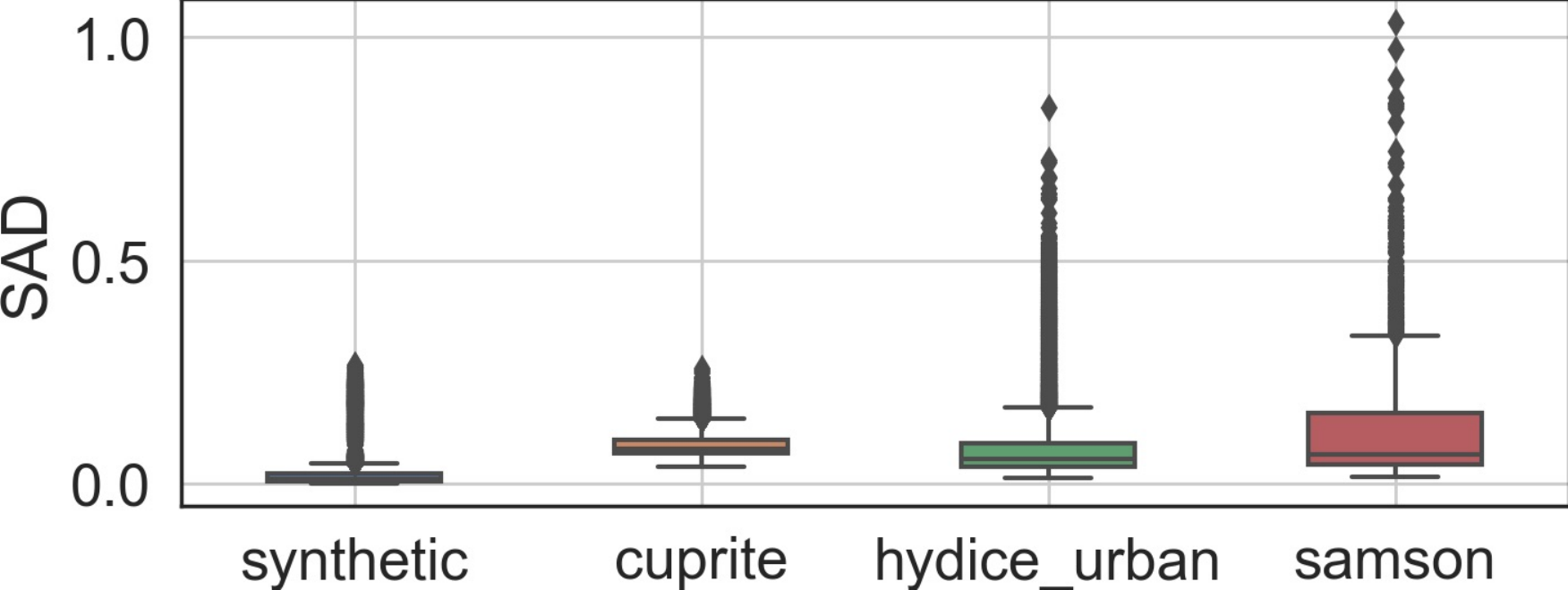}
    \caption{Reconstruction errors for all datasets. SAD and MSE measures capture the differences between pixels and their reconstructions.}
    \label{fig:reconstruction_error_boxplots}
\end{figure}

We study the proposed model on three tasks: 1) signal reconstruction, 2) endmember extraction,
and 3) abundance estimation.  The first task is important to confirm that LDVAE is able to reconstruct
the input spectrum from its latent state.  Tasks 2 and 3 together capture the performance of the proposed
model on the task of hyperspectral pixel unmixing.

\subsection{Spectral Reconstruction}

\begin{figure*}
    \centerline{
        \includegraphics[width=.8\linewidth]{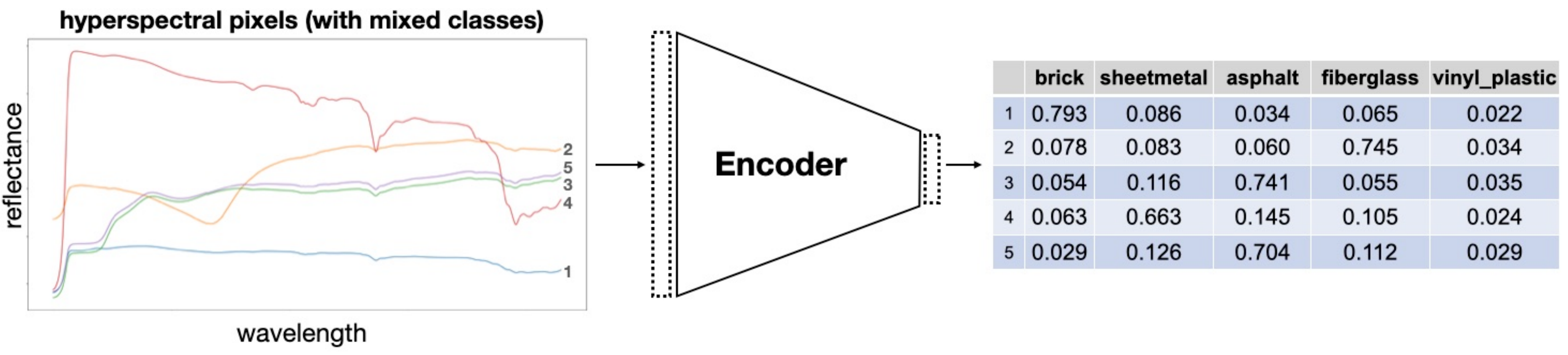}
    }
    \centerline{
        \includegraphics[width=.8\linewidth]{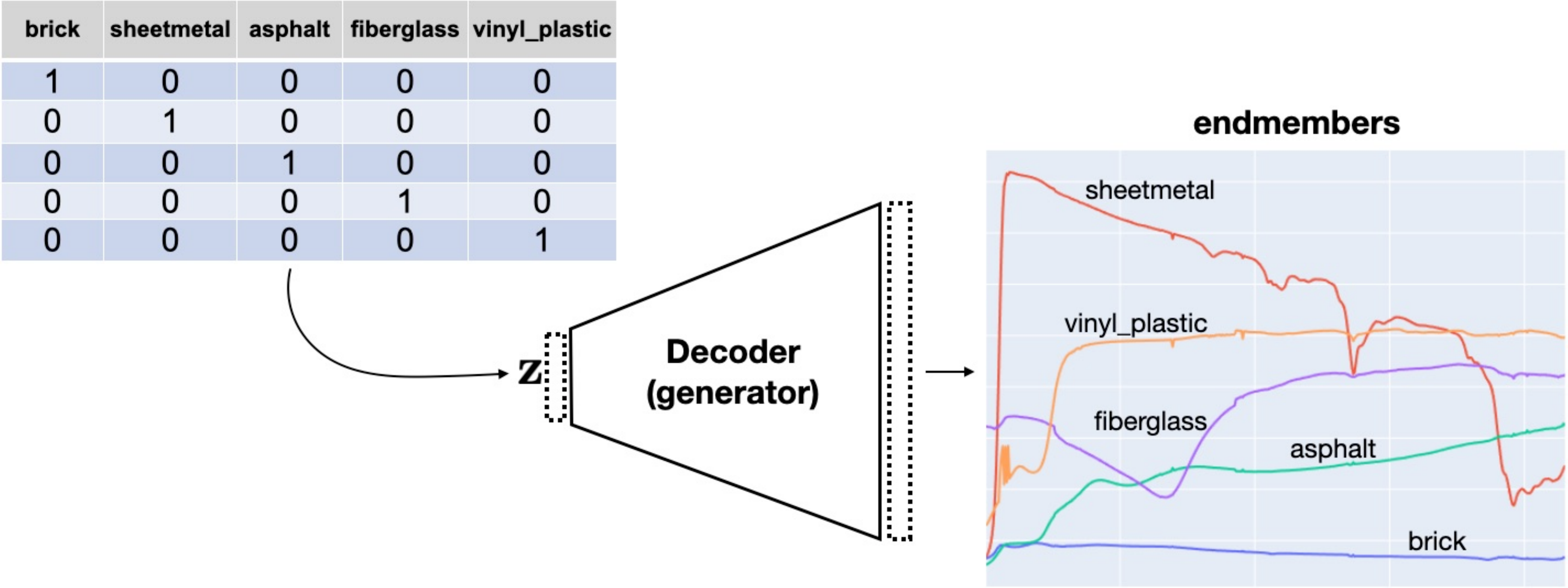}
    }
    \caption{Abundance estimation and endmember extraction using the proposed method. LDVAE encoder estimate abundances for a given input pixel (spectrum) and LDVAE decoder stage extracts endmembers given a on-hot-encoded abundance vector.}
    \label{fig:ldvae-in-action}
\end{figure*}

Table~\ref{table:reconstruction_error_sad_mse} lists reconstruction errors using SAD and MSE metrics on OnTech-HSI-Syn-21, Cuprite, Urban (HYDICE), and Samson datasets. These results confirm that the proposed model is able to reconstruct the input spectra from its latent state.  Recall that the latent state parameterizes a Dirichlet distribution that encodes abundances.  Therefore, we claim that it is possible to use the decoder (stage of trained LDVAE) to generate mixed spectra given abundances.

\subsection{Pixel Unmixing}

Pixel unmixing comprises endmember extraction followed by abundance estimation.
For our method, the encoder (stage of the LDVAE) solves abundance estimation for a given input spectrum.
The decoder (stage) is able to reconstruct the endmember given a one-hot-encoded abundance vector.
Figure~\ref{fig:ldvae-in-action} (top) shows abundance estimation.  Here,
five pixels are passed to the encoder that estimates their corresponding abundances.
Pixel spectra are shown on the left, and each row of the table on the right represents abundances for the
corresponding input spectrum.
Figure~\ref{fig:ldvae-in-action}(bottom) shows endmember extraction.
Five one-hot-encoded abundance vectors are used by the decoder to construct the corresponding endmembers
(shown on the right).
Note that one-hot-encoded abundance vectors represent ``pure materials'' seen in the hyperspectral image.

\subsubsection{Results on OnTech-HSI-Syn-21 Dataset}

\begin{table*}
    \raggedright
\caption{Endmember extraction results for OnTech-HSI-Syn-21 dataset. We use SAD
  metric to evaluate the distance of extracted endmembers from ground truth endmembers.}
\label{table:results-synthetic-nosnr-sad}
\begin{adjustbox}{width=\textwidth}
  \begin{tabular}{lcccccccccccc}
    \toprule
    SNR   & \textbf{LDVAE}   & SSWNMF~\cite{zhang:2021} & SGSNMF~\cite{wang:2017} & TV-RSNMF~\cite{he:2017} & RSNMF~\cite{he:2017} & GLNMF~\cite{lu:2013} & L1/2-NMF~\cite{qian:2011} & VCA+FCLS~\cite{nascimento:2005a} \\
    \midrule
    20 dB & 0.0224 $\pm$0.01 & 0.0636 $\pm$0.40         & 0.0782 $\pm$0.50        & 0.0679 $\pm$0.30        & 0.0731 $\pm$0.50     & 0.0724 $\pm$0.05     & 0.0744 $\pm$0.40          & 0.1358 $\pm$0.30                 \\
    30 dB & 0.0138 $\pm$0.01 & 0.0122 $\pm$0.01         & 0.0176 $\pm$0.03        & 0.0131 $\pm$0.03        & 0.0138 $\pm$0.05     & 0.0144 $\pm$0.04     & 0.0142 $\pm$0.04          & 0.0350 $\pm$0.06                 \\
    40 dB & 0.0081 $\pm$0.00 & 0.0029 $\pm$0.02         & 0.0033 $\pm$0.03        & 0.0036 $\pm$0.02        & 0.0041 $\pm$0.04     & 0.0044 $\pm$0.05     & 0.0037 $\pm$0.04          & 0.0125 $\pm$0.05                 \\
    50 dB & 0.0082 $\pm$0.00 & 0.0012 $\pm$0.02         & 0.0019 $\pm$0.02        & 0.0014 $\pm$0.03        & 0.0020 $\pm$0.04     & 0.0023 $\pm$0.04     & 0.0024 $\pm$0.03          & 0.0049 $\pm$0.06                 \\
    INF   & 0.0069 $\pm$0.00 & \bfseries -              & \bfseries -             & \bfseries -             & \bfseries -          & \bfseries -          & \bfseries -               & \bfseries -                      \\
    \bottomrule
  \end{tabular}
\end{adjustbox}
\end{table*}

\begin{table*}
    \centering
\caption{Abundances estimation results for OnTech-HSI-Syn-21 dataset. We use RMSE
  metric to evaluate the distance of estimated abundances vectors and ground truth abundances vectors.}
\label{table:results-synthetic-nosnr-rmse}
\begin{adjustbox}{width=\textwidth}
  \begin{tabular}{lcccccccccccc}
    \toprule
    \textbf{SNR}   & \textbf{LDVAE}   & SSWNMF~\cite{zhang:2021} & SGSNMF~\cite{wang:2017} & TV-RSNMF~\cite{he:2017} & RSNMF~\cite{he:2017} & GLNMF~\cite{lu:2013} & L1/2-NMF~\cite{qian:2011} & VCA+FCLS~\cite{nascimento:2005a} \\
    \midrule
    \textbf{20 dB} & 0.0052 $\pm$0.00 & 0.1339 $\pm$0.20         & 0.1322 $\pm$0.40        & 0.1342 $\pm$0.30        & 0.1426 $\pm$0.40     & 0.1434 $\pm$0.60     & 0.1430 $\pm$0.50          & 0.1704 $\pm$0.03                 \\
    \textbf{30 dB} & 0.0302 $\pm$0.00 & 0.0386 $\pm$0.20         & 0.0391 $\pm$0.30        & 0.0420 $\pm$0.20        & 0.0426 $\pm$0.30     & 0.0429 $\pm$0.03     & 0.0432 $\pm$0.20          & 0.0548 $\pm$0.20                 \\
    \textbf{40 dB} & 0.0303 $\pm$0.00 & 0.0122 $\pm$0.03         & 0.0148 $\pm$0.05        & 0.0142 $\pm$0.04        & 0.0147 $\pm$0.05     & 0.0150 $\pm$0.04     & 0.0153 $\pm$0.03          & 0.0164 $\pm$0.10                 \\
    \textbf{50 dB} & 0.0303 $\pm$0.00 & 0.0041 $\pm$0.02         & 0.0059 $\pm$0.05        & 0.0050 $\pm$0.03        & 0.0055 $\pm$0.03     & 0.0064 $\pm$0.04     & 0.0061 $\pm$0.04          & 0.0087 $\pm$0.08                 \\
    \textbf{INF  } & 0.0052 $\pm$0.00 & \bfseries -              & \bfseries -             & \bfseries -             & \bfseries -          & \bfseries -          & \bfseries -               & \bfseries -                      \\
    \bottomrule
  \end{tabular}
\end{adjustbox}

\end{table*}

Table~\ref{table:results-synthetic-nosnr-sad} shows endmember extraction results for the
OnTech-HSI-Syn-21 dataset at different Signal-to-Noise Ratios (SNRs).
Similarly, Table~\ref{table:results-synthetic-nosnr-rmse} shows classification results (abundances) for the
OnTech-HSI-Syn-21 dataset at different SNRs.

\subsubsection{Results on Cuprite Dataset}
\label{subsubsec:results_cuprite}

\begin{table*}
    \raggedright

\caption{Endmember extraction results for Cuprite dataset. We use SAD metric to evaluate the distance of extracted endmembers from ground truth endmembers.}
\label{table:results-cuprite-sad}
\begin{adjustbox}{width=\textwidth}
    \begin{tabular}{ccccccccccccc}
        \toprule
                         & \textbf{LDVAE}   & SSWNMF\cite{zhang:2021} & SGSNMF\cite{wang:2017} & TV-RSNMF~\cite{he:2017} & RSNMF~\cite{he:2017} & GLNMF~\cite{lu:2013} & L1/2-NMF~\cite{qian:2011} & VCA+FCLS~\cite{nascimento:2005a} \\
        \midrule
        alunite          & 0.0097 $\pm$0.01 & 0.1497 $\pm$3.97        & 0.1238 $\pm$4.01       & 0.1204 $\pm$4.37        & 0.1189 $\pm$4.39     & 0.1353 $\pm$3.83     & 0.1496 $\pm$3.32          & 0.1574 $\pm$3.71                 \\
        andradite        & 0.0381 $\pm$0.04 & \bfseries -             & \bfseries -            & \bfseries -             & \bfseries -          & \bfseries -          & \bfseries -               & \bfseries -                      \\
        buddingtonite    & 0.0051 $\pm$0.01 & 0.0958 $\pm$4.69        & 0.1021 $\pm$3.47       & 0.0903 $\pm$5.08        & 0.1342 $\pm$4.72     & 0.1437 $\pm$3.62     & 0.1441 $\pm$4.16          & 0.1412 $\pm$3.74                 \\
        dumortierite     & 0.1922 $\pm$0.19 & \bfseries -             & \bfseries -            & \bfseries -             & \bfseries -          & \bfseries -          & \bfseries -               & \bfseries -                      \\
        kaolinite-1      & 0.0258 $\pm$0.03 & 0.0885 $\pm$2.94        & 0.0986 $\pm$3.18       & 0.1097 $\pm$3.47        & 0.0955 $\pm$3.07     & 0.0967 $\pm$4.01     & 0.0825 $\pm$4.66          & 0.0736 $\pm$4.42                 \\
        kaolinite-2      & 0.0699 $\pm$0.07 & 0.1206 $\pm$3.67        & 0.1375 $\pm$3.48       & 0.1213 $\pm$3.82        & 0.1396 $\pm$4.11     & 0.1356 $\pm$3.91     & 0.1402 $\pm$4.18          & 0.1420 $\pm$4.16                 \\
        muscovite        & 0.0064 $\pm$0.01 & 0.1024 $\pm$4.24        & 0.1061 $\pm$3.18       & 0.1131 $\pm$2.88        & 0.0997 $\pm$3.46     & 0.0961 $\pm$3.77     & 0.0889 $\pm$3.03          & 0.1007 $\pm$3.31                 \\
        montmorillonite  & 0.0496 $\pm$0.05 & 0.0651 $\pm$3.08        & 0.0705 $\pm$3.36       & 0.0783 $\pm$3.95        & 0.0744 $\pm$3.12     & 0.0838 $\pm$4.28     & 0.0876 $\pm$2.91          & 0.0974 $\pm$3.39                 \\
        nontronite       & 0.1048 $\pm$0.10 & 0.1138 $\pm$4.15        & 0.1046 $\pm$3.80       & 0.0911 $\pm$3.49        & 0.0832 $\pm$4.18     & 0.0953 $\pm$3.41     & 0.1038 $\pm$4.46          & 0.0772 $\pm$2.10                 \\
        pyrope           & 0.0156 $\pm$0.02 & 0.1106 $\pm$3.32        & 0.1208 $\pm$3.83       & 0.1253 $\pm$3.10        & 0.1469 $\pm$3.12     & 0.1318 $\pm$3.18     & 0.1123 $\pm$4.91          & 0.1437 $\pm$3.76                 \\
        sphene           & 0.0347 $\pm$0.03 & 0.1024 $\pm$3.79        & 0.1179 $\pm$4.02       & 0.1190 $\pm$2.97        & 0.1134 $\pm$2.54     & 0.1291 $\pm$4.21     & 0.1252 $\pm$5.18          & 0.1277 $\pm$4.08                 \\
        chalcedony       & 0.0055 $\pm$0.01 & 0.1496 $\pm$4.12        & 0.1221 $\pm$4.02       & 0.1387 $\pm$4.01        & 0.1224 $\pm$4.19     & 0.1341 $\pm$2.98     & 0.1520 $\pm$3.43          & 0.1514 $\pm$3.83                 \\
        \midrule
        \textit{average} & 0.0465 $\pm$0.05 & 0.1099 $\pm$3.80        & 0.1104 $\pm$3.63       & 0.1107 $\pm$3.71        & 0.1128 $\pm$3.69     & 0.1182 $\pm$3.72     & 0.1186 $\pm$4.02          & 0.1212 $\pm$3.65                 \\
        \bottomrule
    \end{tabular}
\end{adjustbox}

\end{table*}

Table~\ref{table:results-cuprite-sad} lists endmember extraction results for Cuprite dataset.  The results suggest that our approach is able to handle situations where pixel-level abundance data is not available for training by leveraging the {\it transfer learning} paradigm.  Note also that endmember extraction results for our method are similar to those achieved by competing approaches.  These results also confirm that our method is applicable to real-world scenarios where oftentimes pixel-level abundance information is hard to collect.

\subsubsection{Results on Urban (HYDICE) Dataset}
Table~\ref{table:results-hydice-urban-sad} presents the results of endmember extraction obtained from the known Urban Dataset, frequently used as benchmarks for HSI unmixing. Table ~\ref{table:results-hydice-urban-rmse} shows the classification results.
\begin{table*}
    \raggedright
\caption{Endmember extraction results for HYDICE Urban dataset. We use SAD metric to evaluate the distance of extracted endmembers from ground truth endmembers.}
\label{table:results-hydice-urban-sad}

\begin{adjustbox}{width=\textwidth}
  \begin{tabular}{ccccccccccccc}
    \toprule
                 & \textbf{LDVAE}   & SSWNMF~\cite{zhang:2021} & CNNAEU~\cite{palsson:2020} & SGSRNMF~\cite{palsson:2020} & SHDP~\cite{palsson:2020} & MTLAEU~\cite{palsson:2020} & VCA+FCLS~\cite{nascimento:2005a} \\
    \midrule
    asphalt road & 0.4262 $\pm$0.43 & 0.0782 $\pm$3.29         & 0.0575 $\pm$ 0.0058        & 0.2446 $\pm$0.0204          & 0.2658 $\pm$ 0.0751      & 0.0843 $\pm$ 0.0046        & 0.2246 $\pm$3.44                 \\
    grass        & 0.3323 $\pm$0.33 & 0.1490 $\pm$3.58         & 0.0366 $\pm$ 0.0047        & 1.3006 $\pm$0.0444          & 0.5524 $\pm$ 0.3172      & 0.0421 $\pm$ 0.0036        & 0.1981 $\pm$3.39                 \\
    tree         & 0.3177 $\pm$0.32 & 0.1173 $\pm$3.46         & 0.0321 $\pm$ 0.0039        & 0.0967 $\pm$ 0.0113         & 0.0777 $\pm$ 0.0171      & 0.0539 $\pm$ 0.0039        & 0.2137 $\pm$2.41                 \\
    roof         & 0.4393 $\pm$0.44 & 0.0713 $\pm$3.61         & 0.0332 $\pm$ 0.0066        & 0.1916 $\pm$ 0.0862         & 0.4117 $\pm$ 0.1720      & 0.0415 $\pm$ 0.0045        & 0.2673 $\pm$3.77                 \\
    metal        & 0.7004 $\pm$0.70 & 0.1241 $\pm$2.76         & \bfseries -                & \bfseries -                 & \bfseries -              & \bfseries -                & 0.1848 $\pm$3.68                 \\
    dirt         & 0.2806 $\pm$0.28 & 0.0802 $\pm$3.17         & \bfseries -                & \bfseries -                 & \bfseries -              & \bfseries -                & 0.1992 $\pm$3.43                 \\
    \midrule
    average      & 0.4161 $\pm$0.42 & 0.1034 $\pm$3.31         & 0.0398 $\pm$ 0.0030        & 0.4584 $\pm$ 0.0148         & 0.3269 $\pm$ 0.0555      & 0.0555$\pm$ 0.0019         & 0.2146 $\pm$3.35                 \\
    \bottomrule
  \end{tabular}
\end{adjustbox}

\end{table*}

\begin{table*}
    \raggedright
\caption{Abundances estimation results for Hydice Urban dataset. We use RMSE metric to evaluate the distance of estimated abundances vectors and ground truth abundances vectors.}
\label{table:results-hydice-urban-rmse}

\begin{adjustbox}{width=\textwidth}
  \begin{tabular}{ccccccccccccc}
    \toprule
                 & \textbf{LDVAE}    & SSWNMF~\cite{zhang:2021} & CNNAEU~\cite{palsson:2020} & SGSRNMF~\cite{palsson:2020} & SHDP~\cite{palsson:2020} & MTLAEU~\cite{palsson:2020} & VCA+FCLS~\cite{nascimento:2005a} \\
    \midrule
    asphalt road & 0.2289 $\pm$ 0.00 & \bfseries -              & 0.1249 $\pm$ 0.0400        & 0.2857 $\pm$ 0.0762         & 0.3015 $\pm$ 0.1200      & 0.151658  $\pm$ 0.0316     & \bfseries -                      \\
    grass        & 0.1832 $\pm$ 0.00 & \bfseries -              & 0.1256 $\pm$ 0.0400        & 0.4467 $\pm$ 0.1015         & 0.3847 $\pm$ 0.2691      & 0.15      $\pm$ 0.0400     & \bfseries -                      \\
    tree         & 0.1737 $\pm$ 0.00 & \bfseries -              & 0.0854 $\pm$ 0.0387        & 0.2674 $\pm$ 0.1308         & 0.2886 $\pm$ 0.1533      & 0.0824621 $\pm$ 0.0300     & \bfseries -                      \\
    roof         & 0.1250 $\pm$ 0.00 & \bfseries -              & 0.0854 $\pm$ 0.0387        & 0.1892 $\pm$ 0.0424         & 0.2729 $\pm$ 0.2385      & 0.0888819 $\pm$ 0.0283     & \bfseries -                      \\
    metal        & 0.2599 $\pm$ 0.00 & \bfseries -              & \bfseries -                & \bfseries -                 & \bfseries -              & \bfseries -                & \bfseries -                      \\
    dirt         & 0.1334 $\pm$ 0.00 & \bfseries -              & \bfseries -                & \bfseries -                 & \bfseries -              & \bfseries -                & \bfseries -                      \\
    \midrule
    average      & 0.1840 $\pm$ 0.00 & 0.0048 $\pm$0.72         & 0.1072 $\pm$ 0.0316        & 0.3116 $\pm$ 0.0922         & 0.3150 $\pm$ 0.1428      & 0.122474 $\pm$ 0.0283      & 0.0119 $\pm$ 0.66                \\
    \bottomrule
  \end{tabular}
\end{adjustbox}

\end{table*}

\subsubsection{Results on Samson Dataset}
Table~\ref{table:results-samson-sad} presents the results of endmember extraction and Table ~\ref{table:results-samson-rmse} shows the classification results obtained from Samson Dataset.

\begin{table*}
    \raggedright
\caption{Endmember extraction results for Samson dataset. We use SAD metric to evaluate distances between extracted and ground truth endmembers.}
\label{table:results-samson-sad}
\begin{adjustbox}{width=\textwidth}
  \begin{tabular}{ccccccccccccc}
    \toprule
            & \textbf{LDVAE}   & SSWNMF~\cite{zhang:2021} & CNNAEU~\cite{palsson:2020} & SGSRNMF~\cite{palsson:2020} & SHDP~\cite{palsson:2020} & MTLAEU~\cite{palsson:2020} \\
    \midrule
    soil    & 0.0959 $\pm$0.10 & \bfseries -              & 0.0373 $\pm$ 0.0210        & 0.0086 $\pm$ 0.0001         & 0.2147 $\pm$ 0.3299      & 0.0225 $\pm$ 0.0060        \\
    tree    & 1.2788 $\pm$1.28 & \bfseries -              & 0.0397 $\pm$ 0.0038        & 0.0395 $\pm$ 0.0019         & 0.0375 $\pm$ 0.0004      & 0.0371 $\pm$ 0.0028        \\
    water   & 0.4022 $\pm$0.40 & \bfseries -              & 0.0430 $\pm$ 0.0092        & 0.0923 $\pm$ 0.0024         & 0.2064 $\pm$ 0.0916      & 0.0338 $\pm$ 0.0031        \\
    \midrule
    average & 0.5923 $\pm$0.59 & \bfseries -              & 0.0400 $\pm$ 0.0067        & 0.0468 $\pm$ 0.0003         & 0.1527 $\pm$ 0.1390      & 0.0311 $\pm$ 0.0017        \\
    \bottomrule
  \end{tabular}
\end{adjustbox}

\end{table*}

\begin{table*}
    \raggedright
\caption{Abundances estimation results for Samson dataset. We use RMSE metric to evaluate the distance of estimated abundances vectors and ground truth abundances vectors.}
\label{table:results-samson-rmse}

\begin{adjustbox}{width=\textwidth}
  \begin{tabular}{ccccccccccccc}
    \toprule
            & \textbf{LDVAE}   & CNNAEU              & SGSRNMF             & SHDP                & MTLAEU              & VCA+FCLS~\cite{nascimento:2005a} & PLMM~\cite{borsoi:2019} & ELMM~\cite{borsoi:2019} & GLMM~\cite{borsoi:2019} & DeepGUn~\cite{borsoi:2019} \\
    \midrule
    soil    & 0.2609 $\pm$0.00 & 0.2766 $\pm$ 0.2600 & 0.1778 $\pm$ 0.0200 & 0.2853 $\pm$ 0.1697 & 0.0872 $\pm$ 0.0374 & \bfseries -                      & \bfseries -             & \bfseries -             & \bfseries -             & \bfseries -                \\
    tree    & 0.3461 $\pm$0.00 & 0.2512 $\pm$ 0.2659 & 0.2400 $\pm$ 0.0557 & 0.2496 $\pm$ 0.1204 & 0.0608 $\pm$ 0.0265 & \bfseries -                      & \bfseries -             & \bfseries -             & \bfseries -             & \bfseries -                \\
    water   & 0.3165 $\pm$0.00 & 0.1288 $\pm$ 0.1153 & 0.3503 $\pm$ 0.0583 & 0.3948 $\pm$ 0.0949 & 0.0539 $\pm$ 0.0316 & \bfseries -                      & \bfseries -             & \bfseries -             & \bfseries -             & \bfseries -                \\
    \midrule
    average & 0.3078 $\pm$0.00 & 0.2283 $\pm$ 0.2119 & 0.2657 $\pm$ 0.0458 & 0.3160 $\pm$ 0.1095 & 0.0693 $\pm$ 0.0283 & 0.0545                           & 0.0239                  & 0.0119                  & 0.0006                  & 0.0862                     \\
    \bottomrule
  \end{tabular}
\end{adjustbox}

\end{table*}

\subsection{Discussion}
Unmixing algorithms perform two tasks: endmember extraction and pixel-level abundance estimation. The proposed method is able to extract endmembers by using its decoder to generate spectra corresponding to one-hot encoded abundance vector.  Recall that one-hot encoded abundance vectors correspond to pure pixels (see Figure~\ref{fig:ldvae-in-action} bottom).  The proposed method performs pixel-level abundance estimation using its encoder that takes in a pixel spectra and outputs the latent space that represents abundances (Figure~\ref{fig:ldvae-in-action} top).   The proposed architecture is inherently non-linear, consequently, we surmise that, it is able to capture the non-linear effects present in the unmixing task.  These effects are more prevalent in micro-spectroscopy images where each material is composed of several elements.  Similarly, these effects are present in low-resolution HSI images captured in a remote sensing setup.

\begin{figure*}
    \centering
    \includegraphics[width=0.103\textwidth,trim={0cm 0cm 0cm 1cm},clip]{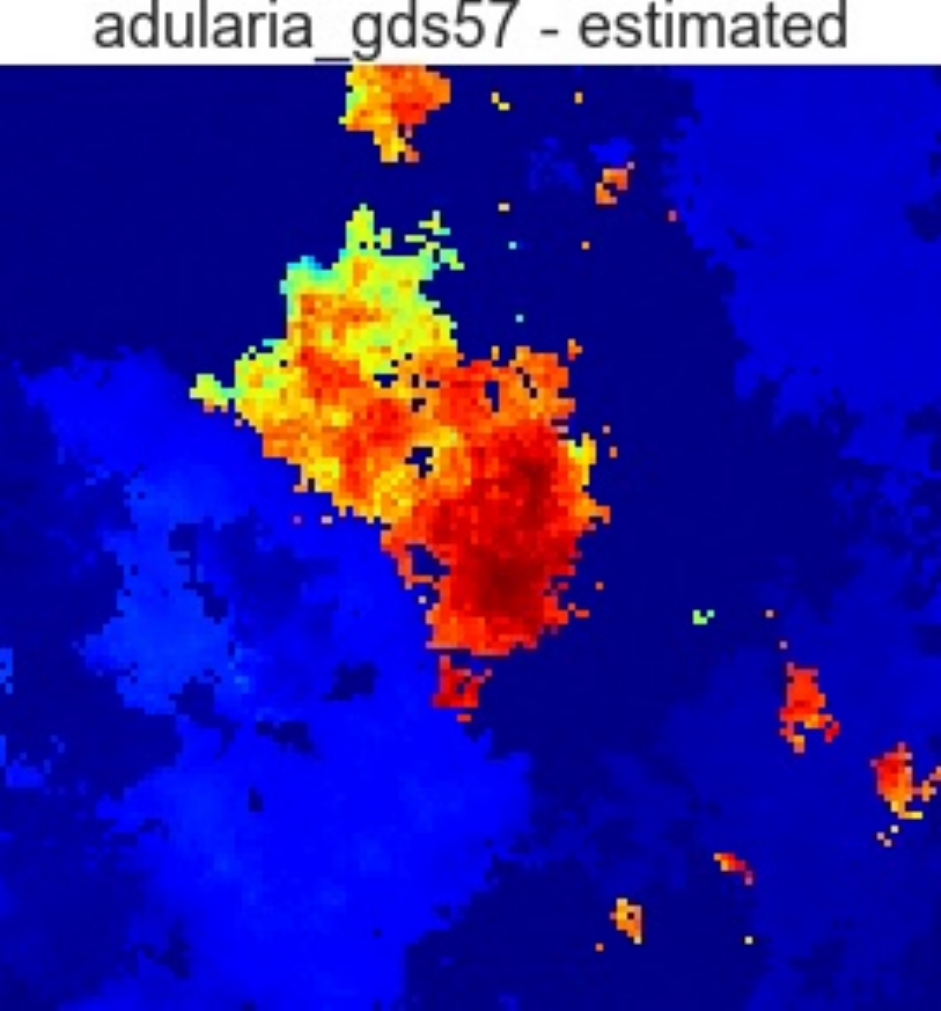}
    \includegraphics[width=0.103\textwidth,trim={0cm 0cm 0cm 1cm},clip]{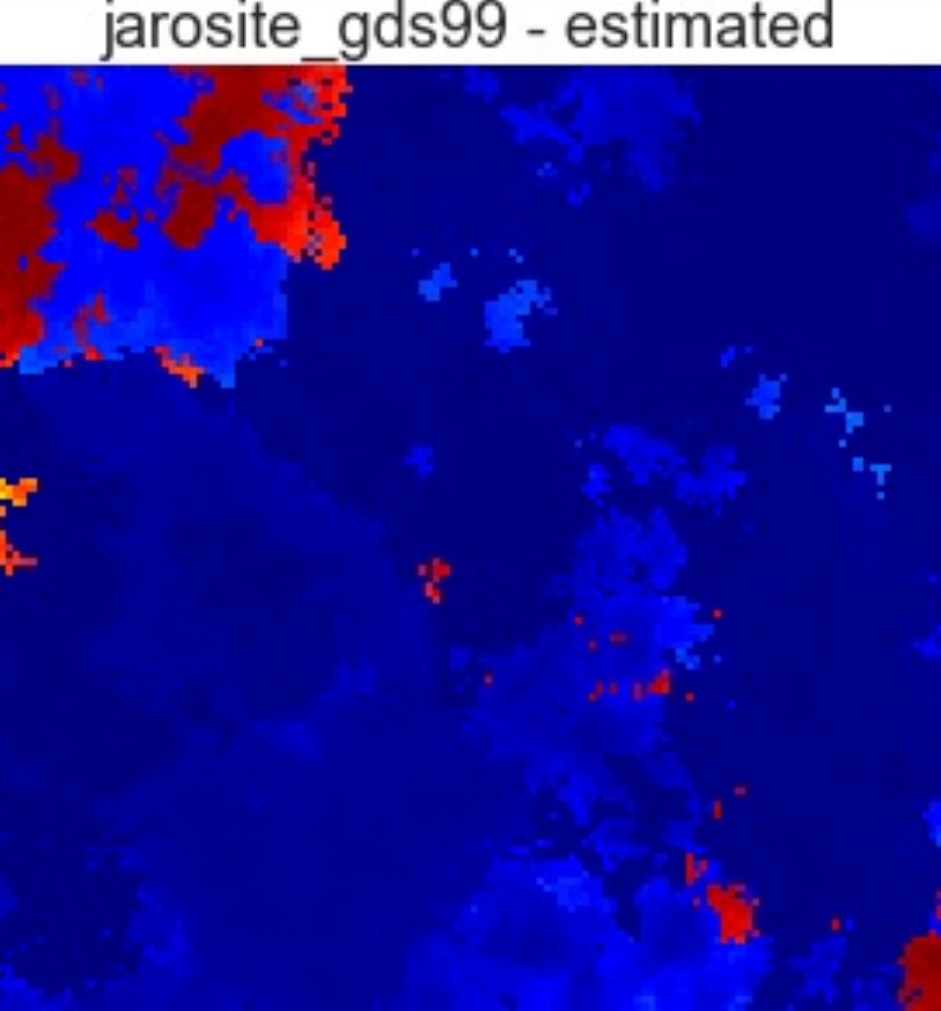}
    \includegraphics[width=0.103\textwidth,trim={0cm 0cm 0cm 1cm},clip]{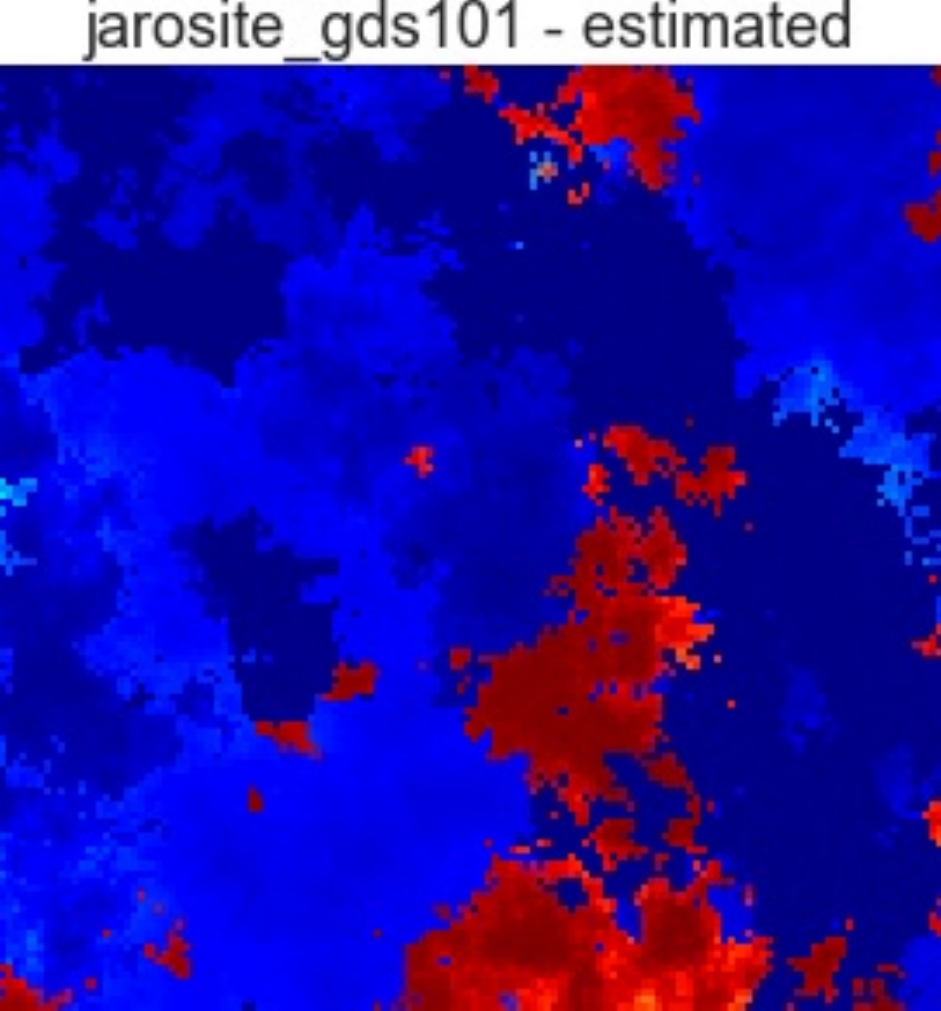}
    \includegraphics[width=0.103\textwidth,trim={0cm 0cm 0cm 1cm},clip]{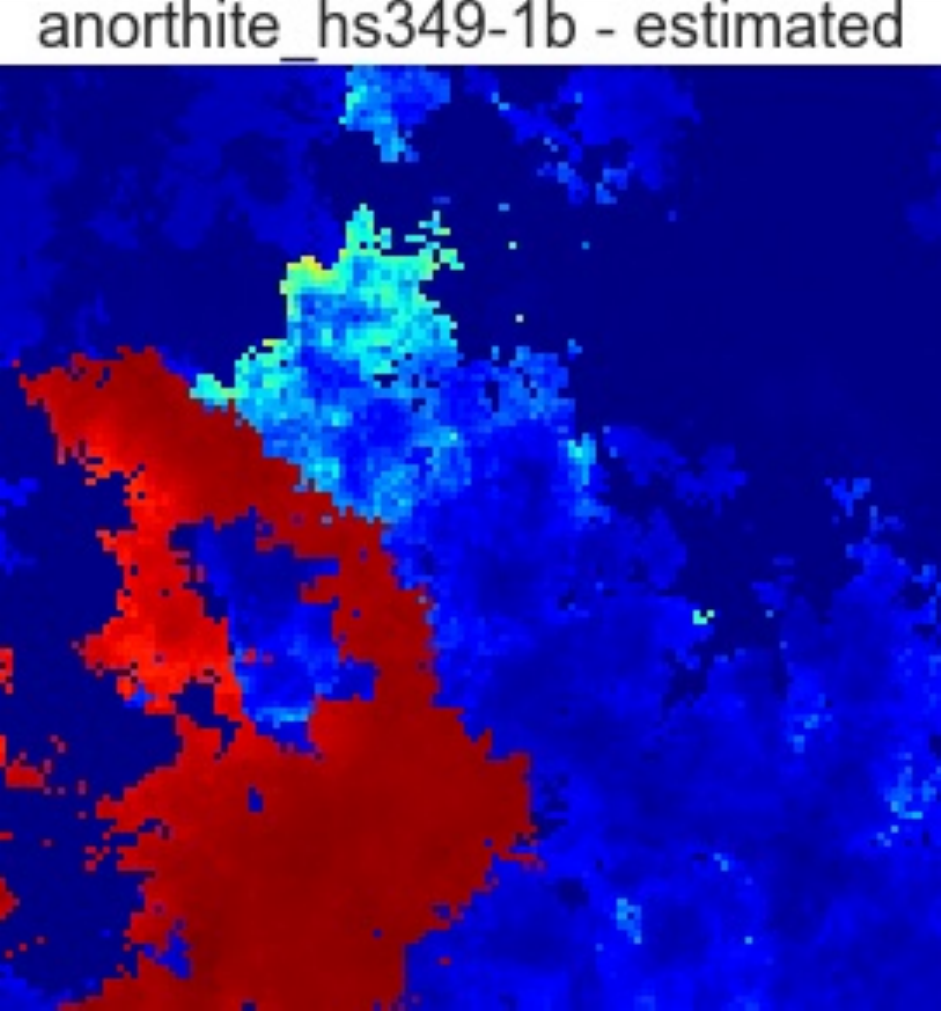}
    \includegraphics[width=0.103\textwidth,trim={0cm 0cm 0cm 1cm},clip]{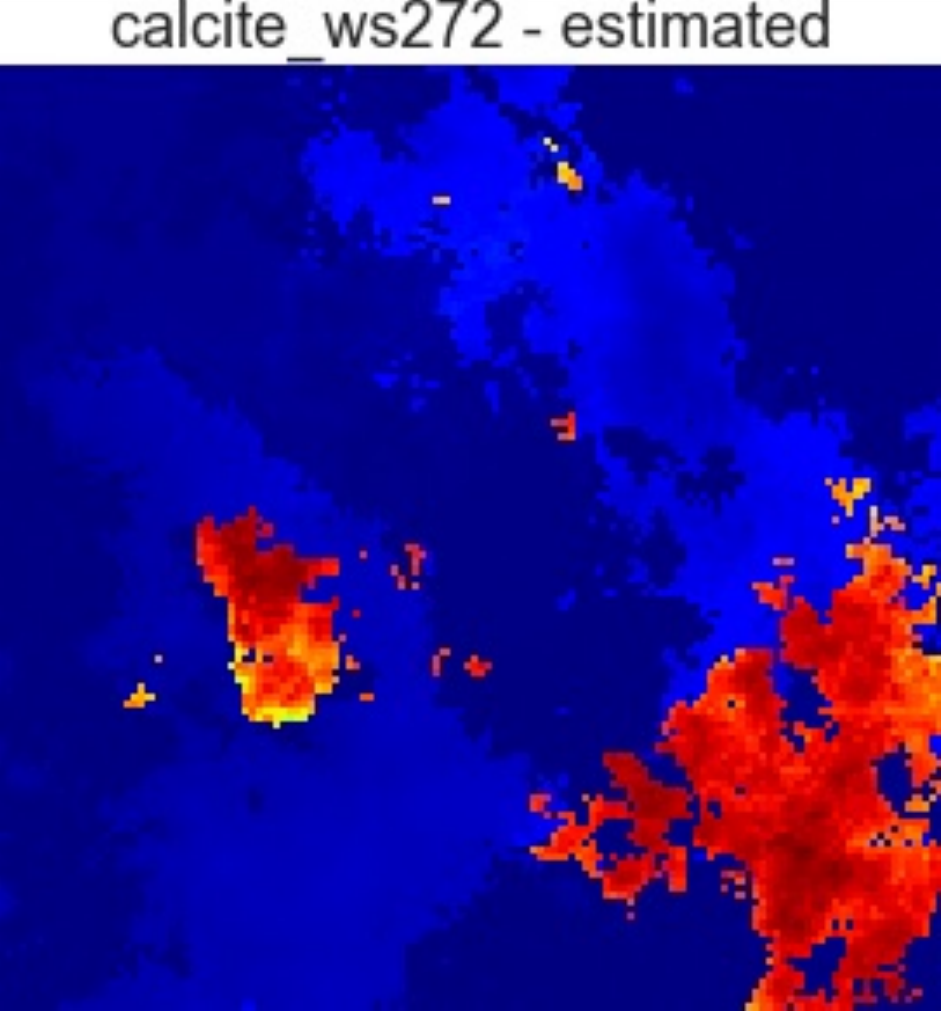}
    \includegraphics[width=0.103\textwidth,trim={0cm 0cm 0cm 1cm},clip]{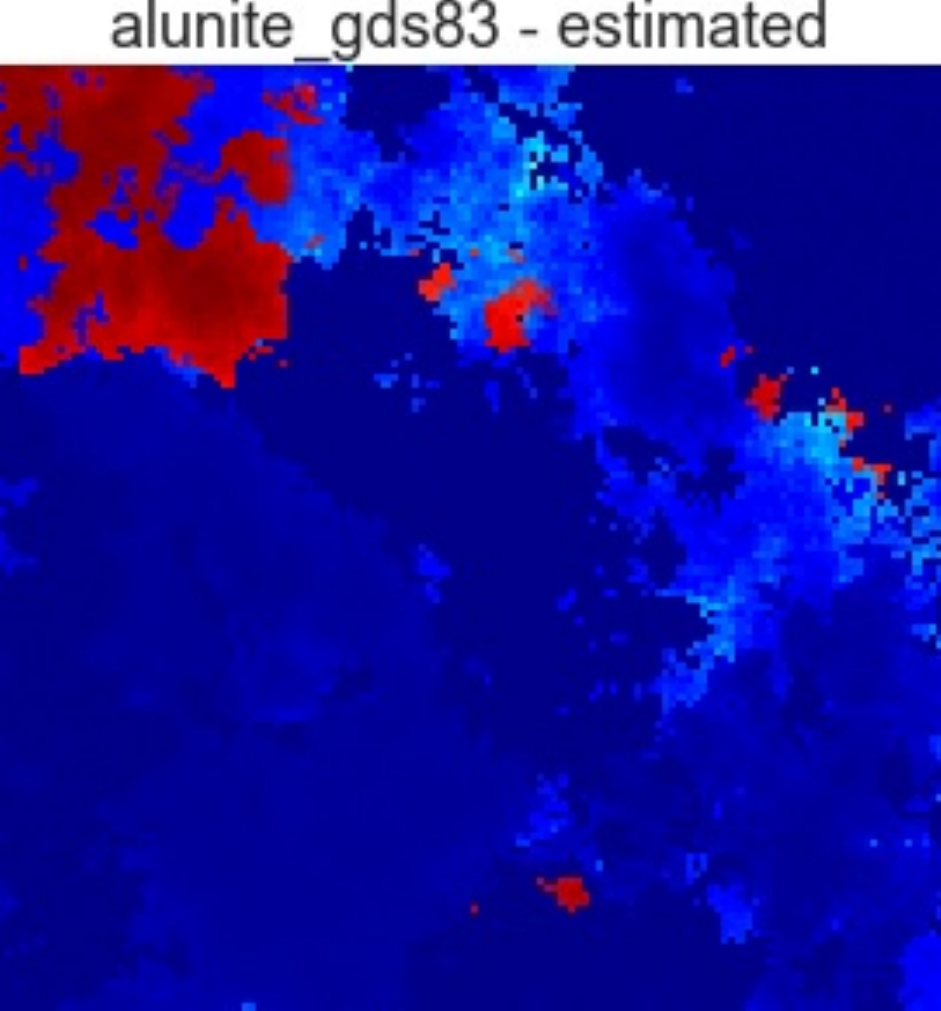}
    \includegraphics[width=0.103\textwidth,trim={0cm 0cm 0cm 1cm},clip]{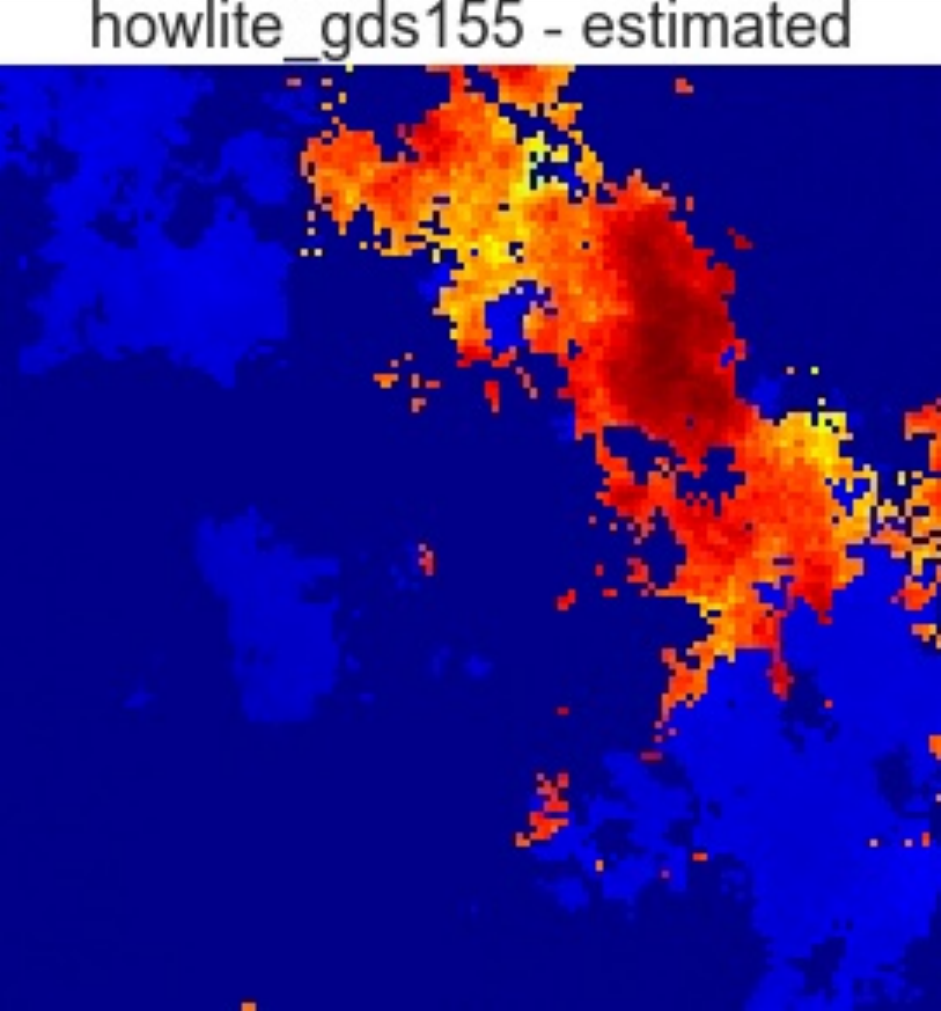}
    \includegraphics[width=0.103\textwidth,trim={0cm 0cm 0cm 1cm},clip]{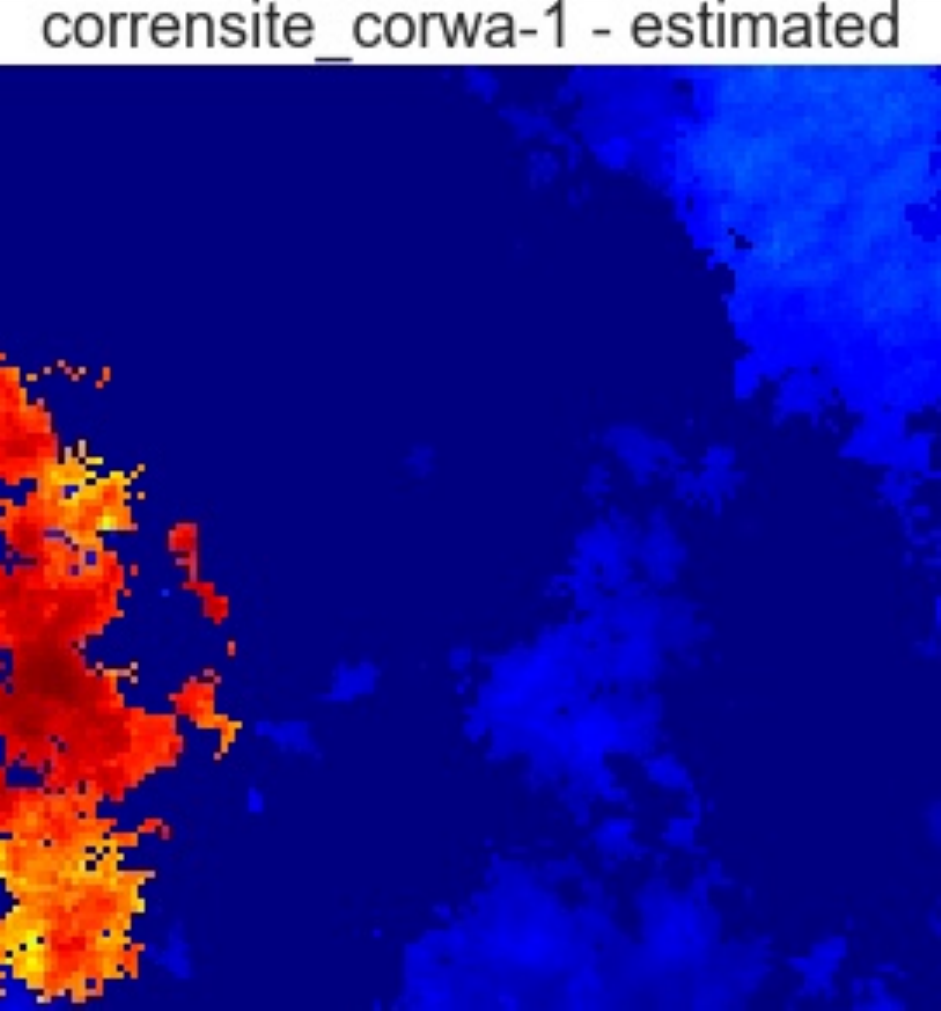}
    \includegraphics[width=0.103\textwidth,trim={0cm 0cm 0cm 1cm},clip]{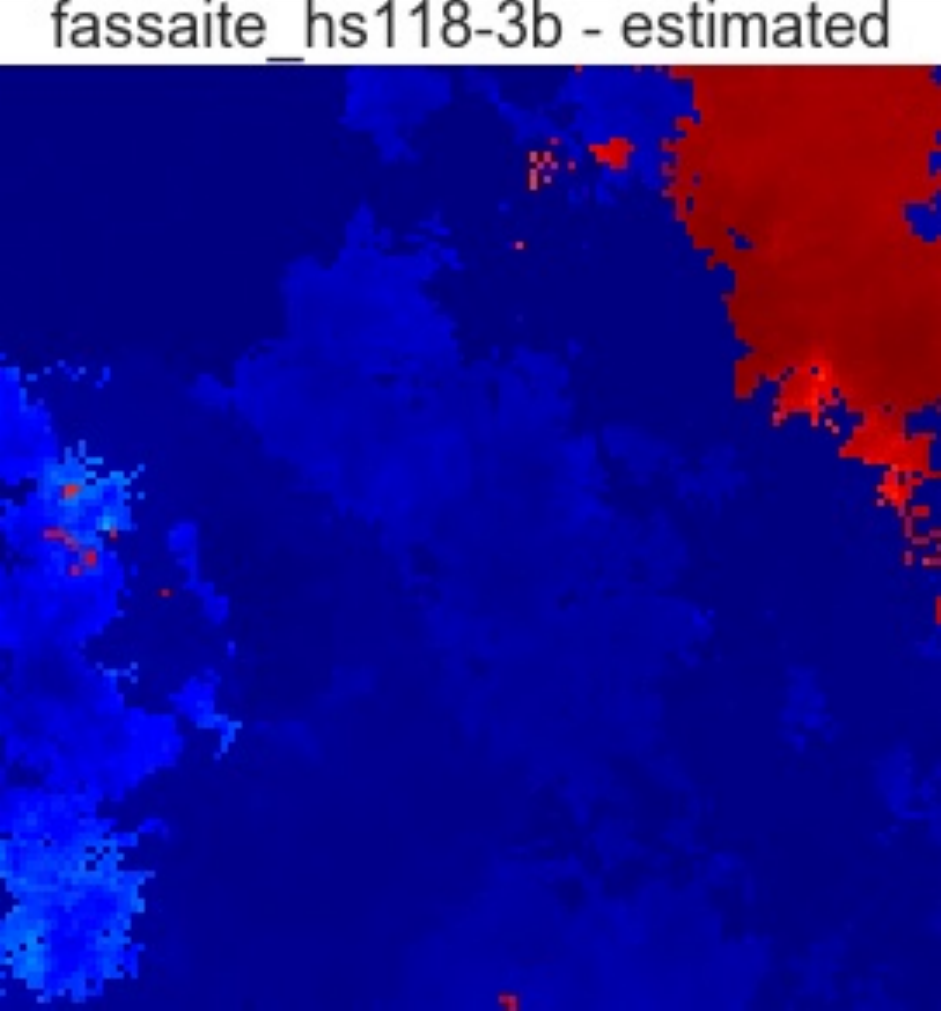}
    \linebreak
    \includegraphics[width=0.103\textwidth,trim={0cm 0cm 0cm 1cm},clip]{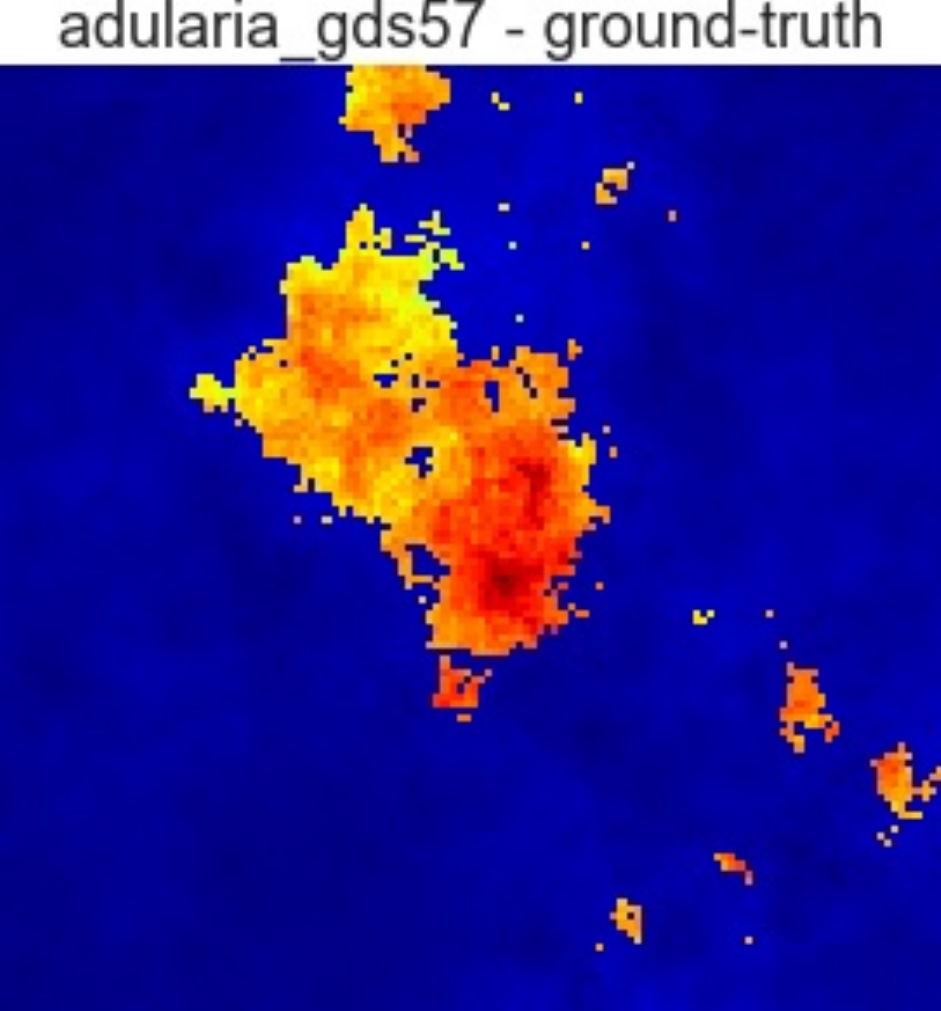}
    \includegraphics[width=0.103\textwidth,trim={0cm 0cm 0cm 1cm},clip]{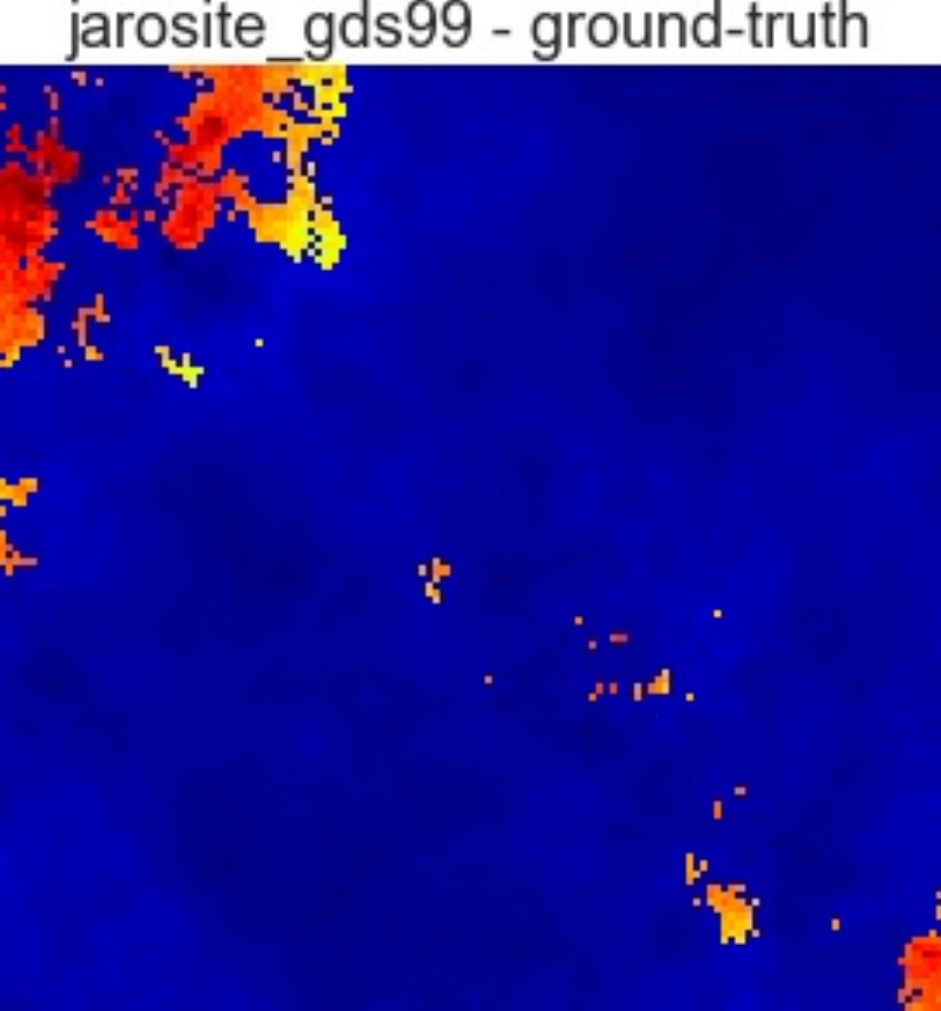}
    \includegraphics[width=0.103\textwidth,trim={0cm 0cm 0cm 1cm},clip]{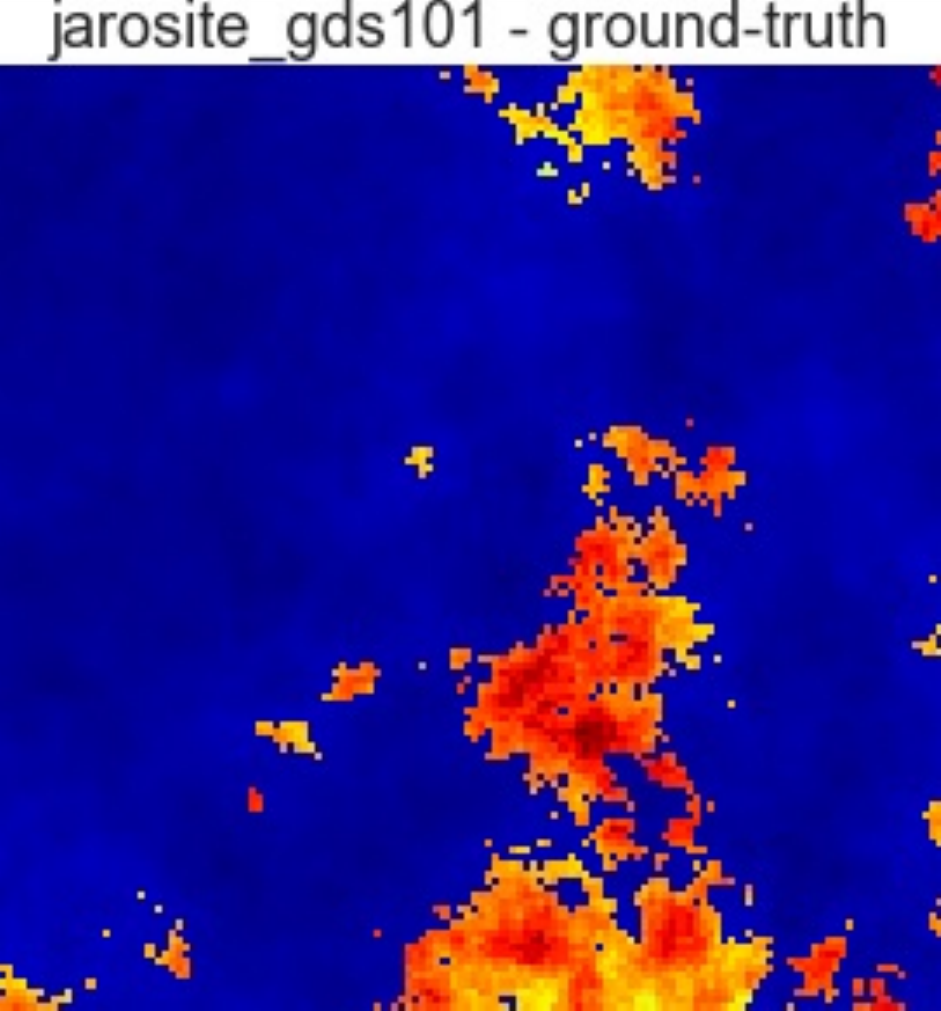}
    \includegraphics[width=0.103\textwidth,trim={0cm 0cm 0cm 1cm},clip]{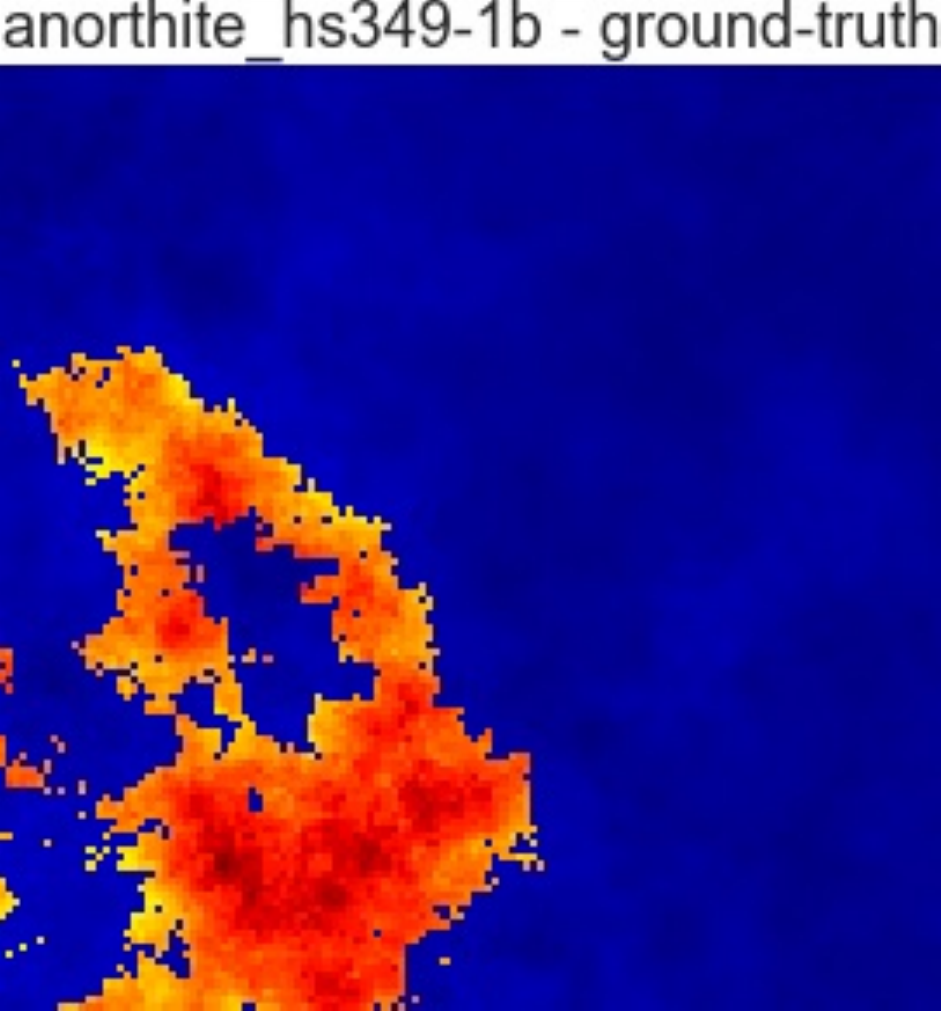}
    \includegraphics[width=0.103\textwidth,trim={0cm 0cm 0cm 1cm},clip]{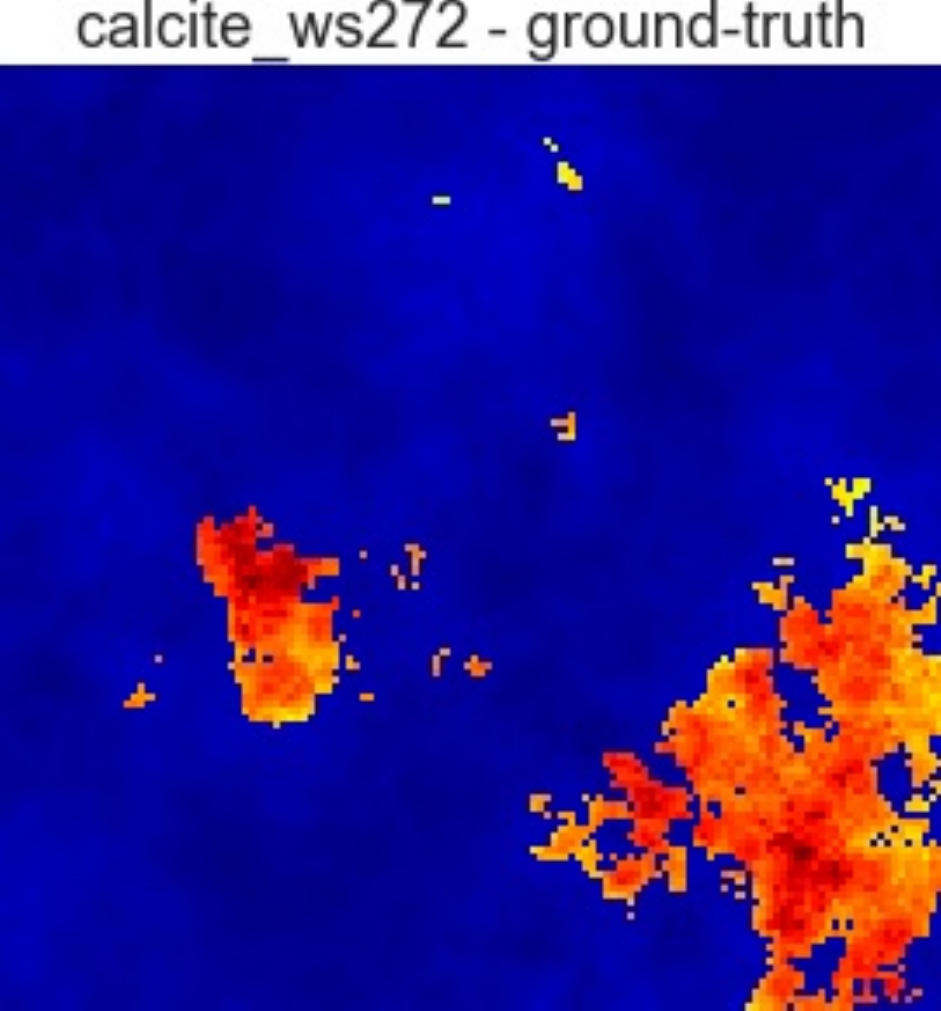}
    \includegraphics[width=0.103\textwidth,trim={0cm 0cm 0cm 1cm},clip]{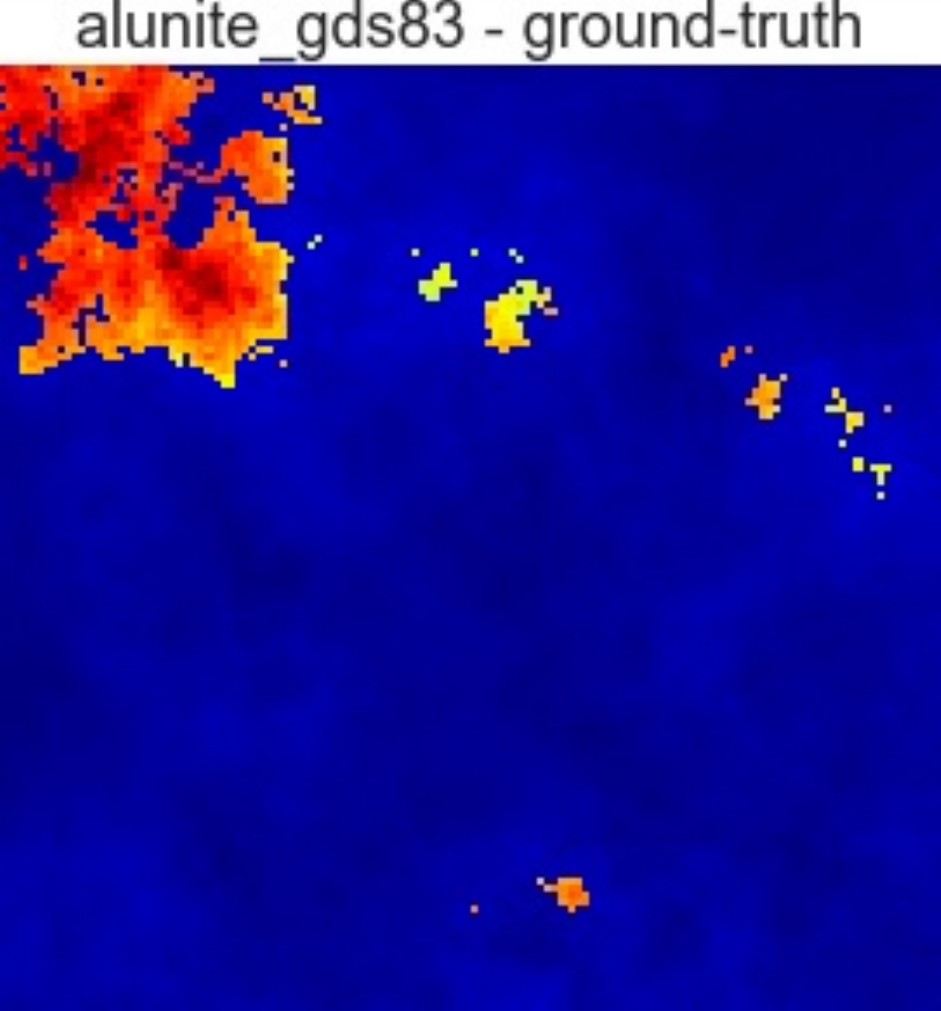}
    \includegraphics[width=0.103\textwidth,trim={0cm 0cm 0cm 1cm},clip]{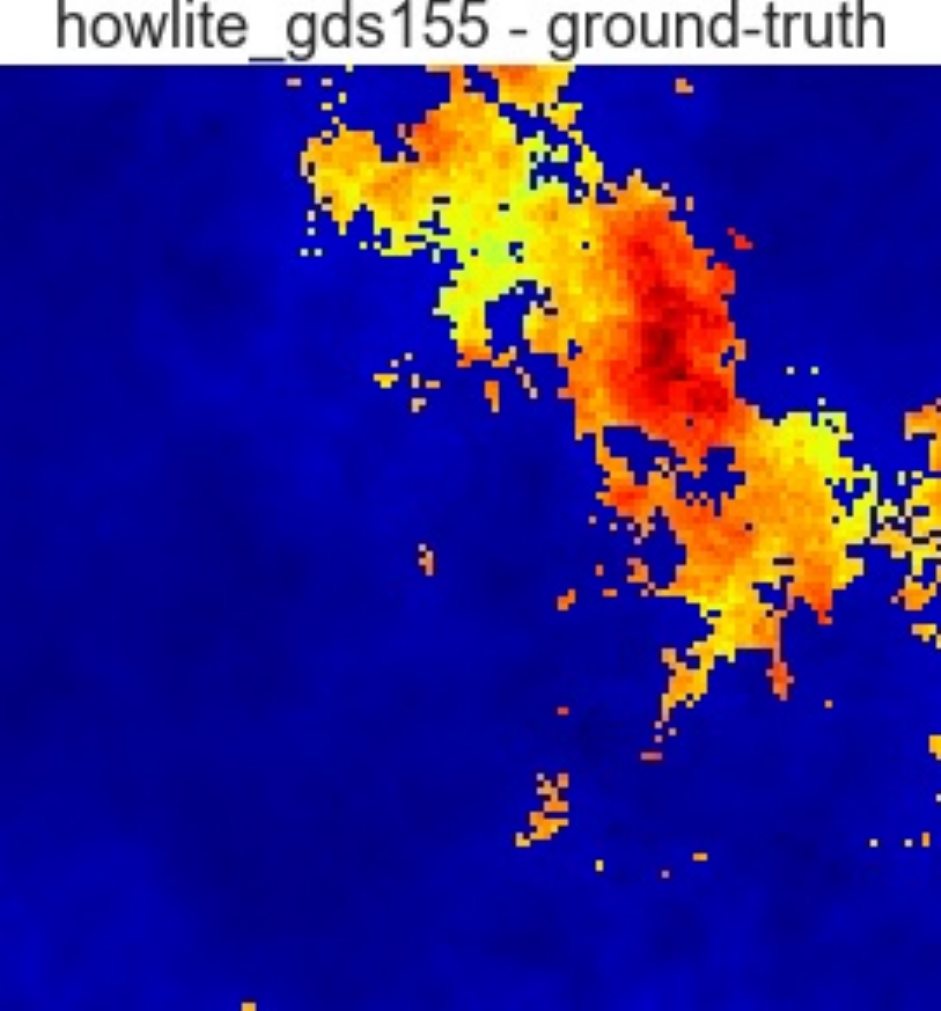}
    \includegraphics[width=0.103\textwidth,trim={0cm 0cm 0cm 1cm},clip]{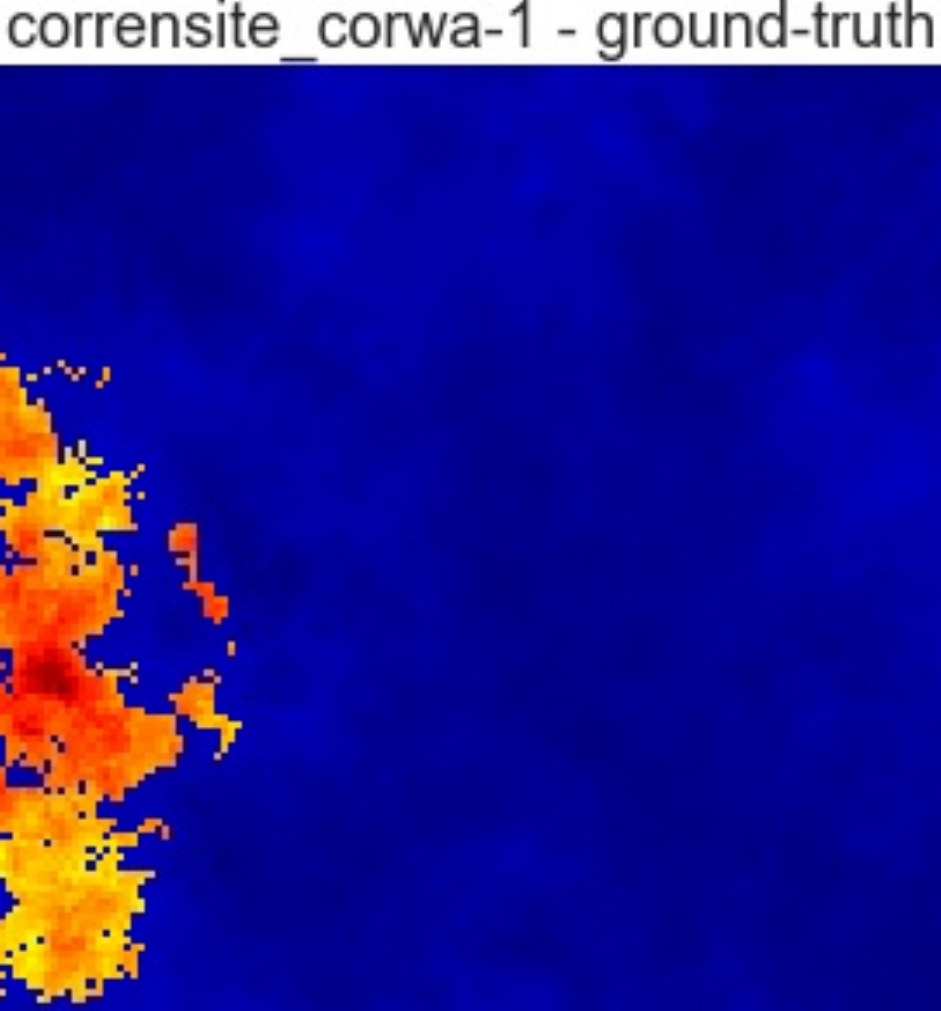}
    \includegraphics[width=0.103\textwidth,trim={0cm 0cm 0cm 1cm},clip]{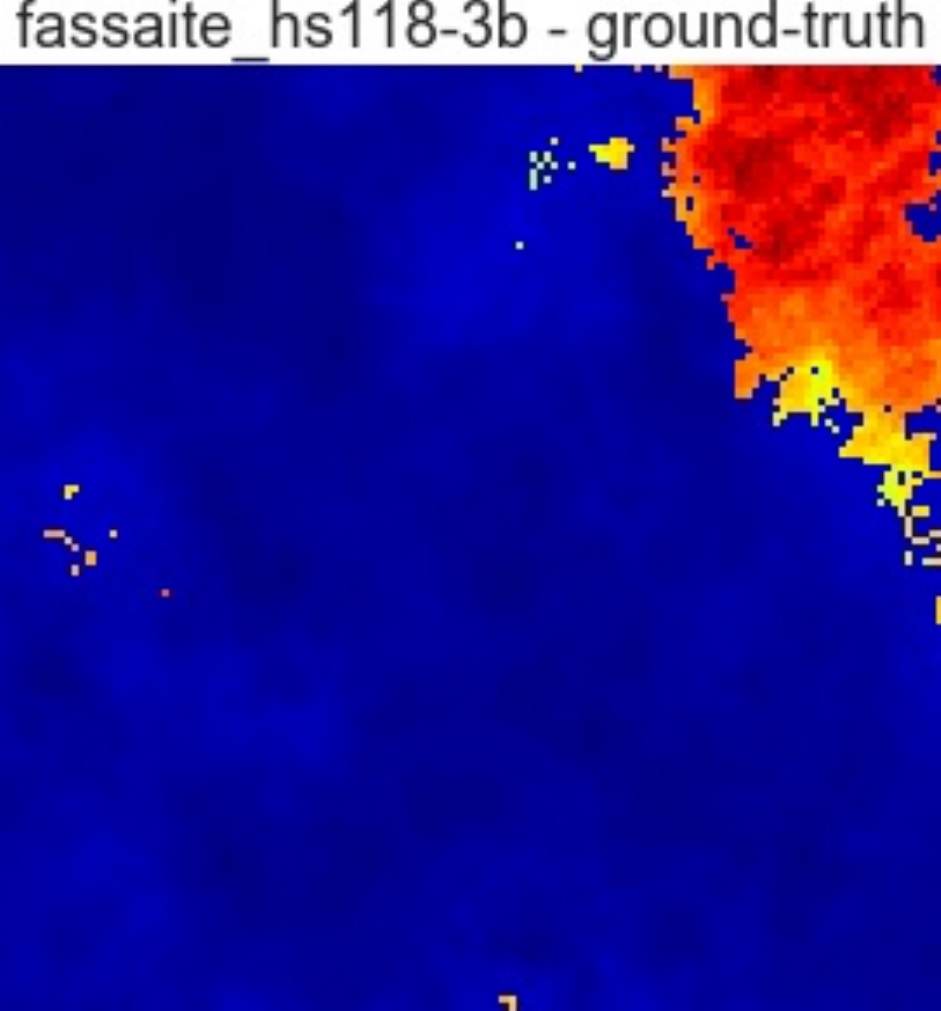}
    \caption{Abundances maps of the synthetic dataset (top row: LDVAE, bottom row: ground truth).  From left to right: Adularia,
        Jarosite gds99, Jarosite gds101, Anorthite, Calcite, Alunite, Howlite, Corrensite, Fassaite.}
    \label{fig:synthetic_abundances}
\end{figure*}

The results show that LDVAE performed well in all scenarios.
The results on synthetic data are aggregated over all classes, as we focused on the robustness to the noise present in the signal.
The LDVAE model shows consistent performance at all noise levels; however, sometimes it does not achieve the best results.
The results on tables~\ref{table:results-synthetic-nosnr-sad} and~\ref{table:results-synthetic-nosnr-rmse} demonstrate the LDVAE performed similarly to the other methods, however LDVAE's performance does not degrade when SNR decrease.
We attribute this behavior to the fact that LDVAE learned the small variability on pixel values due to the probabilistic approach of variational autoencoders.
However, this variability also affected the absolute values of the pixels during reconstruction.
In other words, the LDVAE managed to average out the noise, which also explains the reduced reconstruction metrics, but holding up classification performance (Table ~\ref{table:results-synthetic-nosnr-sad}).
The images in Figure~\ref{fig:synthetic_abundances} show results of classification results compared to the ground truth.

\begin{figure*}
    \centering
    \includegraphics[width=0.36\textwidth,trim={  0cm 1.5cm 0cm 0cm},clip]{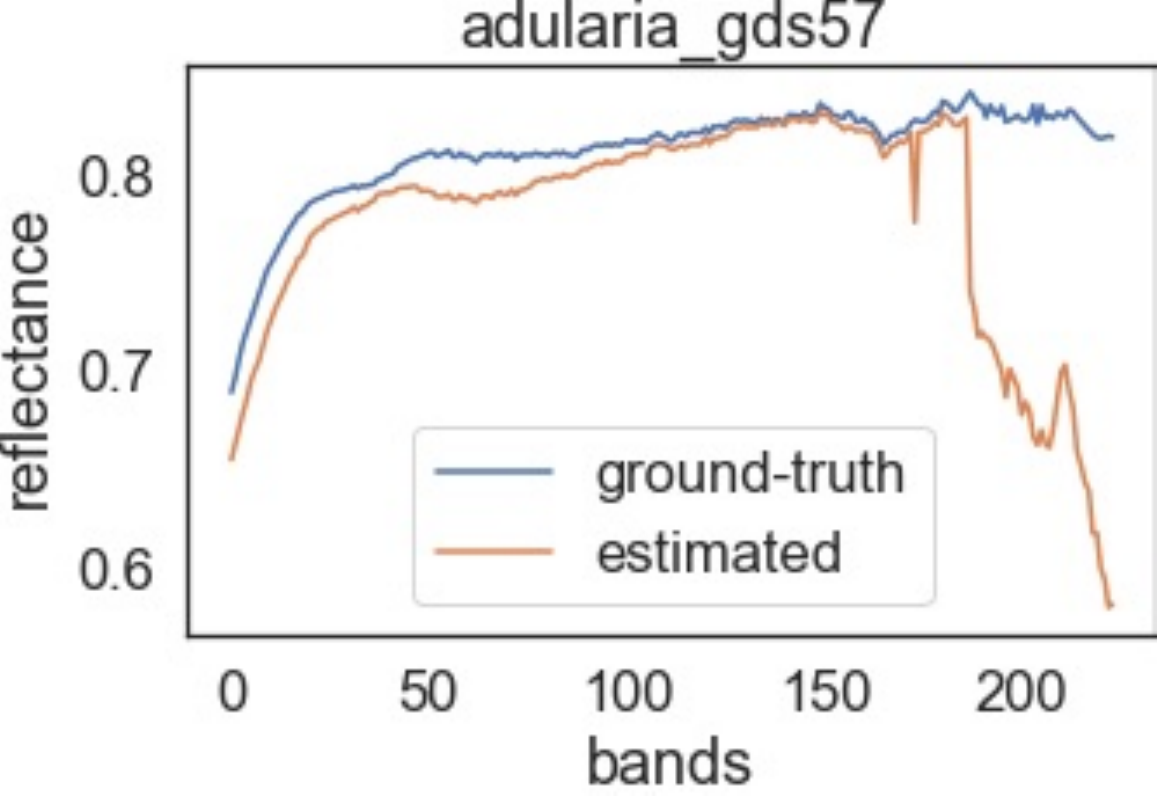}
    \includegraphics[width=0.31\textwidth,trim={1.6cm 1.5cm 0cm 0cm},clip]{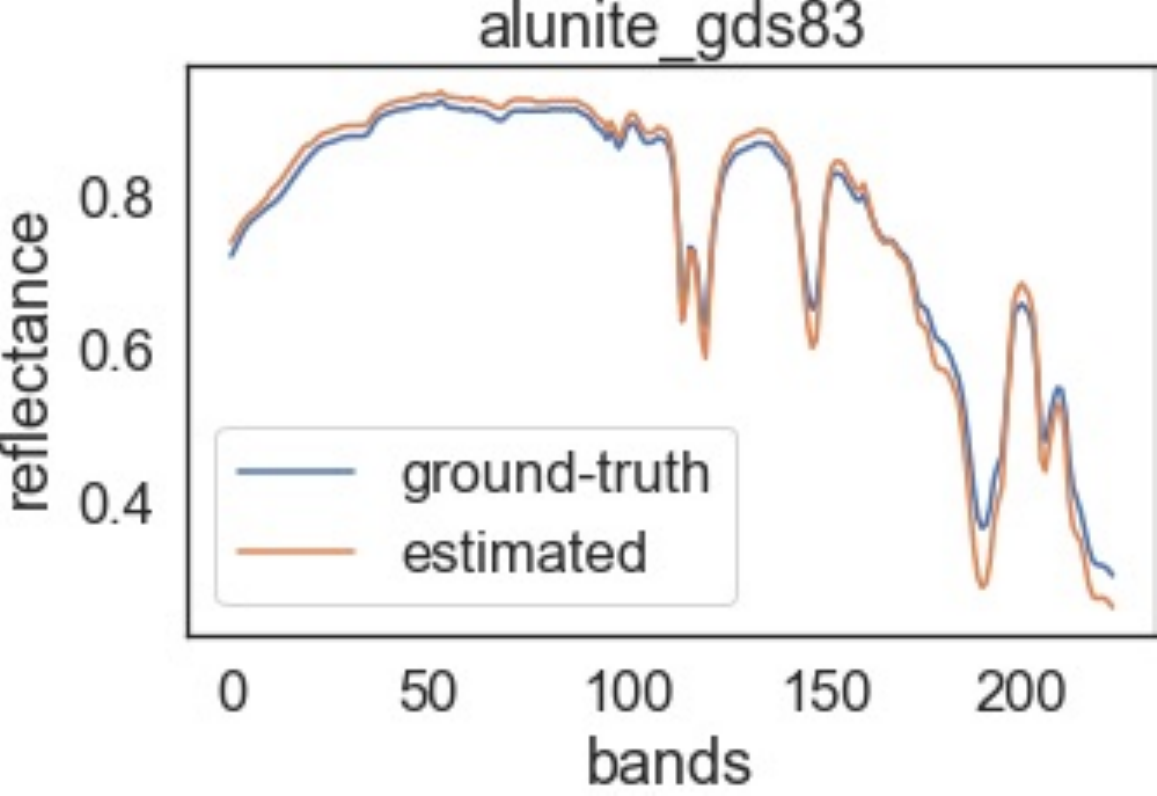}
    \includegraphics[width=0.31\textwidth,trim={1.6cm 1.5cm 0cm 0cm},clip]{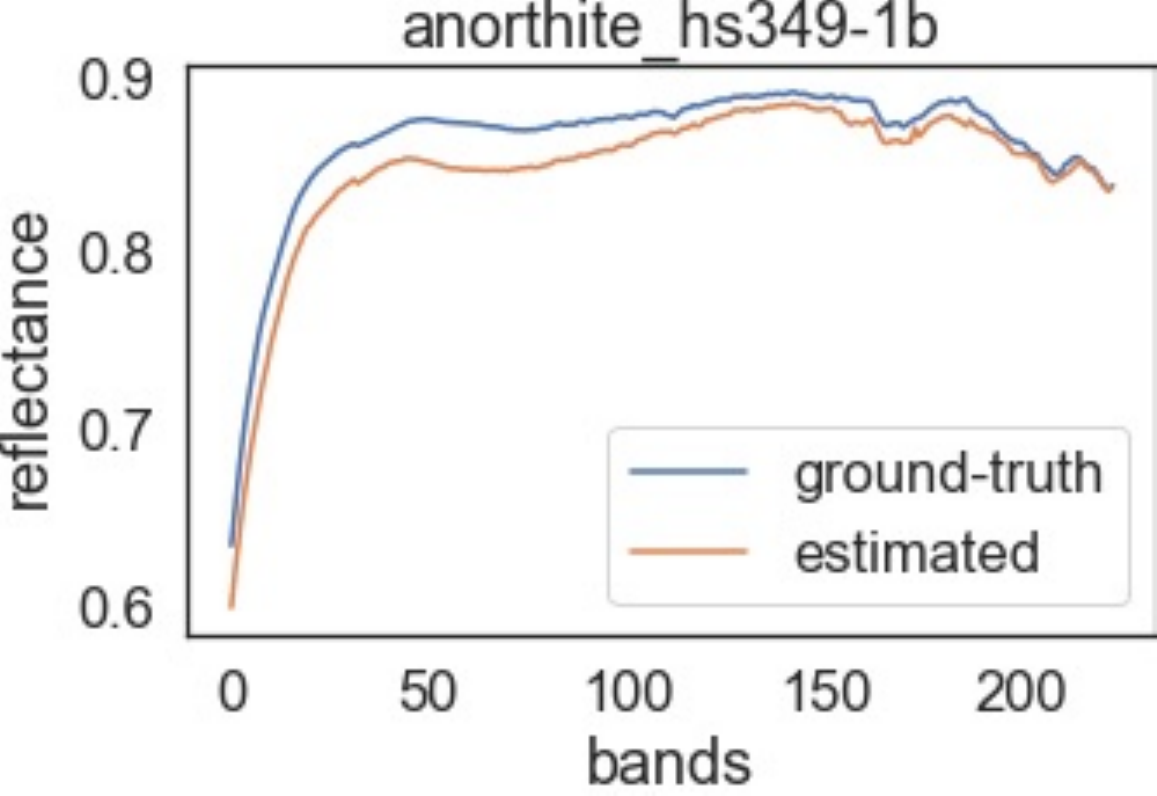}\linebreak
    \includegraphics[width=0.36\textwidth,trim={  0cm 1.5cm 0cm 0cm},clip]{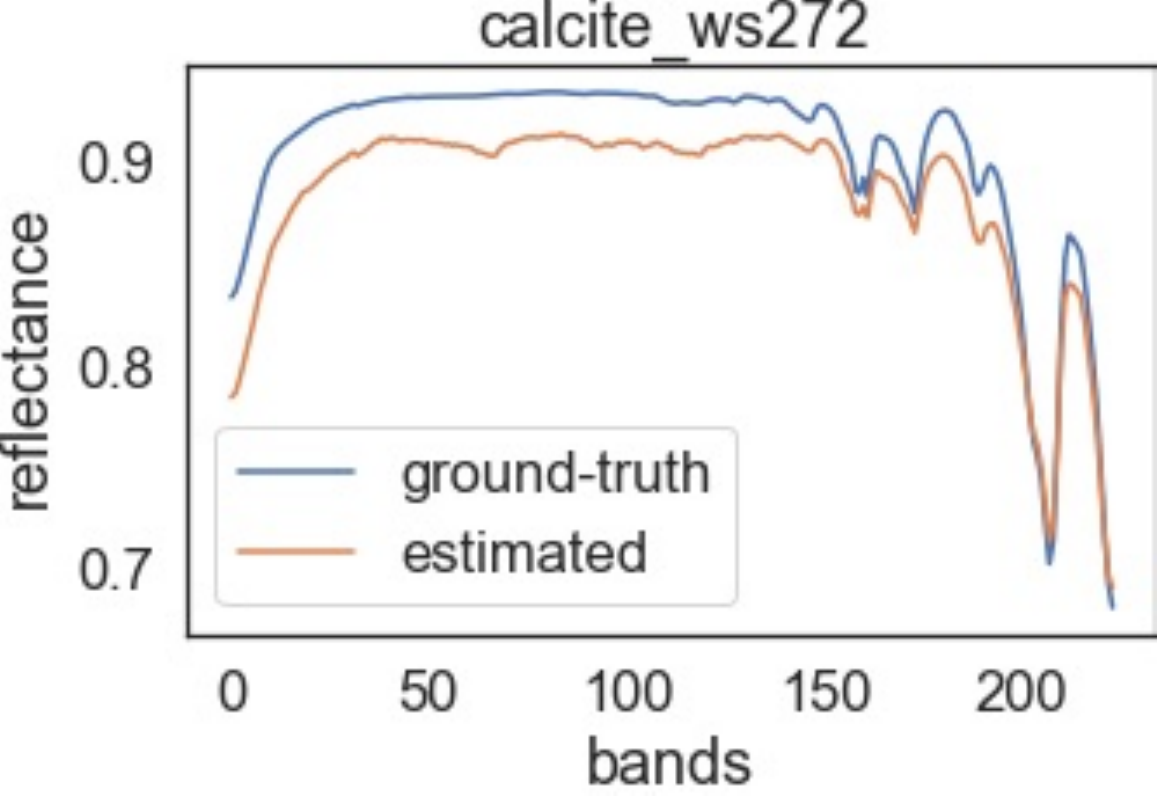}
    \includegraphics[width=0.31\textwidth,trim={1.6cm 1.5cm 0cm 0cm},clip]{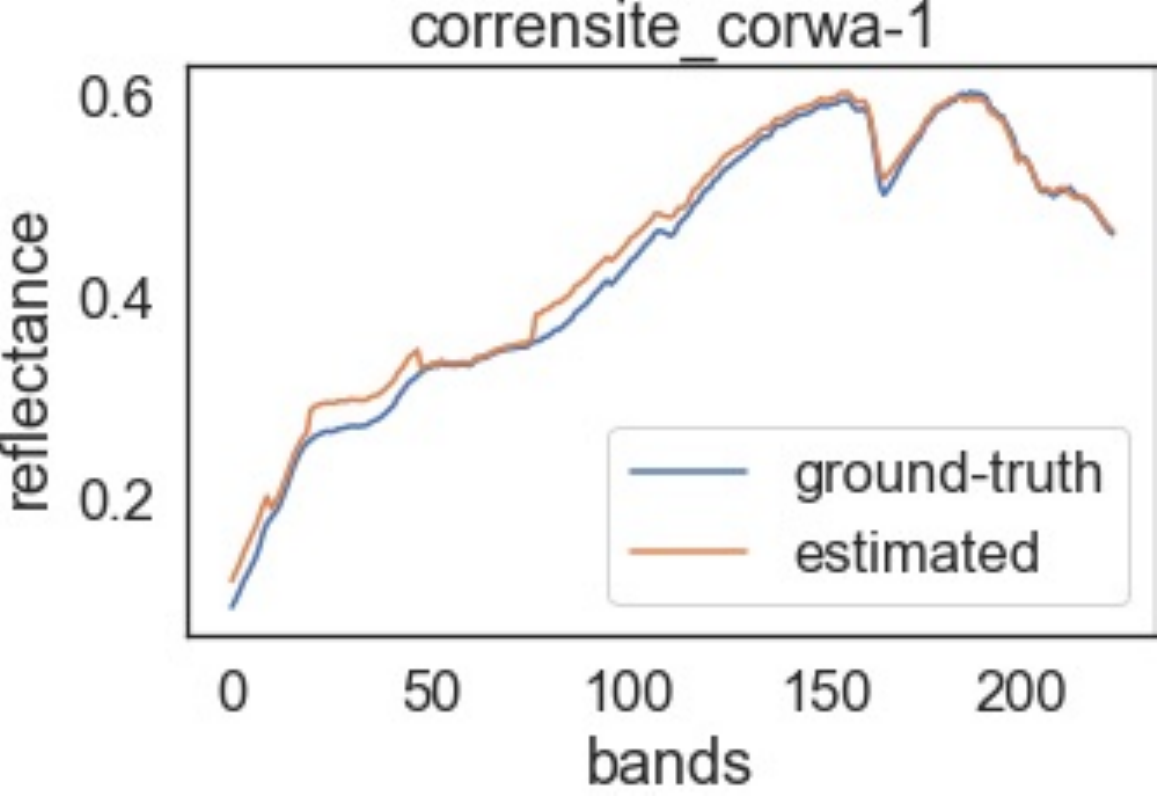}
    \includegraphics[width=0.31\textwidth,trim={1.6cm 1.5cm 0cm 0cm},clip]{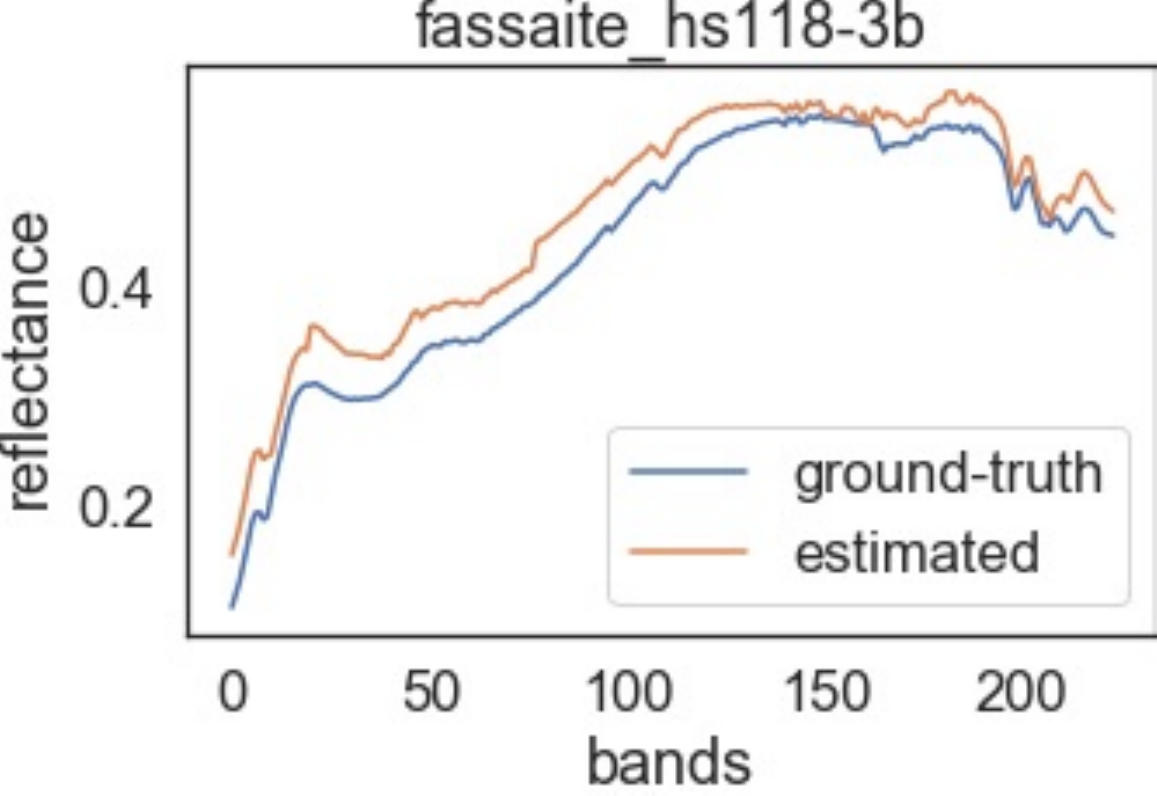}\linebreak
    \includegraphics[width=0.36\textwidth,trim={  0cm 0cm 0cm 0cm},clip]{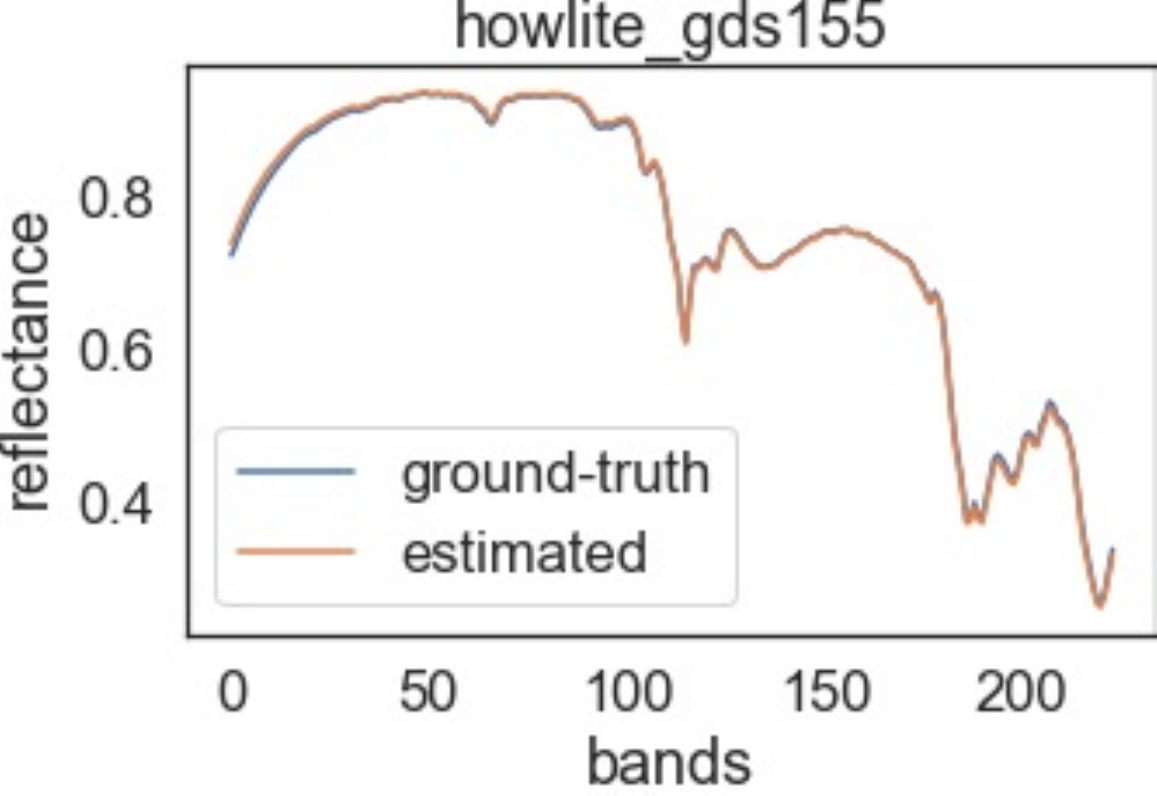}
    \includegraphics[width=0.31\textwidth,trim={2.0cm 0cm 0cm 0cm},clip]{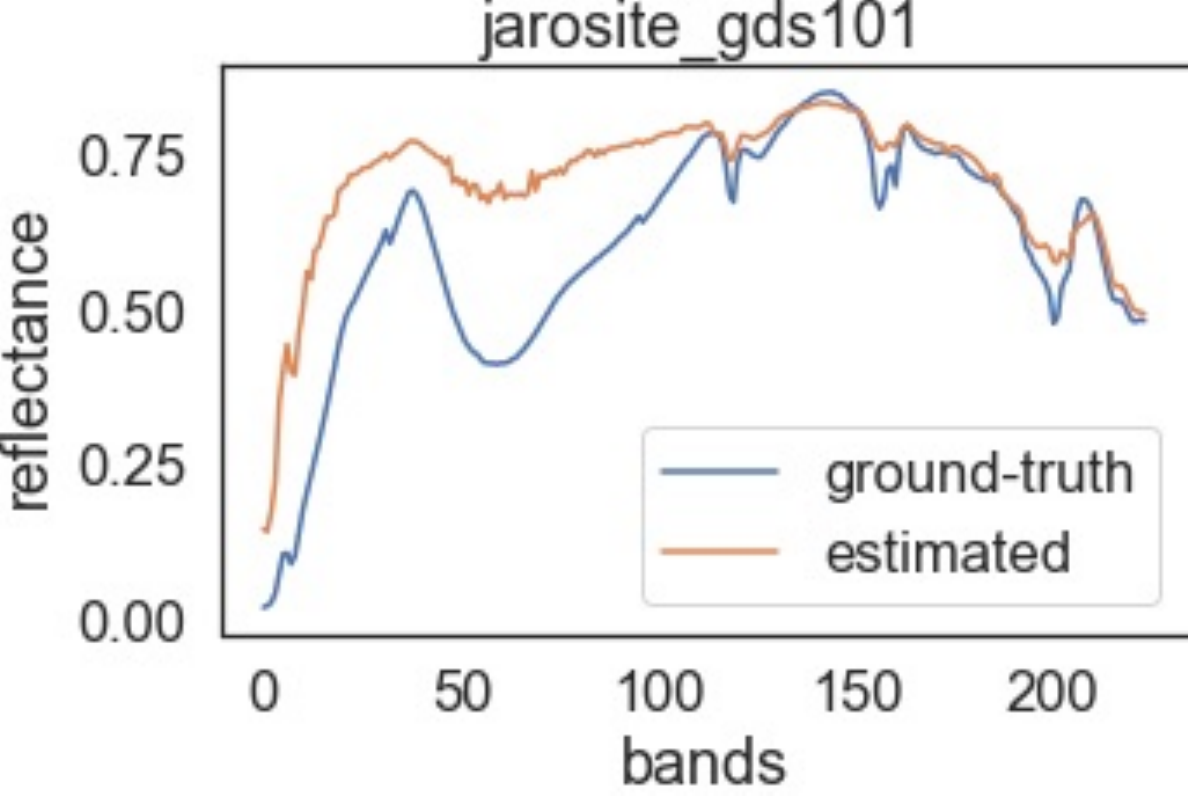}
    \includegraphics[width=0.31\textwidth,trim={2.0cm 0cm 0cm 0cm},clip]{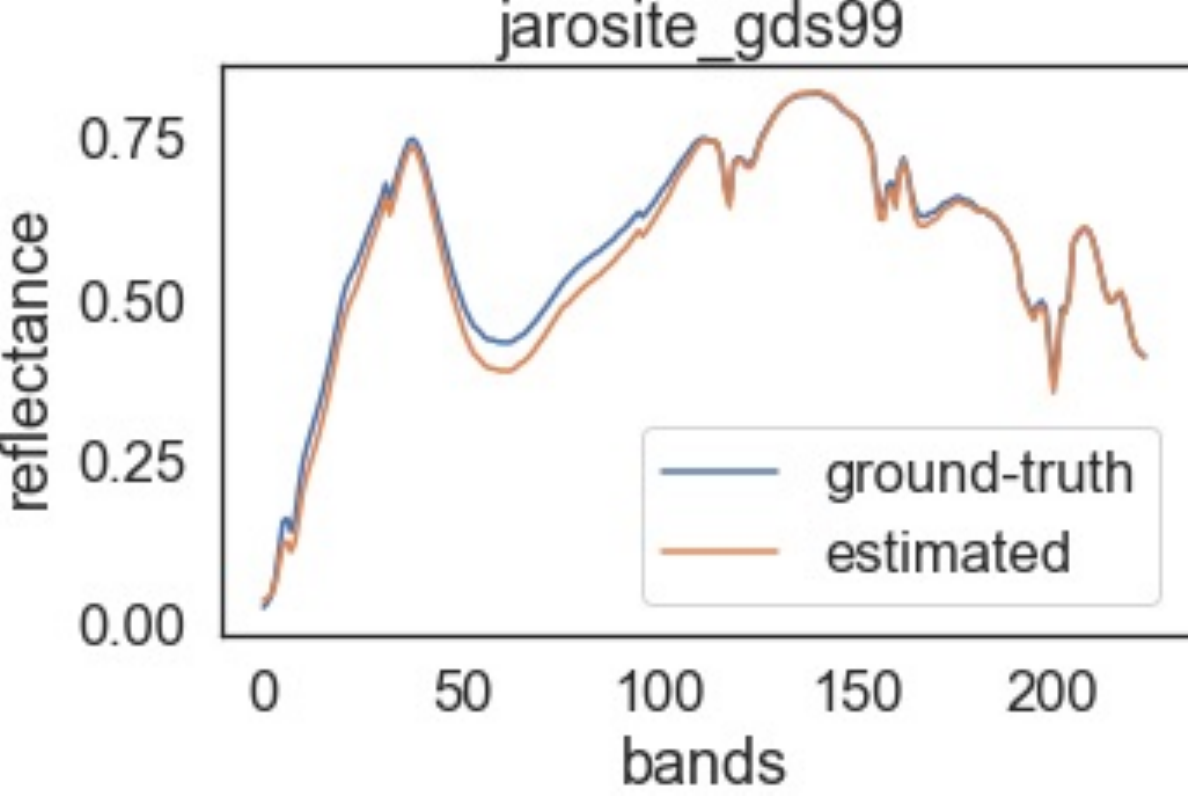}
    \caption{Endmembers of the Synthetic dataset generated by LDVAE: comparison with  ground truth.}
    \label{fig:endmembers_synthetic}
\end{figure*}

\begin{figure*}
    \centering
    \includegraphics[width=0.36\textwidth,trim={  0cm 1.6cm 0cm 0cm},clip]{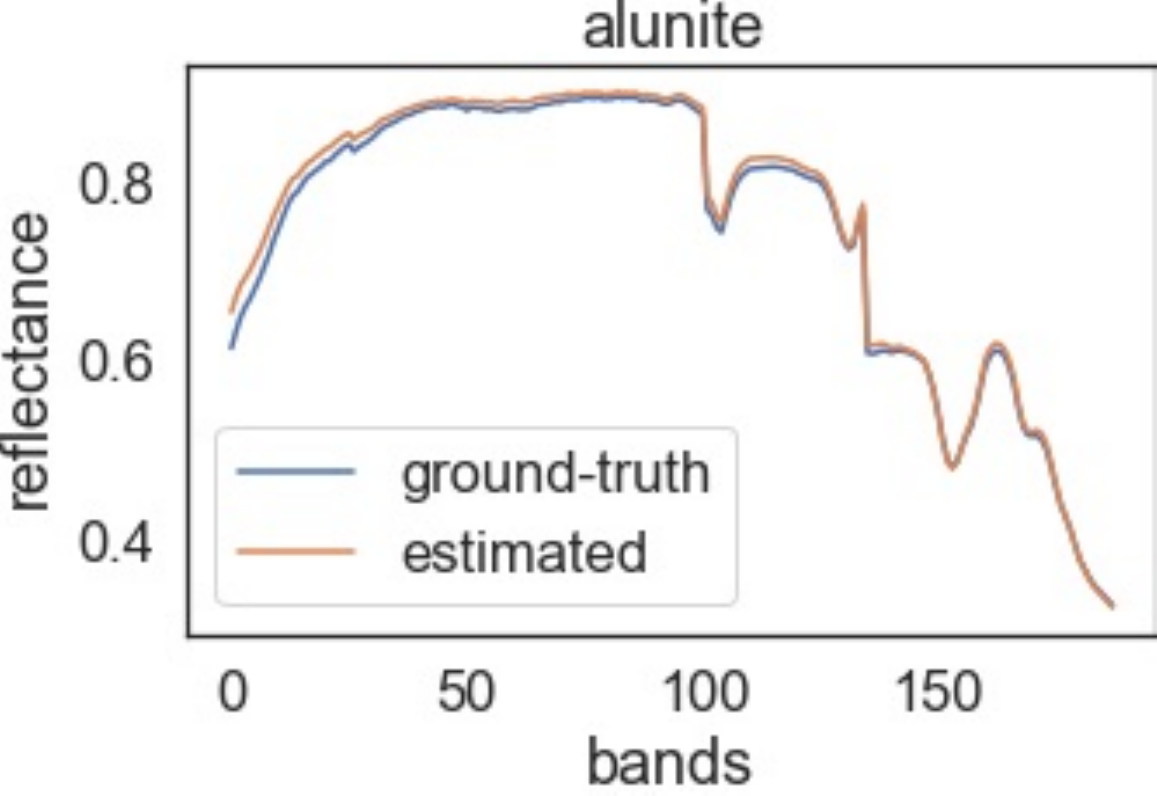}
    \includegraphics[width=0.31\textwidth,trim={1.7cm 1.6cm 0cm 0cm},clip]{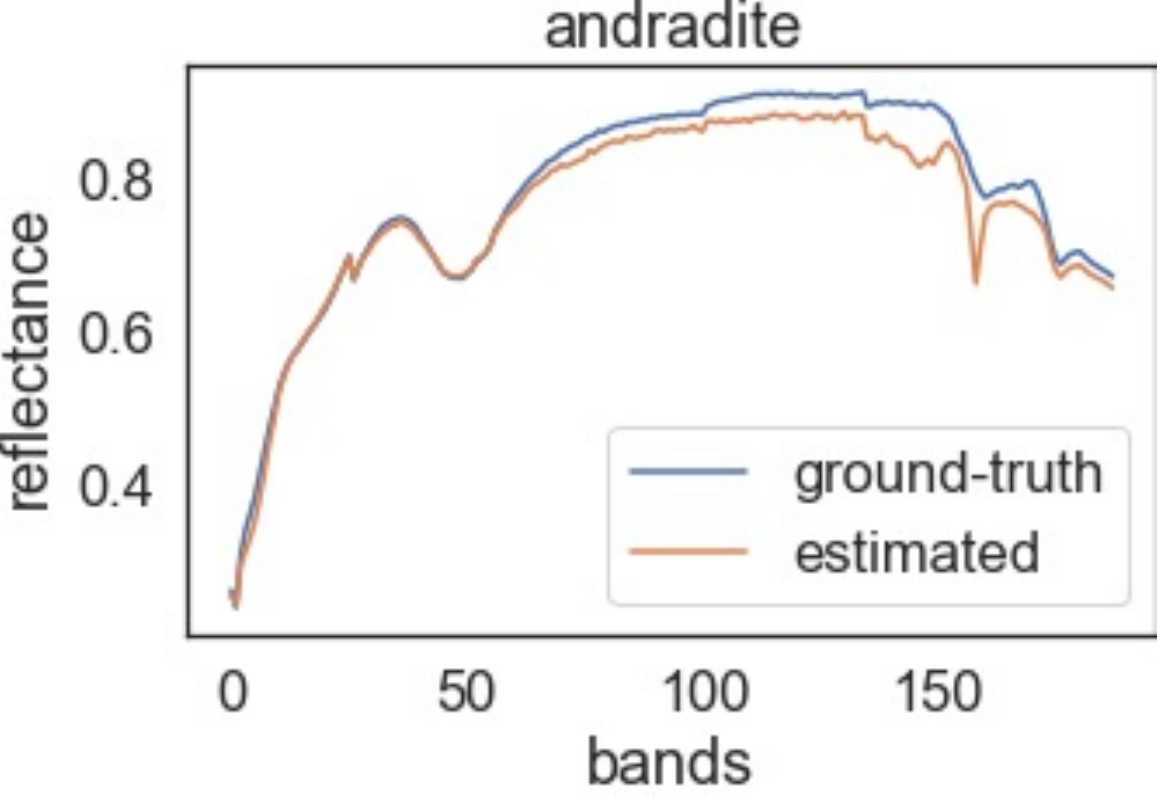}
    \includegraphics[width=0.31\textwidth,trim={1.7cm 1.6cm 0cm 0cm},clip]{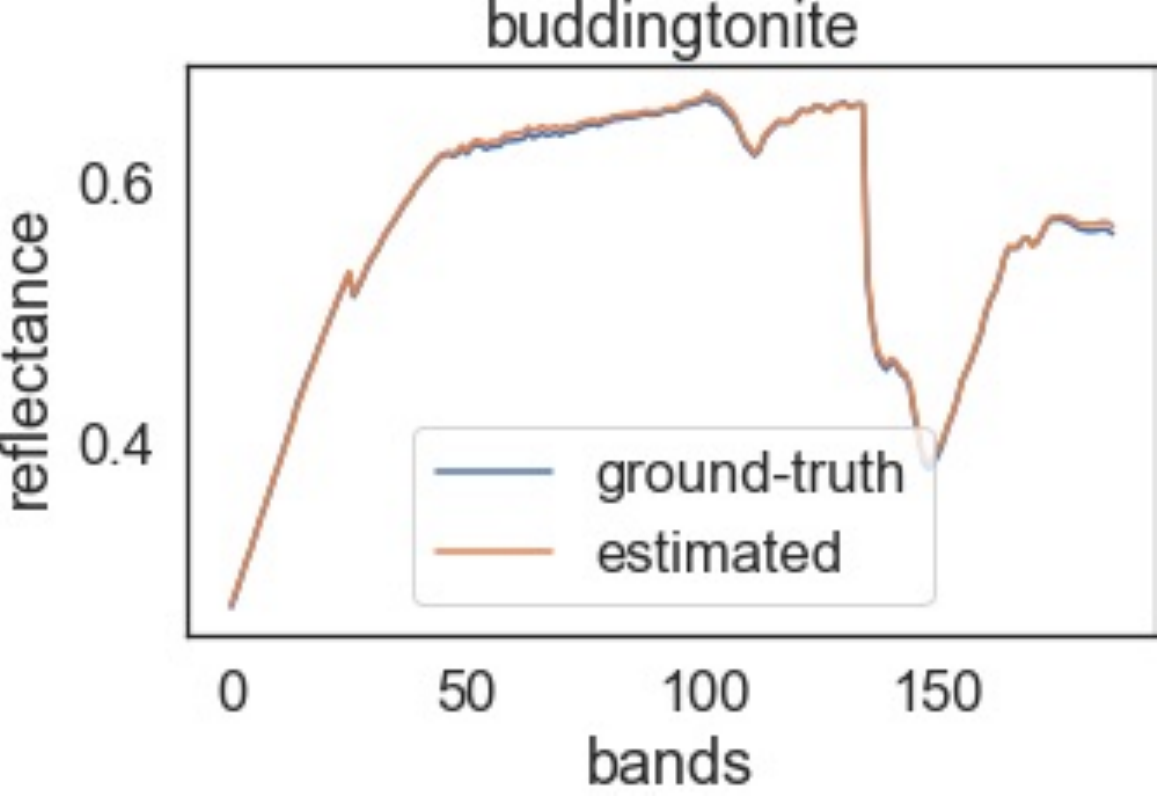}\linebreak
    \includegraphics[width=0.36\textwidth,trim={  0cm 1.6cm 0cm 0cm},clip]{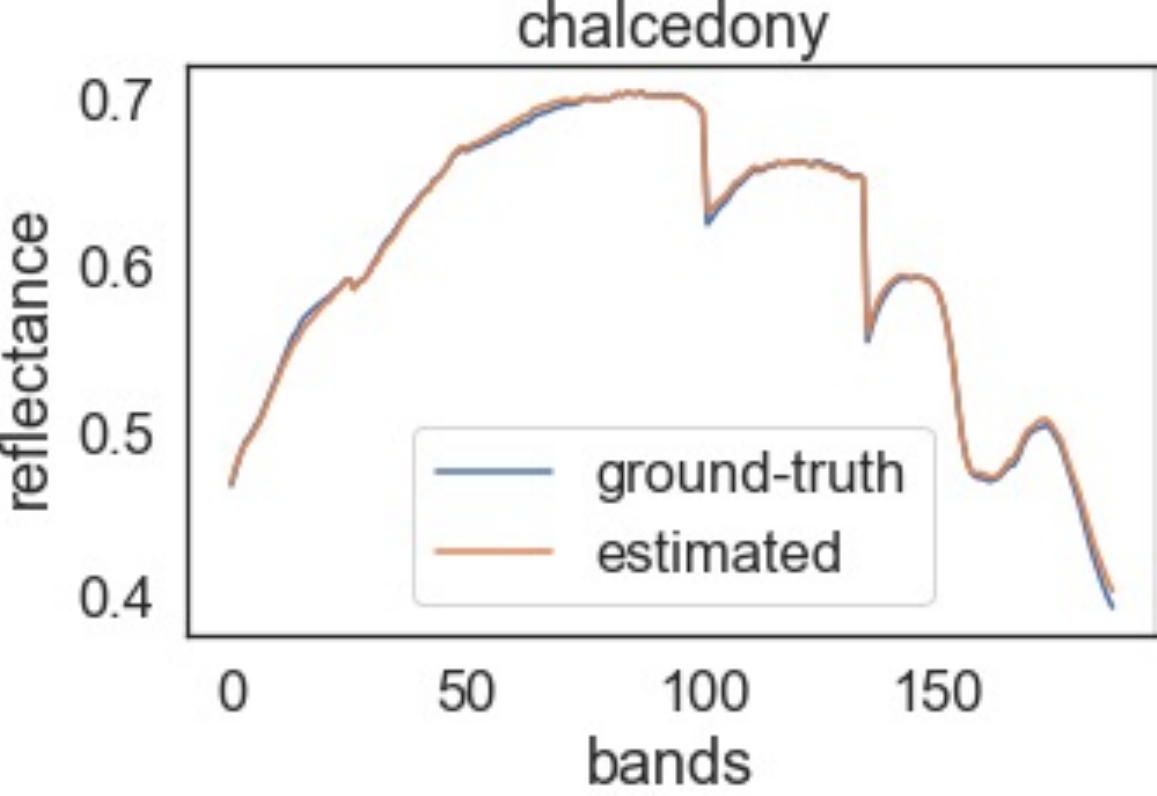}
    \includegraphics[width=0.31\textwidth,trim={1.7cm 1.6cm 0cm 0cm},clip]{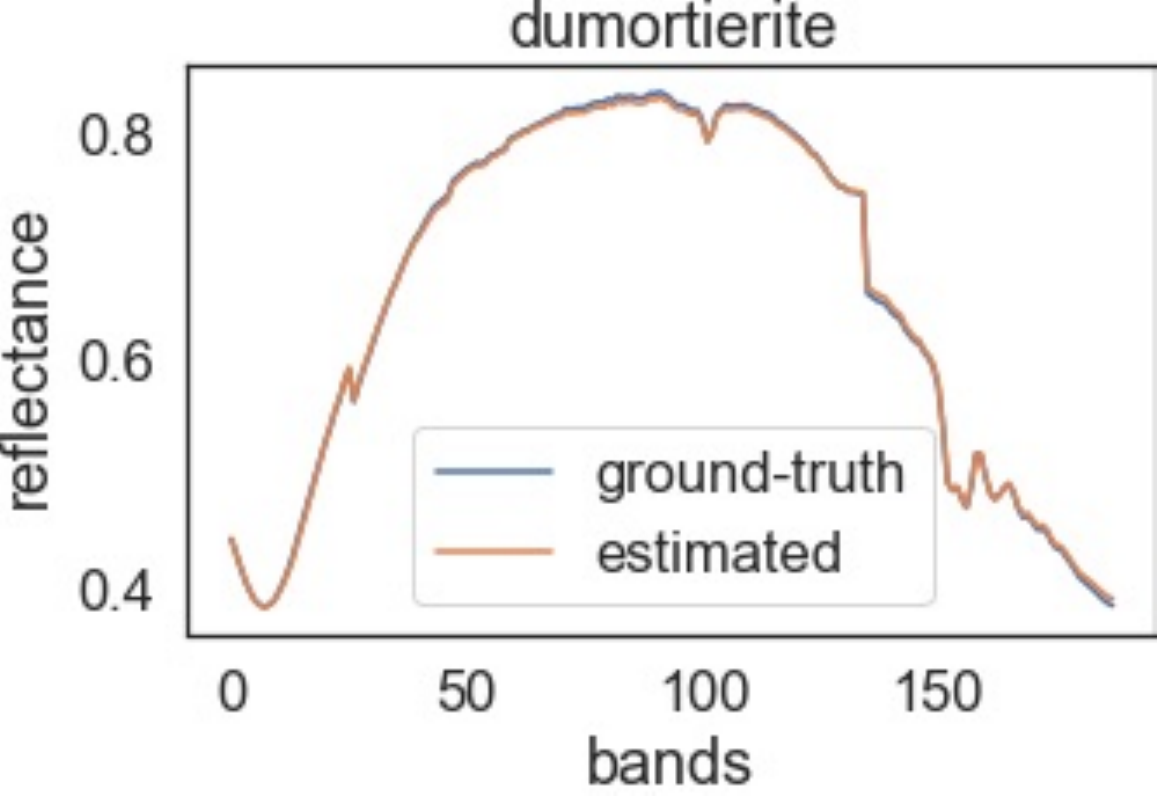}
    \includegraphics[width=0.31\textwidth,trim={1.7cm 1.6cm 0cm 0cm},clip]{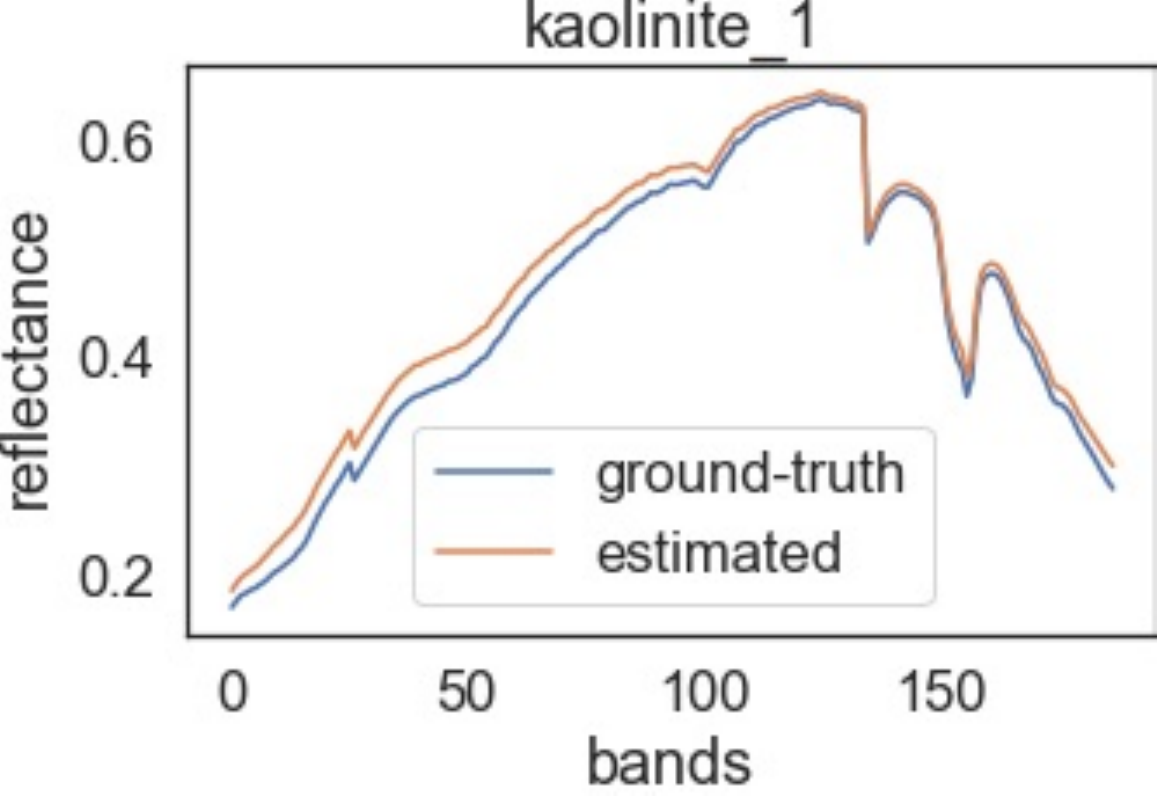}\linebreak
    \includegraphics[width=0.36\textwidth,trim={  0cm 1.6cm 0cm 0cm},clip]{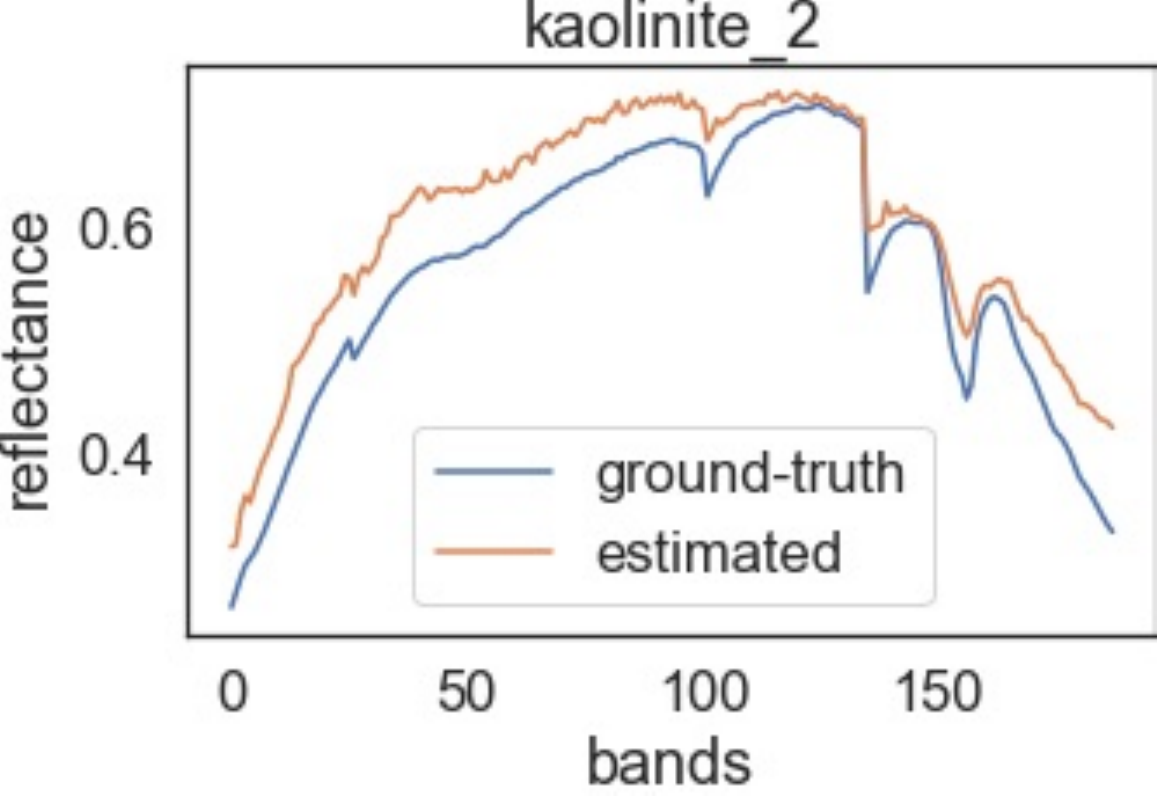}
    \includegraphics[width=0.31\textwidth,trim={1.7cm 1.6cm 0cm 0cm},clip]{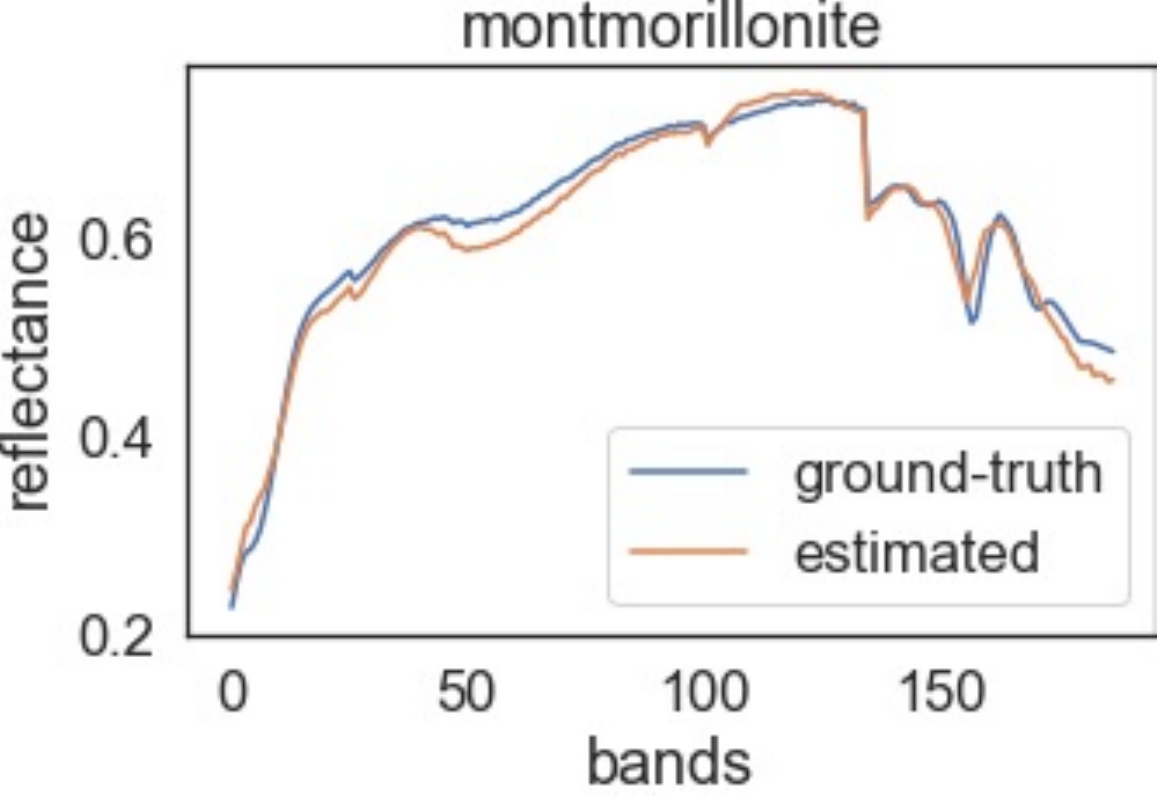}
    \includegraphics[width=0.31\textwidth,trim={1.7cm 1.6cm 0cm 0cm},clip]{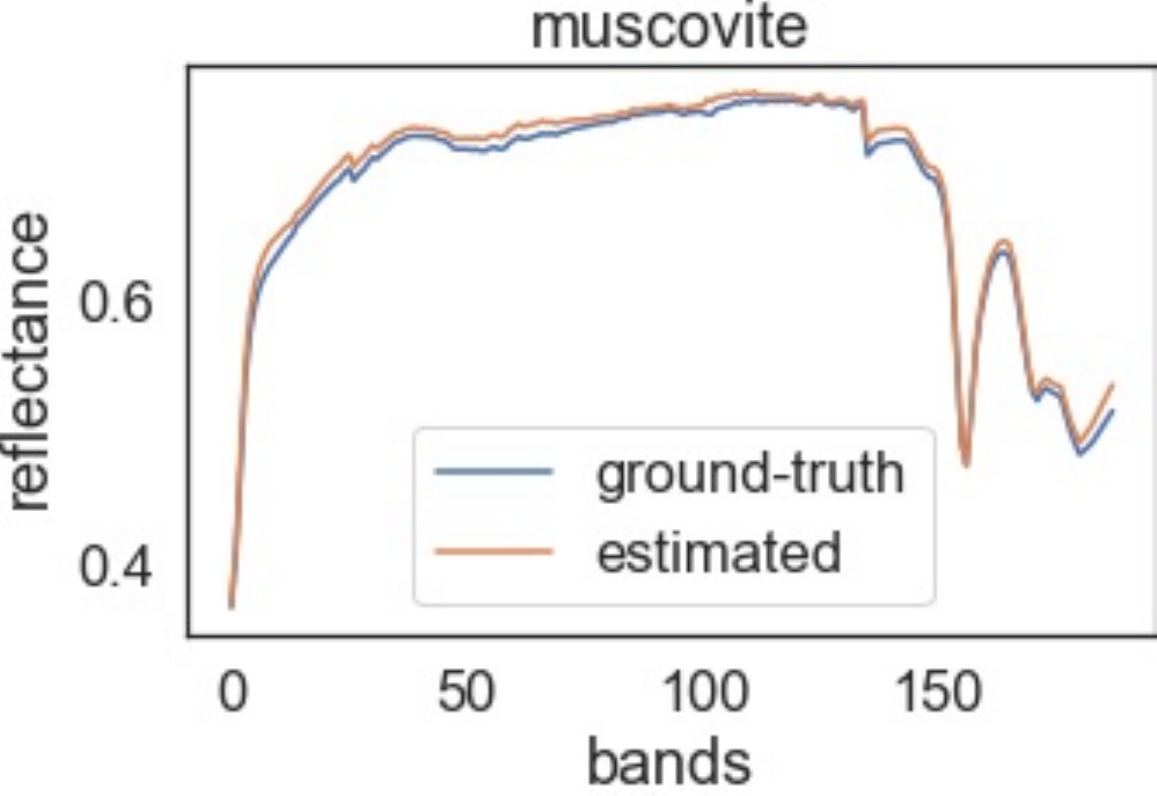}\linebreak
    \includegraphics[width=0.36\textwidth,trim={  0cm   0cm 0cm 0cm},clip]{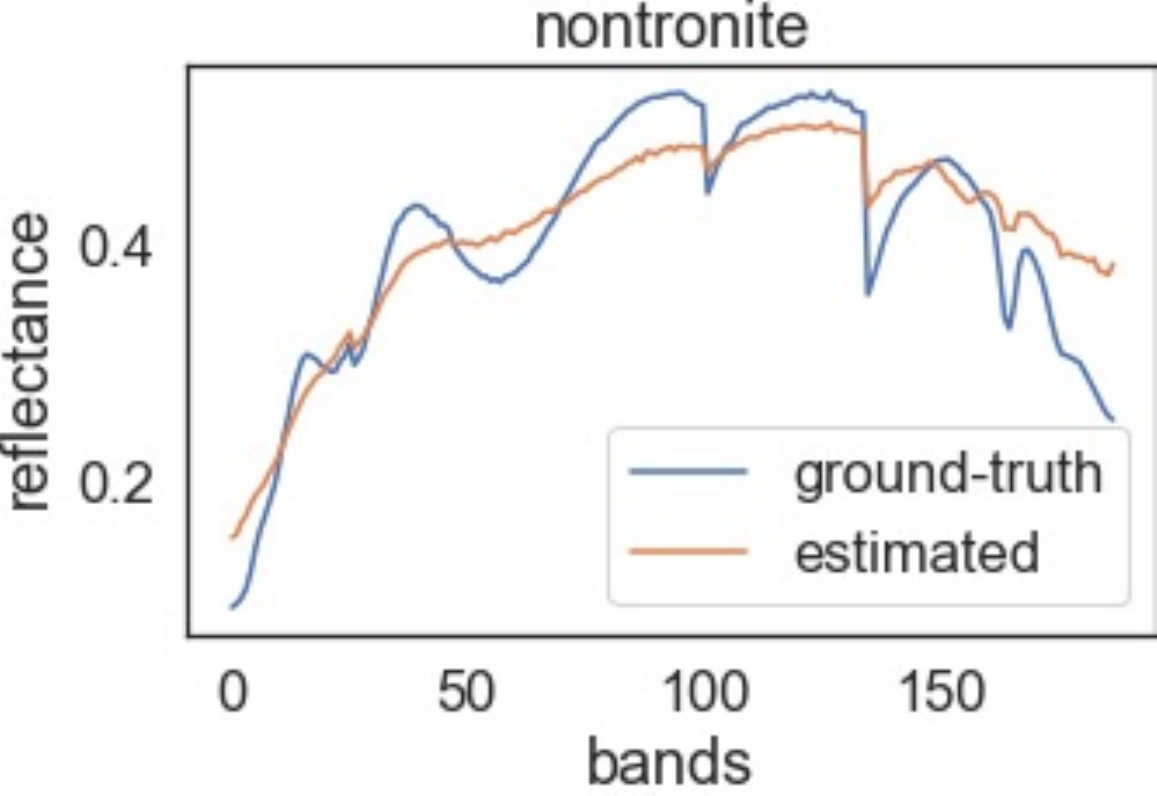}
    \includegraphics[width=0.31\textwidth,trim={1.7cm   0cm 0cm 0cm},clip]{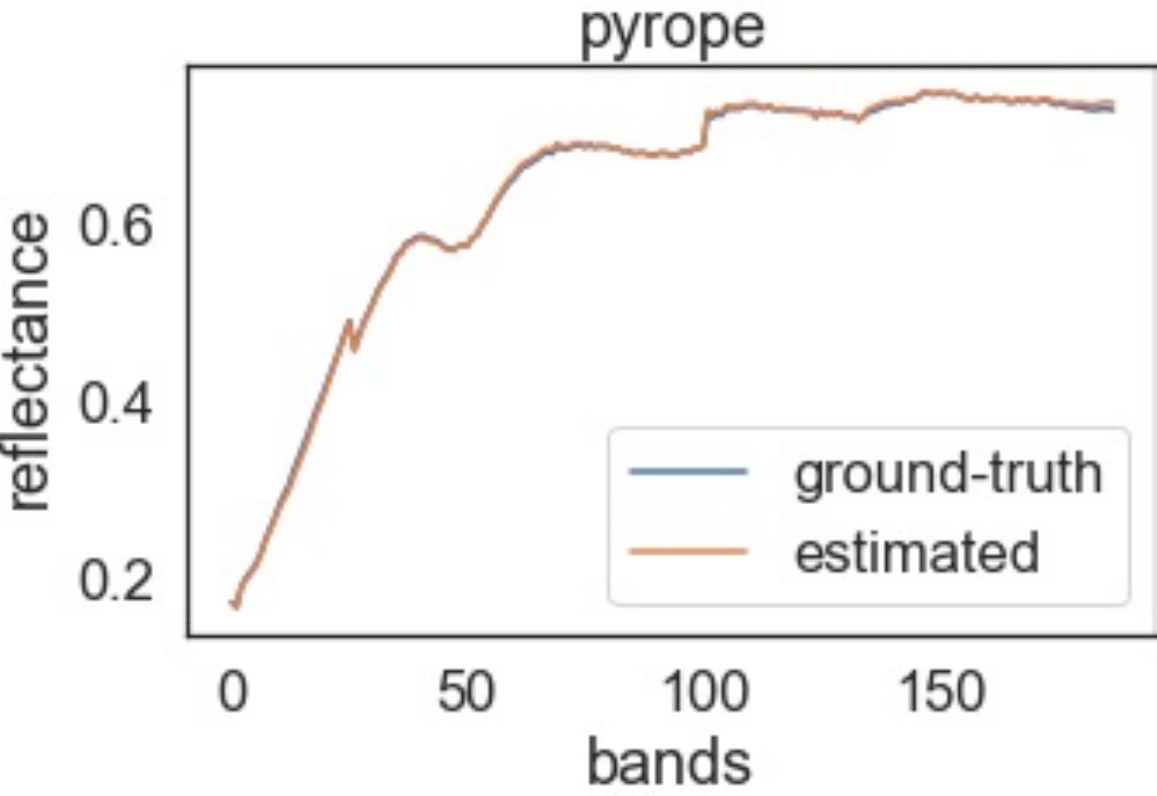}
    \includegraphics[width=0.31\textwidth,trim={1.7cm   0cm 0cm 0cm},clip]{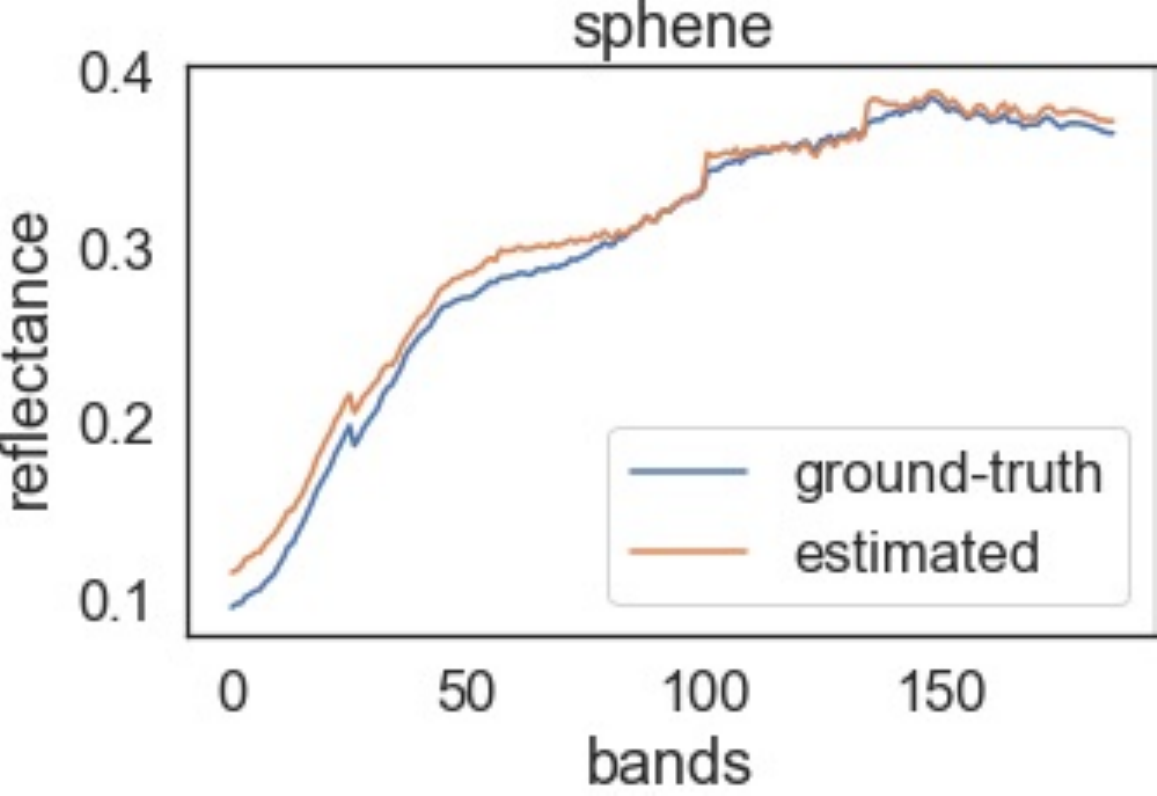}
    \caption{Endmembers of the Cuprite dataset generated by LDVAE: comparison with ground truth.}
    \label{fig:endmembers_cuprite}
\end{figure*}

Table~\ref{table:results-cuprite-sad} indicates that LDVAE successfully performs endmember extraction on the Cuprite dataset (see also  Figure~\ref{fig:endmembers_cuprite}).  Recall that the model was never trained on the Cuprite dataset.  Rather the model was trained on Cuprite Synthetic dataset.
Considering the lack of ground truth and the applicability of a synthetic dataset for model training, we observed several opportunities for further research and improvements as will be discussed in Section ~\ref{sec:conclusion}.

\begin{figure*}
    \centering
    \includegraphics[width=0.158\textwidth, trim={0cm 0cm 0cm 1cm},clip]{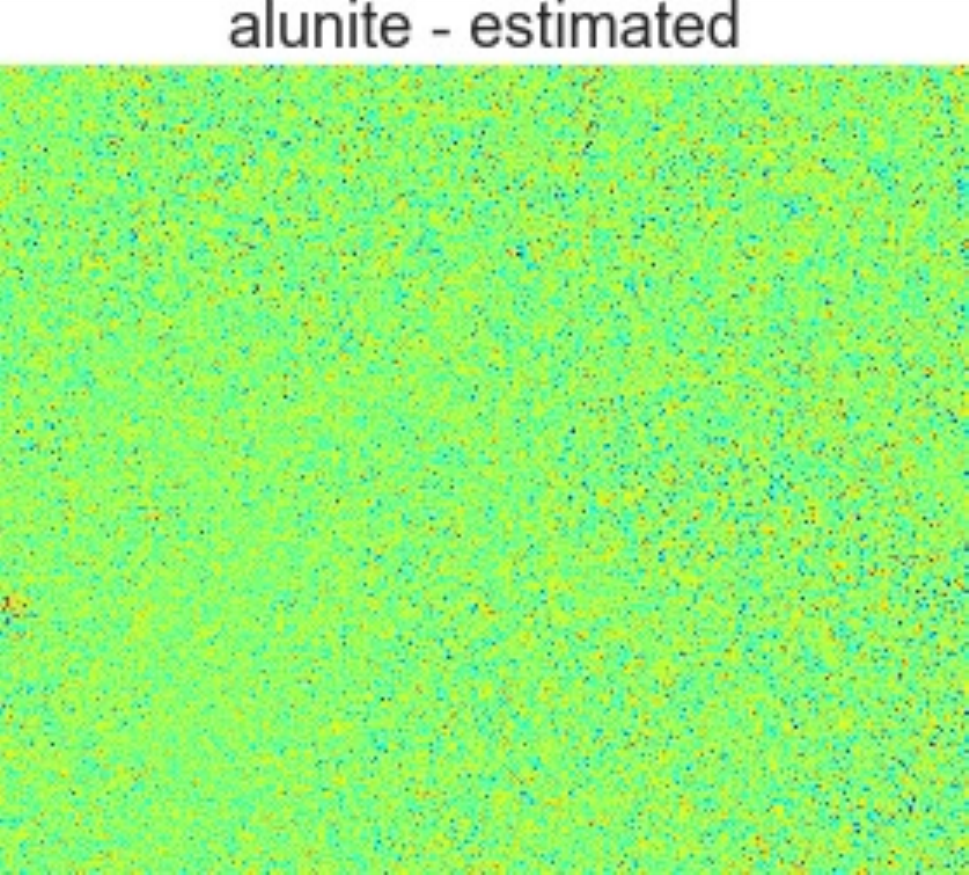}
    \includegraphics[width=0.158\textwidth, trim={0cm 0cm 0cm 1cm},clip]{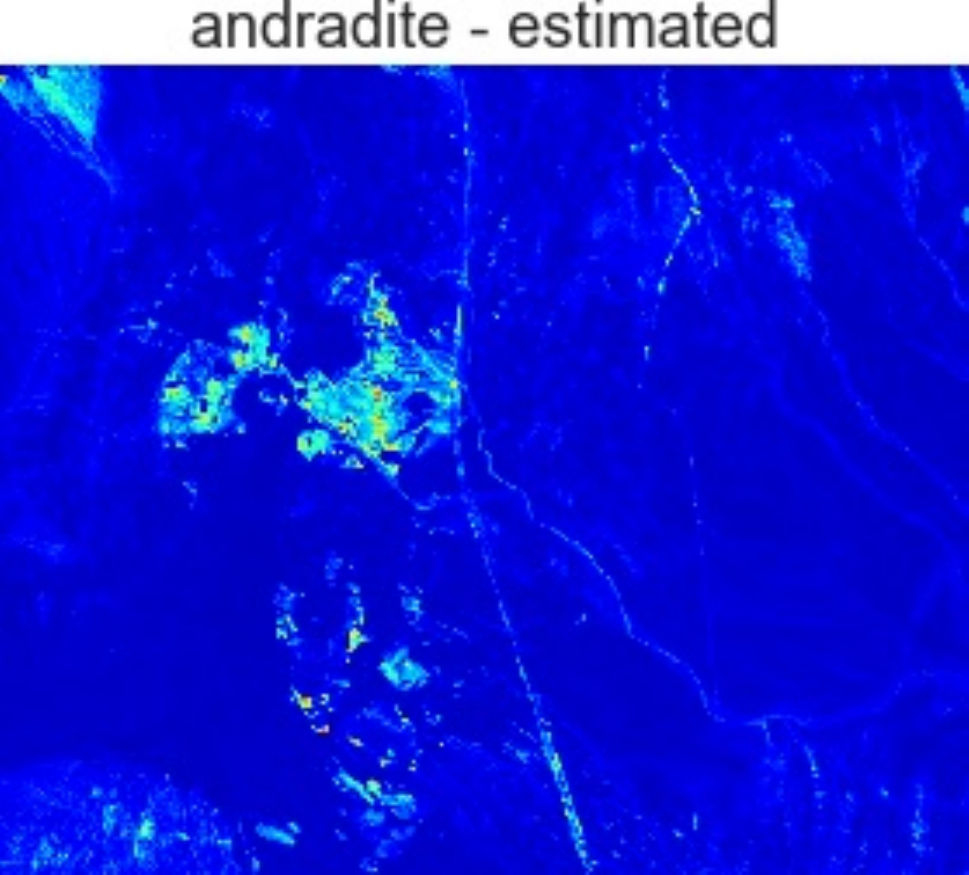}
    \includegraphics[width=0.158\textwidth, trim={0cm 0cm 0cm 1cm},clip]{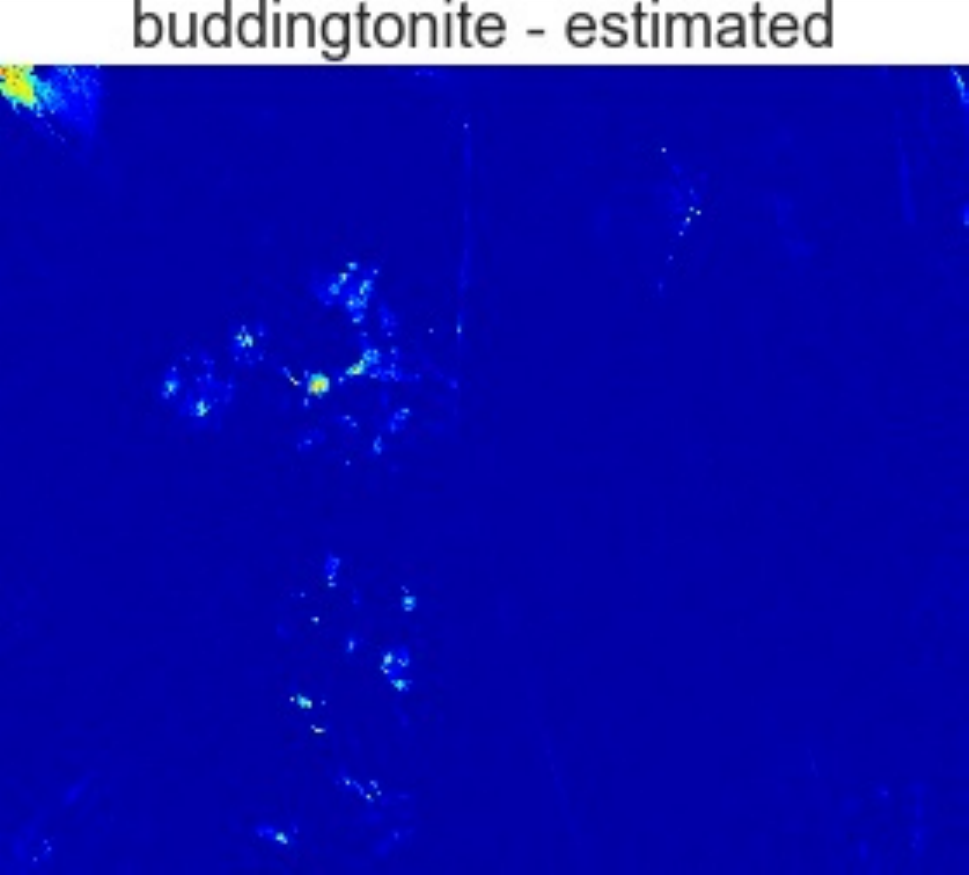}
    \includegraphics[width=0.158\textwidth, trim={0cm 0cm 0cm 1cm},clip]{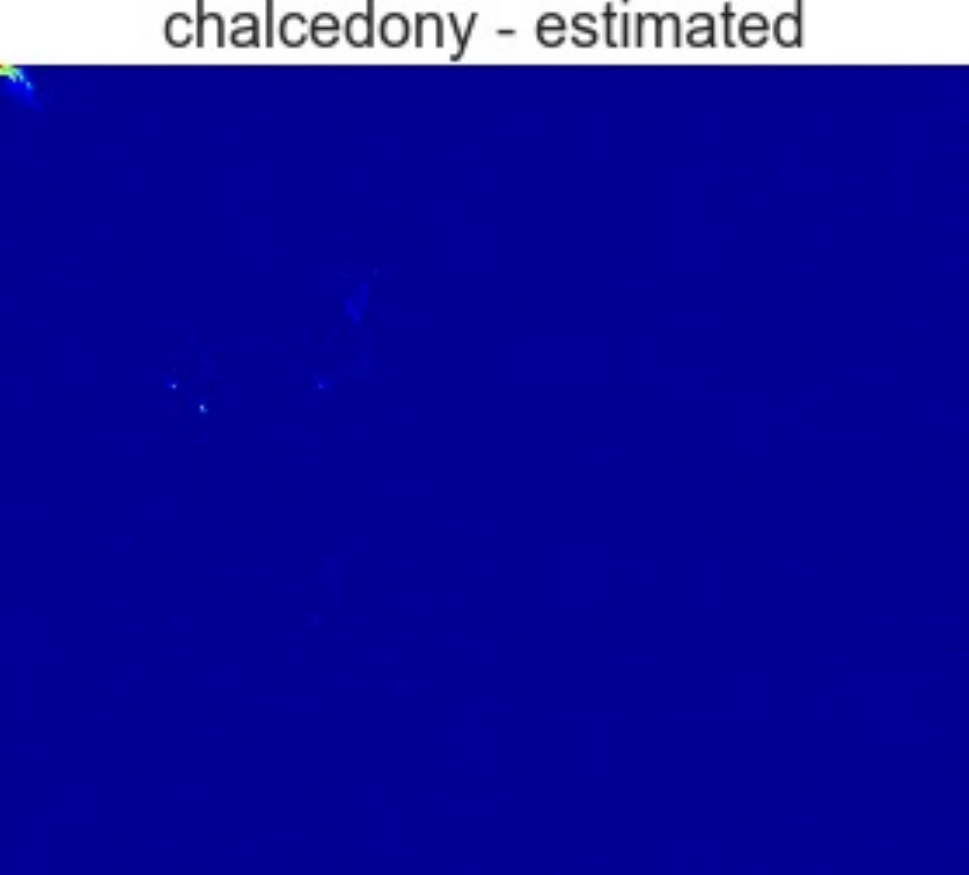}
    \includegraphics[width=0.158\textwidth, trim={0cm 0cm 0cm 1cm},clip]{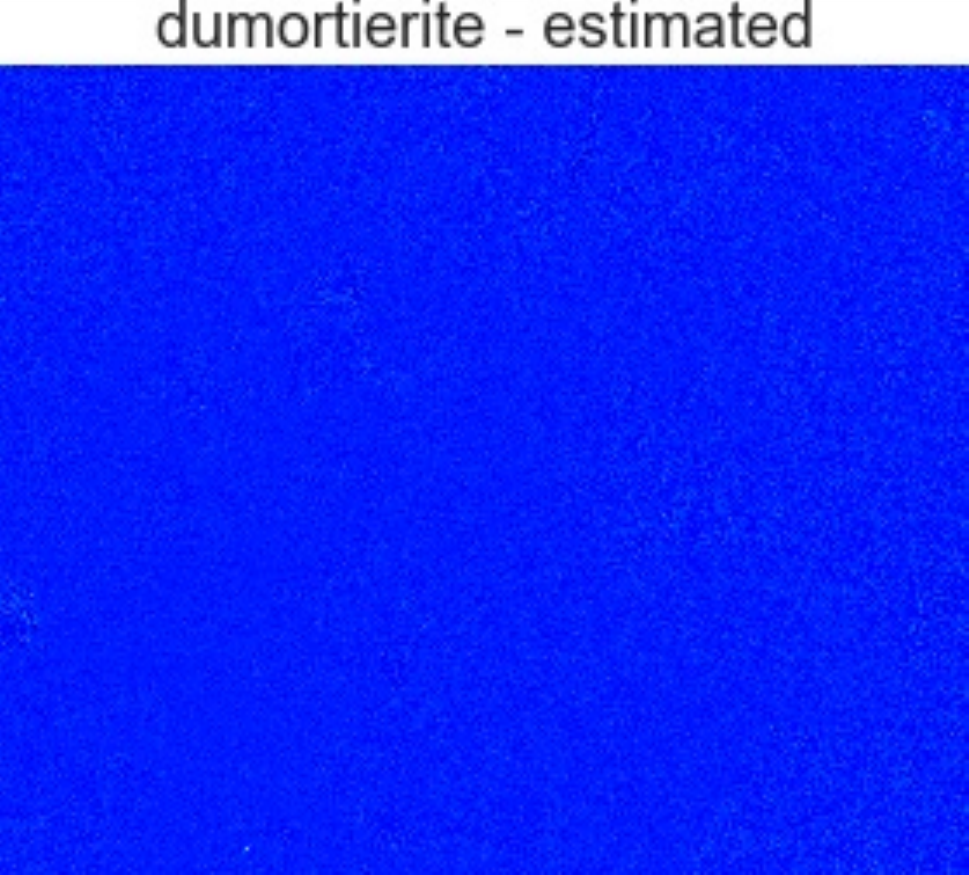}
    \includegraphics[width=0.158\textwidth, trim={0cm 0cm 0cm 1cm},clip]{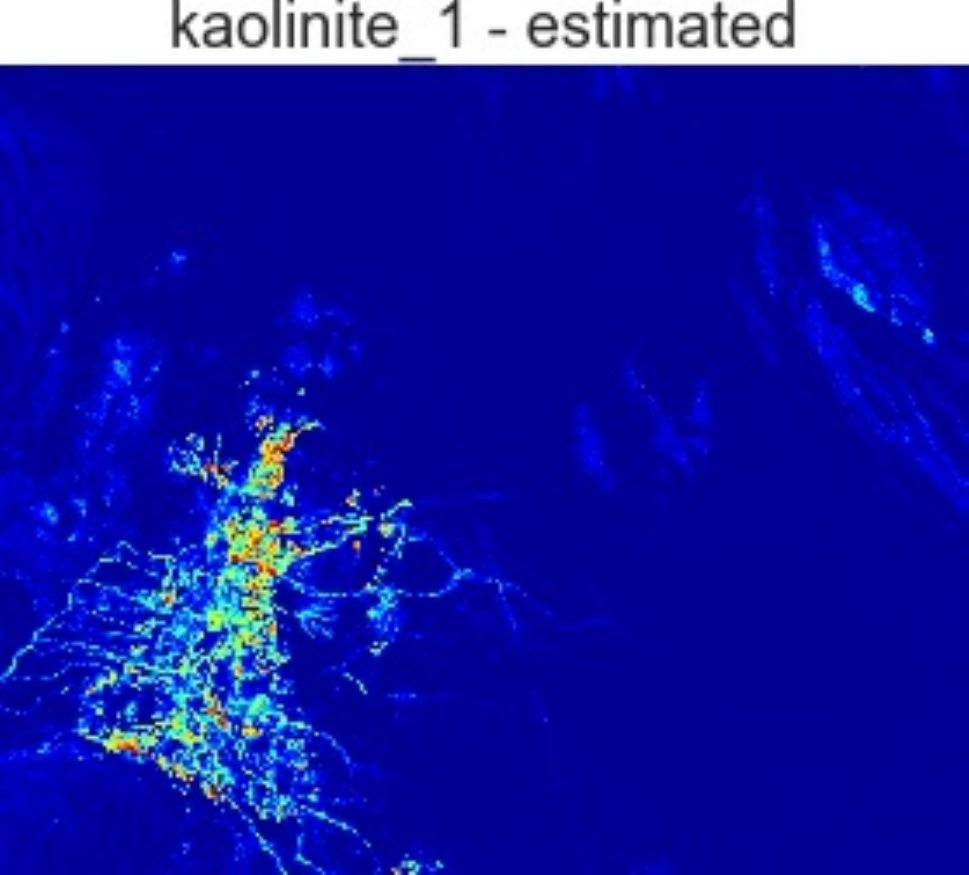}
    \includegraphics[width=0.158\textwidth, trim={0cm 0cm 0cm 1cm},clip]{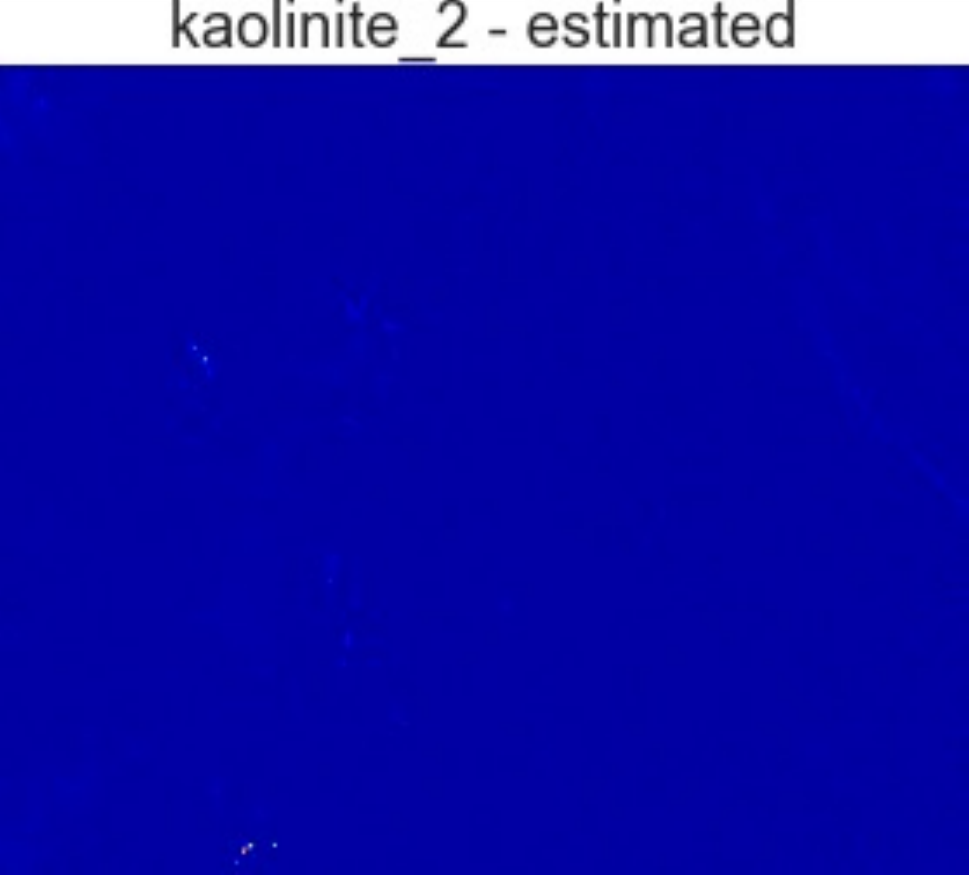}
    \includegraphics[width=0.158\textwidth, trim={0cm 0cm 0cm 1cm},clip]{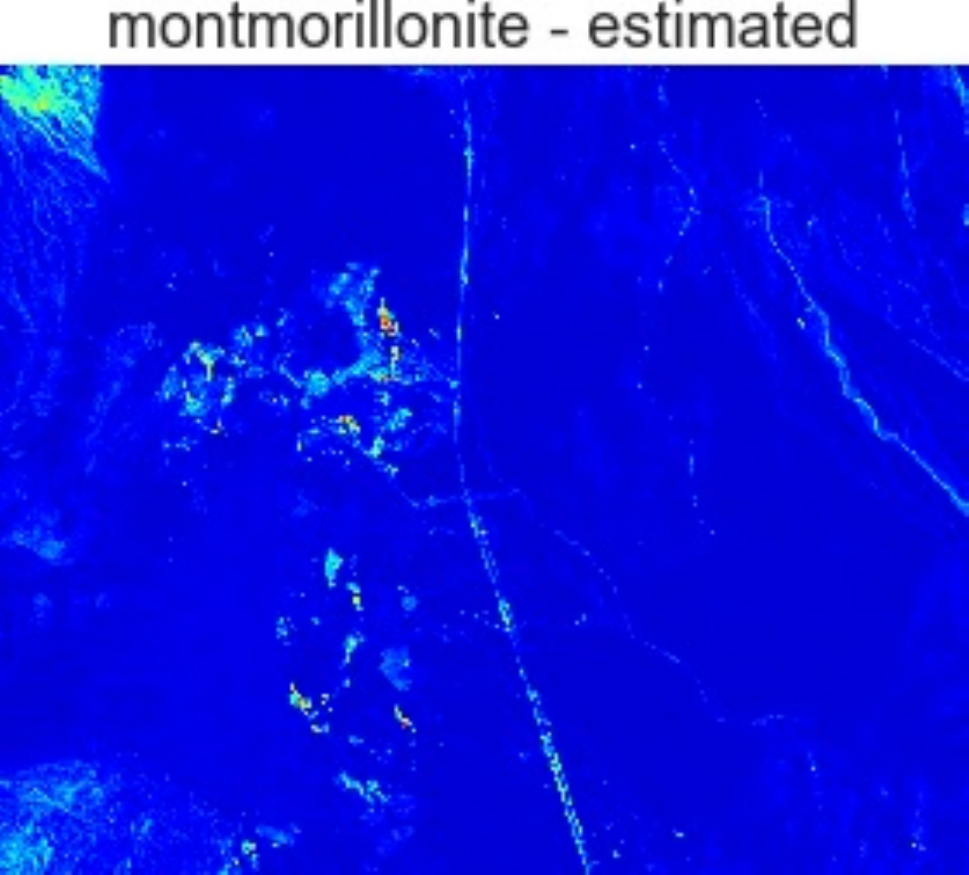}
    \includegraphics[width=0.158\textwidth, trim={0cm 0cm 0cm 1cm},clip]{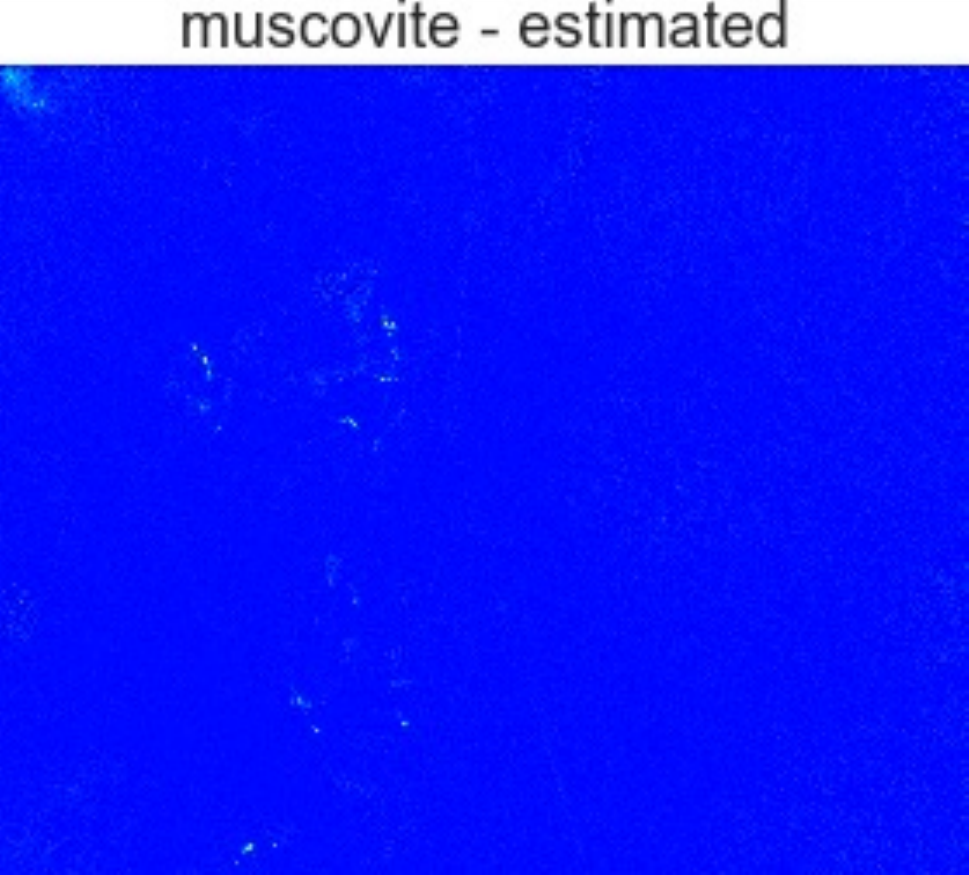}
    \includegraphics[width=0.158\textwidth, trim={0cm 0cm 0cm 1cm},clip]{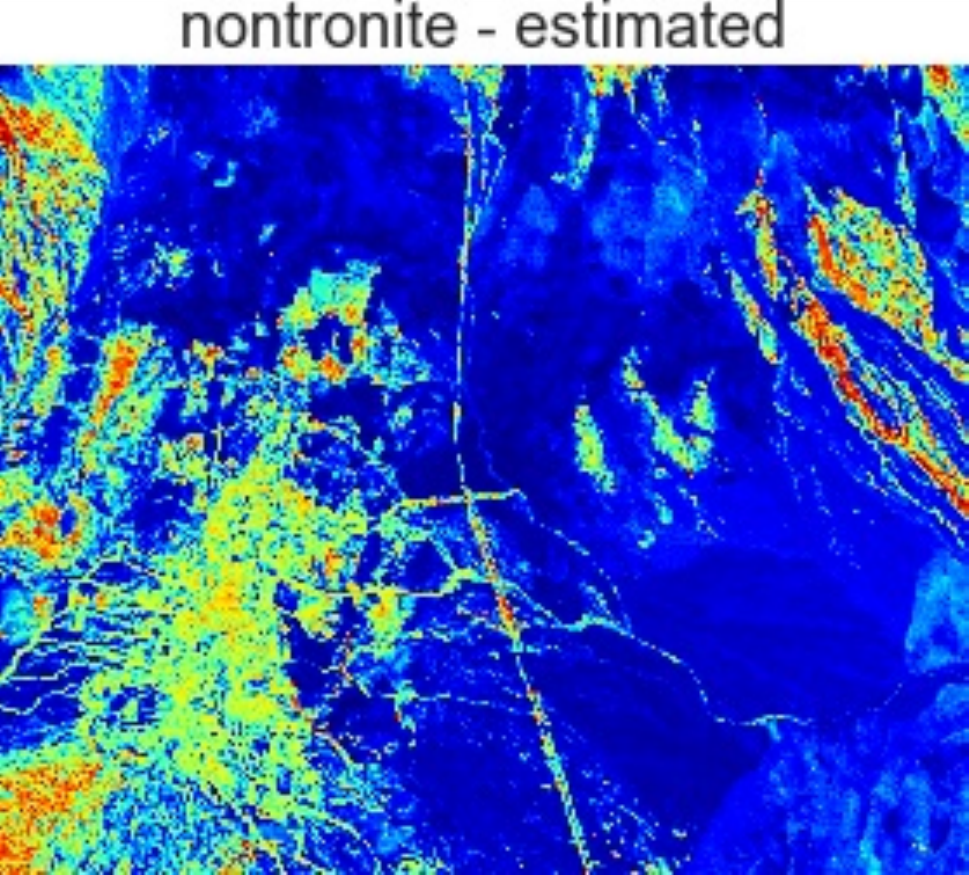}
    \includegraphics[width=0.158\textwidth, trim={0cm 0cm 0cm 1cm},clip]{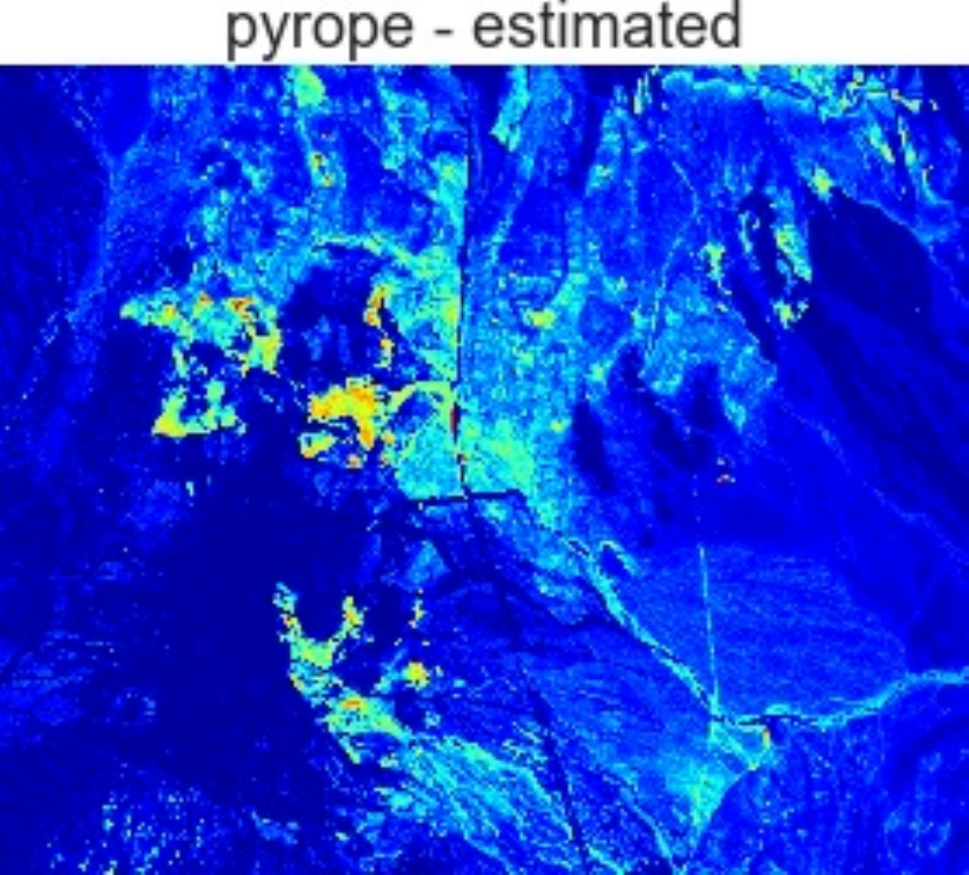}
    \includegraphics[width=0.158\textwidth, trim={0cm 0cm 0cm 1cm},clip]{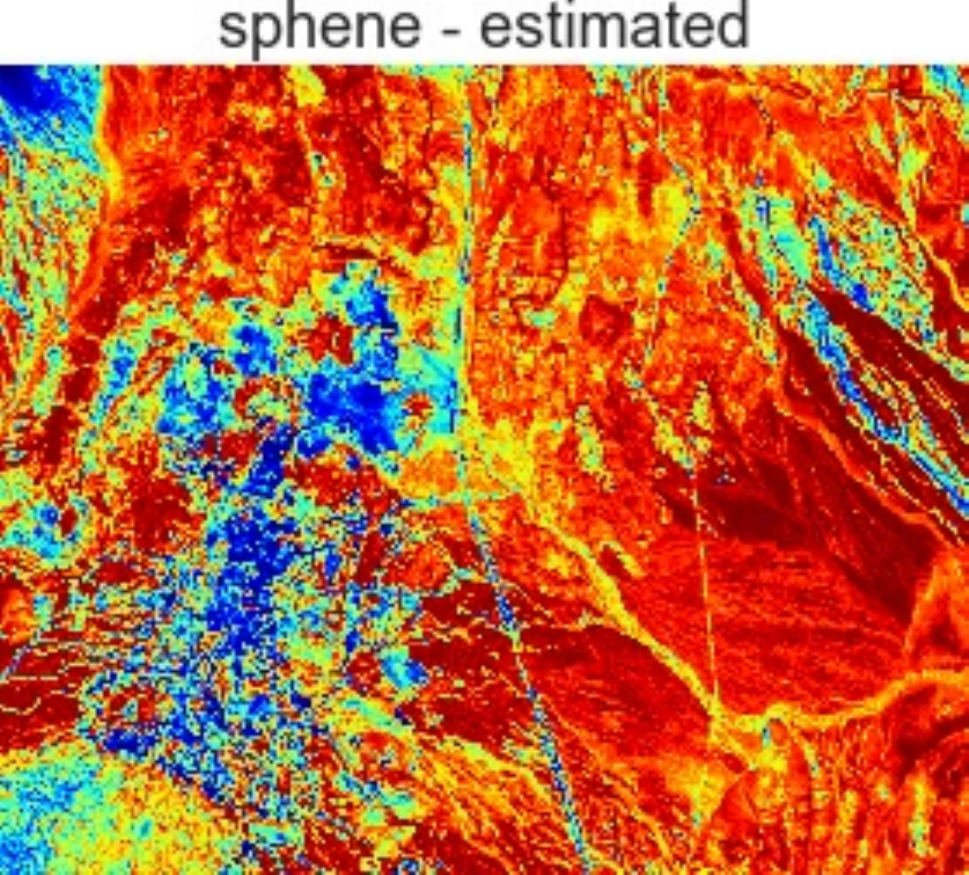}
    \caption{Abundances maps of the Cuprite dataset estimated by LDVAE (ground truth not available). From left to right:
        Alunite, Andradite, Buddingtonite, Chalcedony, Dumortierite, Kaolinite1, Kaolinite2, Montmorillonite, Muscovite
        Nontronite, Pyrope, and Sphene.}
    \label{fig:cuprite_abundances}
\end{figure*}

The worst performance of LDVAE was on the HYDICE Urban dataset, even though it had the larger training set.
We used the 50/50 split strategy for training and evaluation of this model.
We started with an initial random separation; then, we manually fixed some unbalanced classes. It is noticeable that all machine learning approaches suffer from these issues. We plan to investigate limitations and performance issues surrounding unbalanced input data at another time.

\begin{figure*}
    \centering
    \includegraphics[width=0.155\textwidth, trim={0cm 0cm 0cm 1cm},clip]{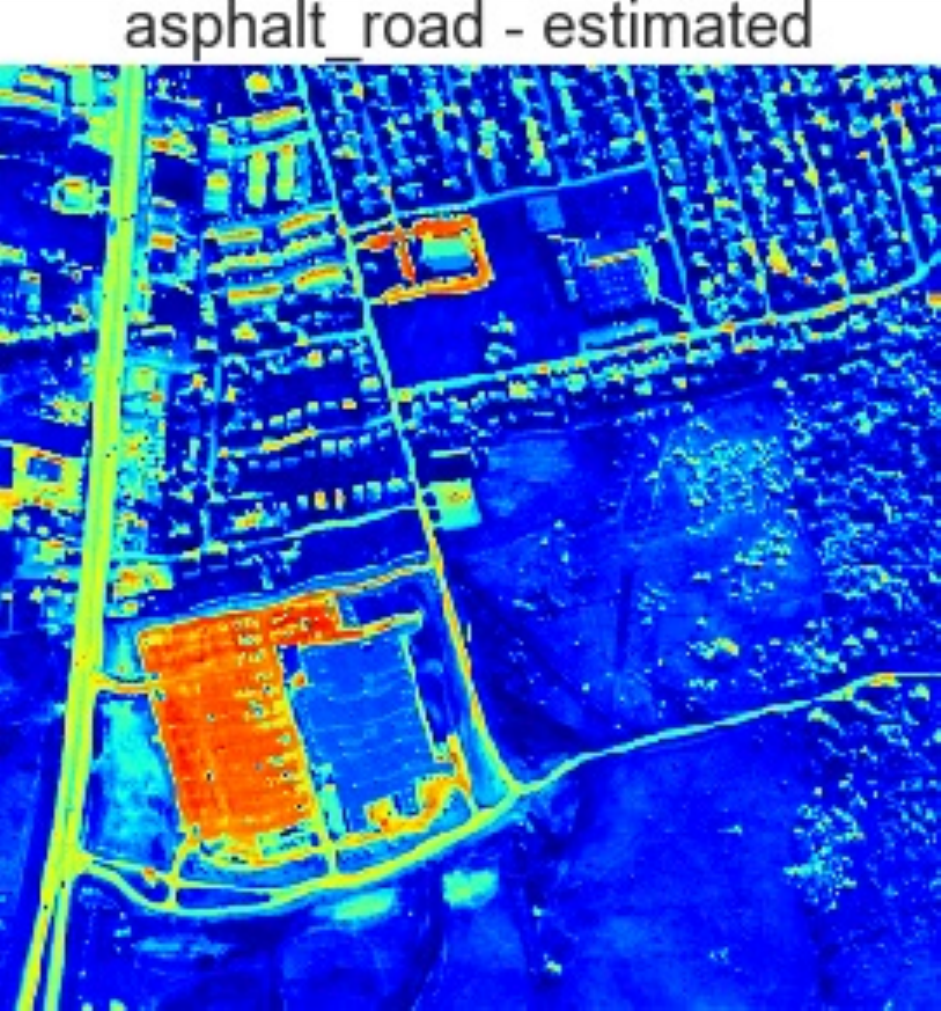}
    \includegraphics[width=0.155\textwidth, trim={0cm 0cm 0cm 1cm},clip]{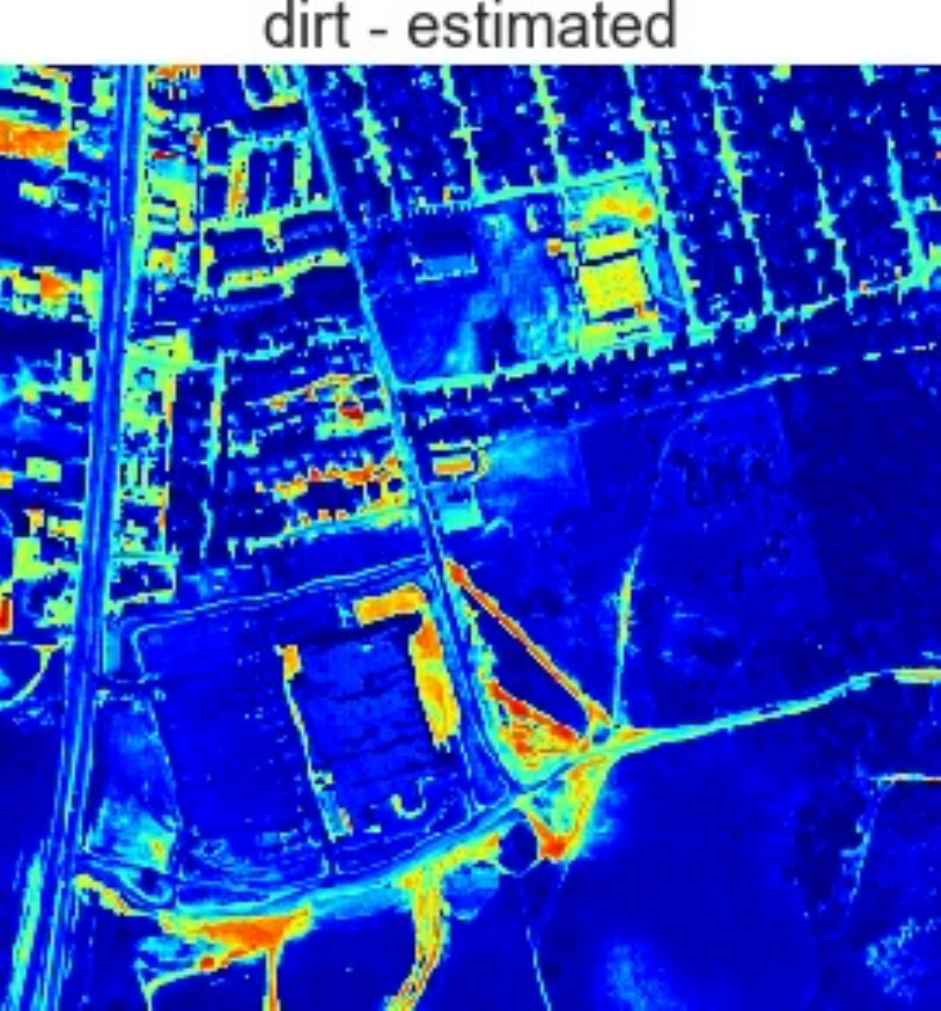}
    \includegraphics[width=0.155\textwidth, trim={0cm 0cm 0cm 1cm},clip]{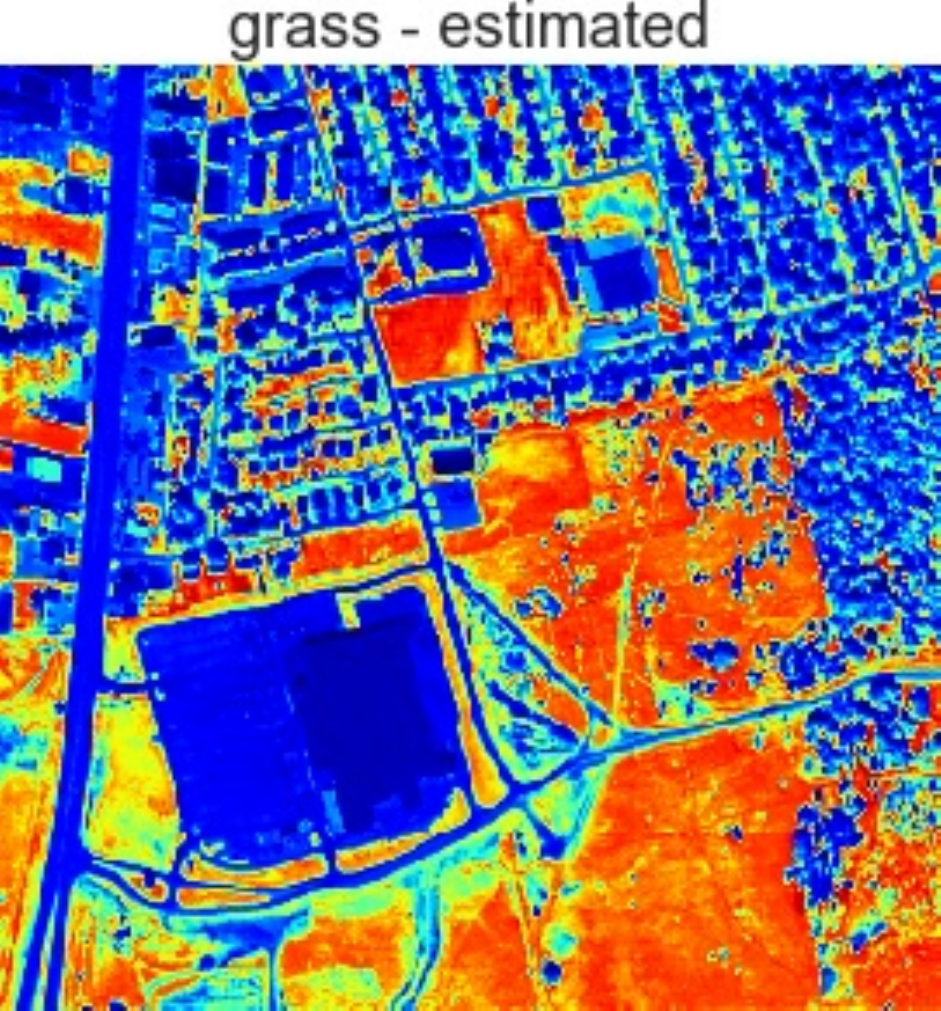}
    \includegraphics[width=0.155\textwidth, trim={0cm 0cm 0cm 1cm},clip]{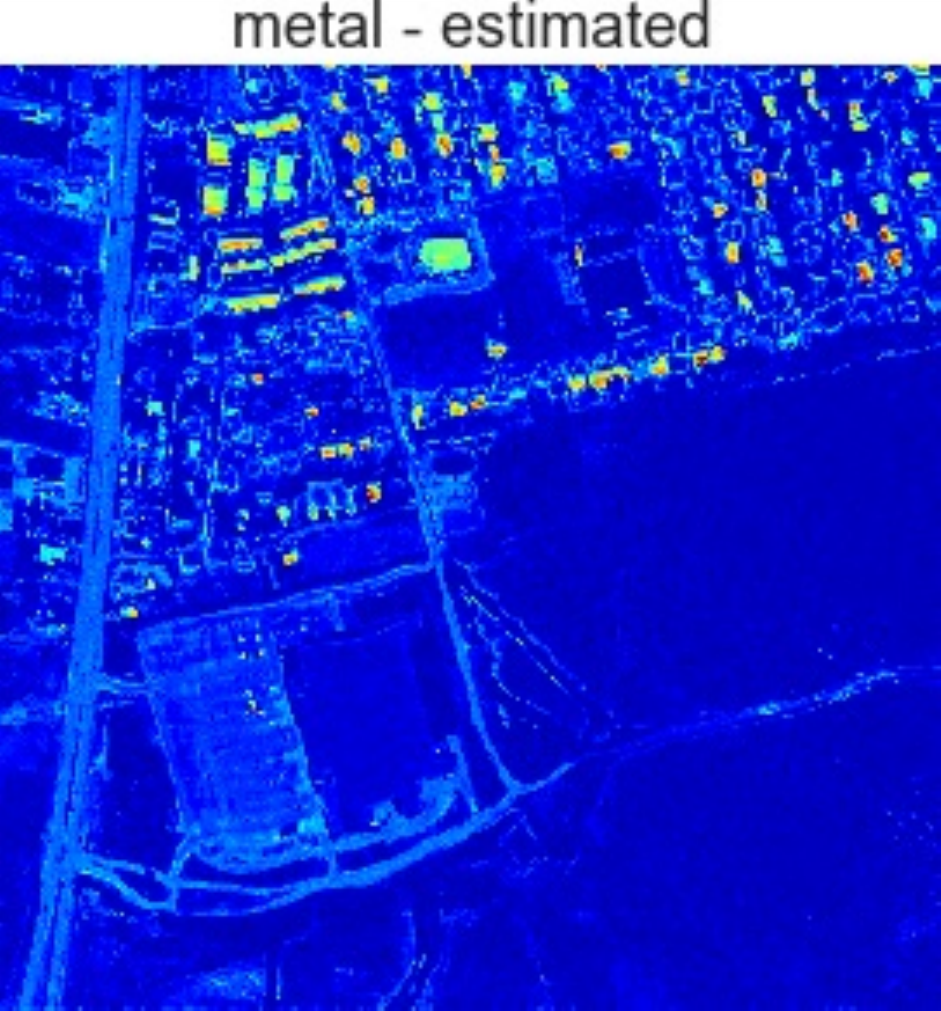}
    \includegraphics[width=0.155\textwidth, trim={0cm 0cm 0cm 1cm},clip]{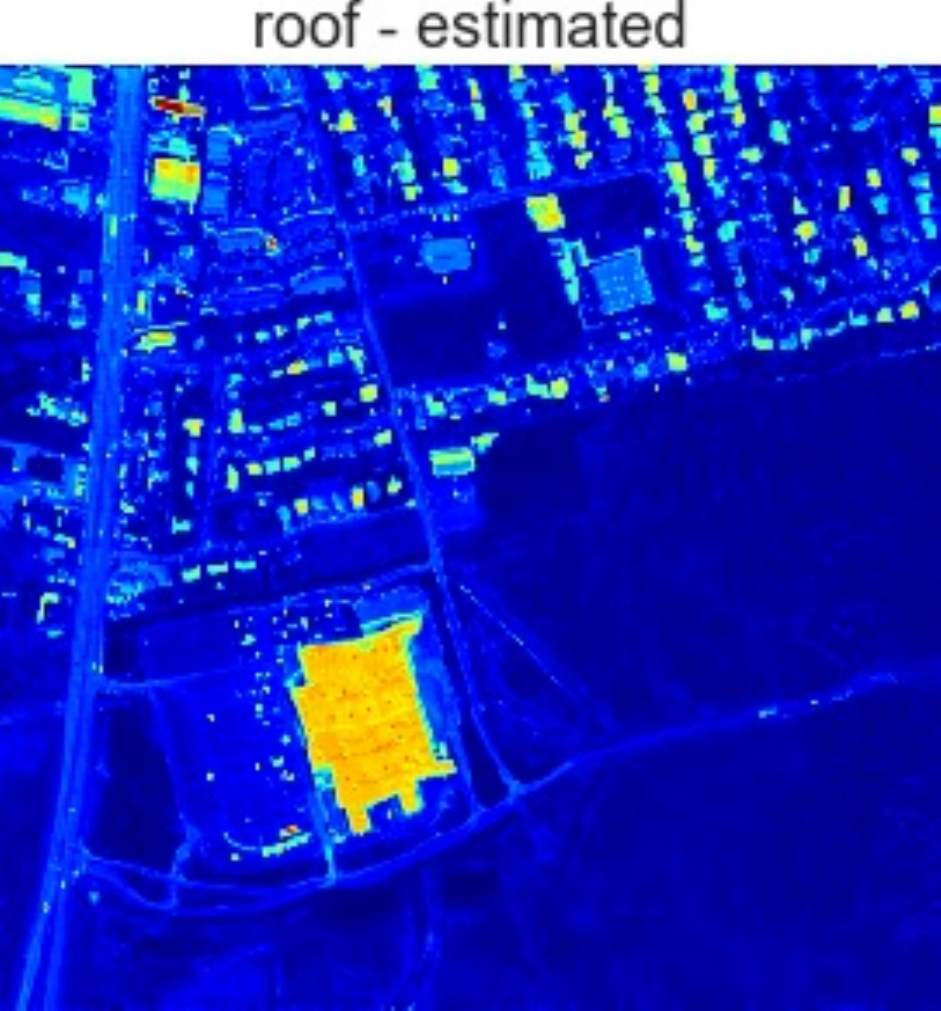}
    \includegraphics[width=0.155\textwidth, trim={0cm 0cm 0cm 1cm},clip]{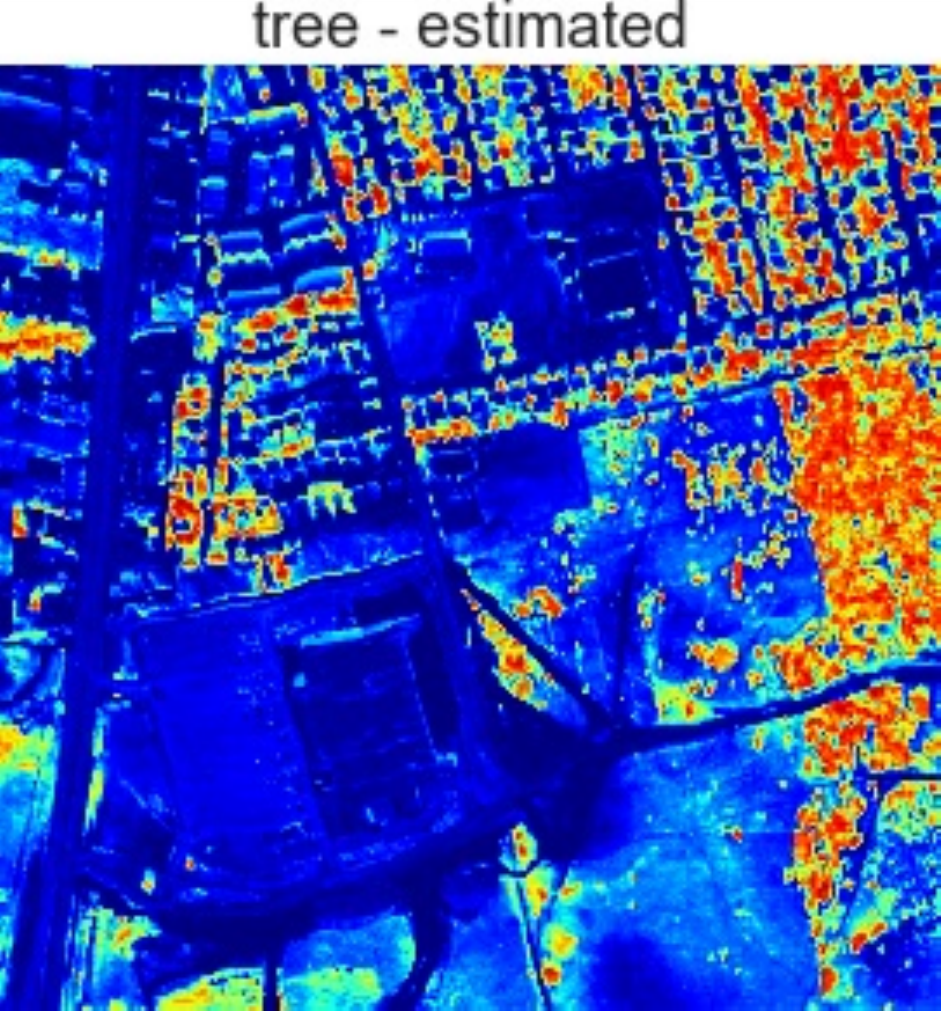} \linebreak
    \includegraphics[width=0.155\textwidth, trim={0cm 0cm 0cm 1cm},clip]{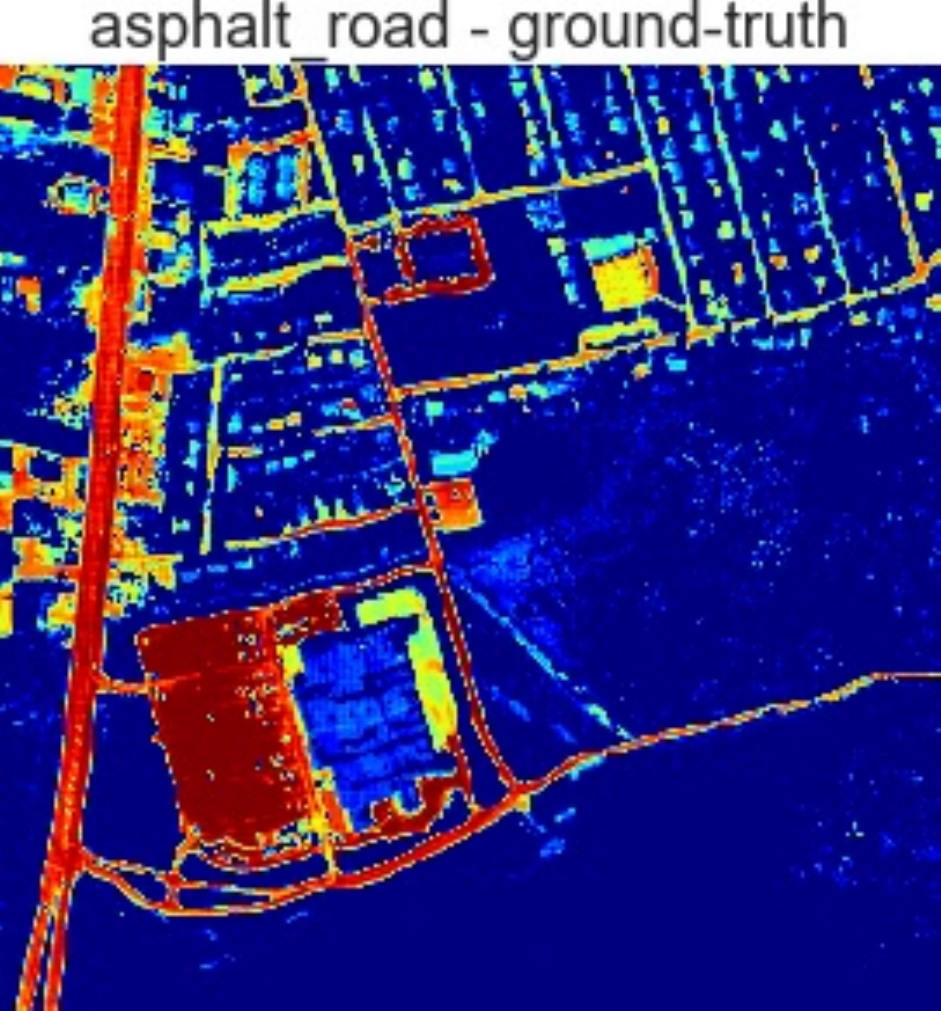}
    \includegraphics[width=0.155\textwidth, trim={0cm 0cm 0cm 1cm},clip]{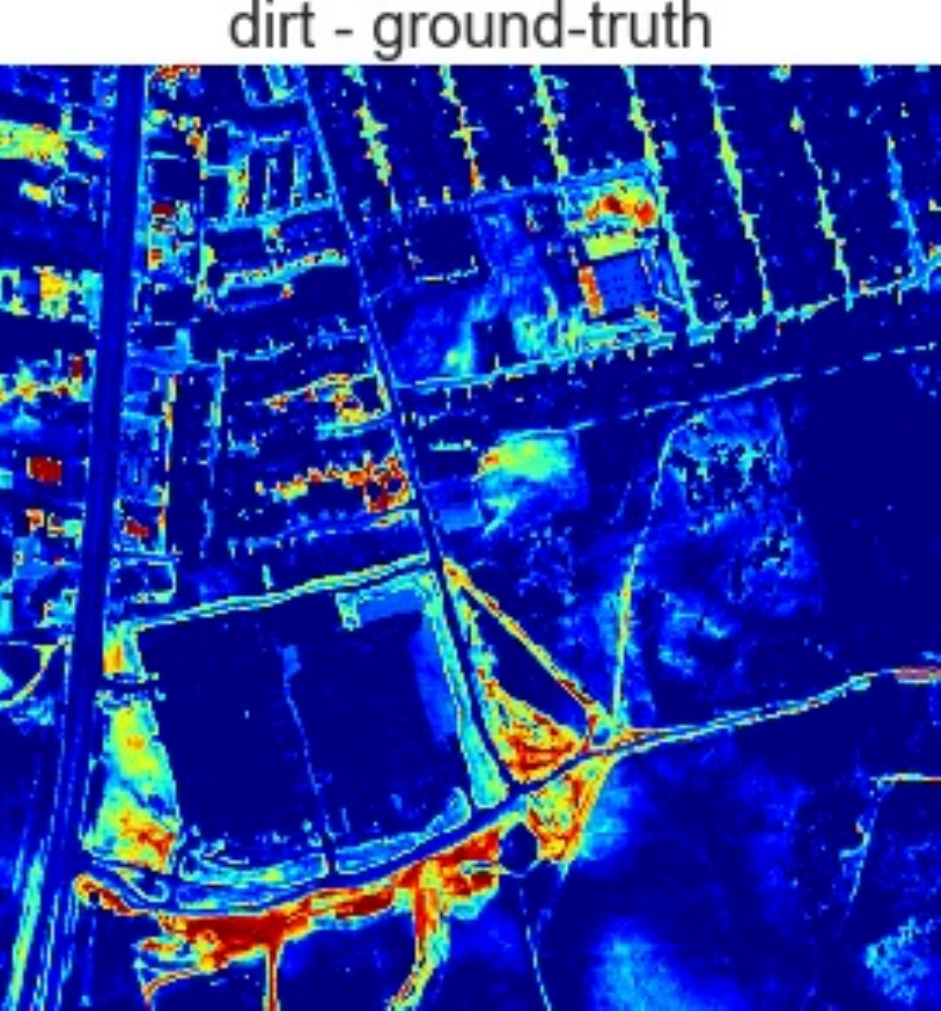}
    \includegraphics[width=0.155\textwidth, trim={0cm 0cm 0cm 1cm},clip]{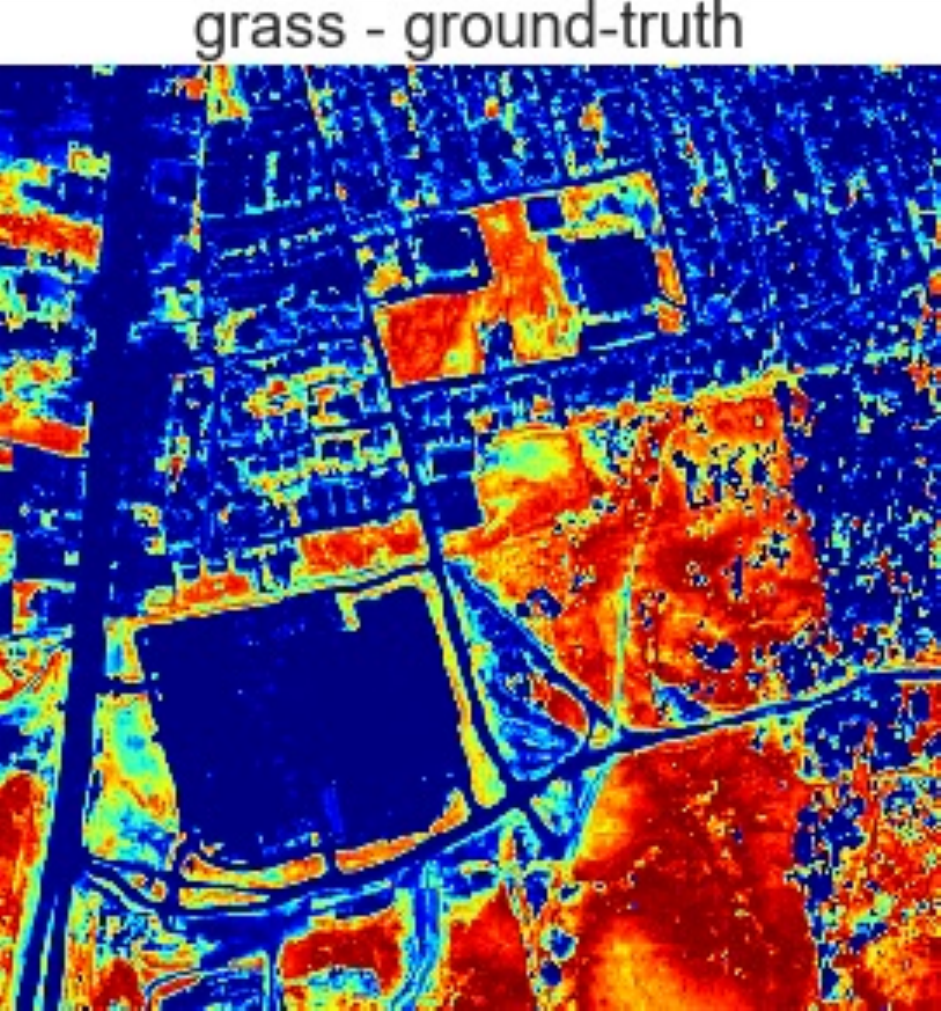}
    \includegraphics[width=0.155\textwidth, trim={0cm 0cm 0cm 1cm},clip]{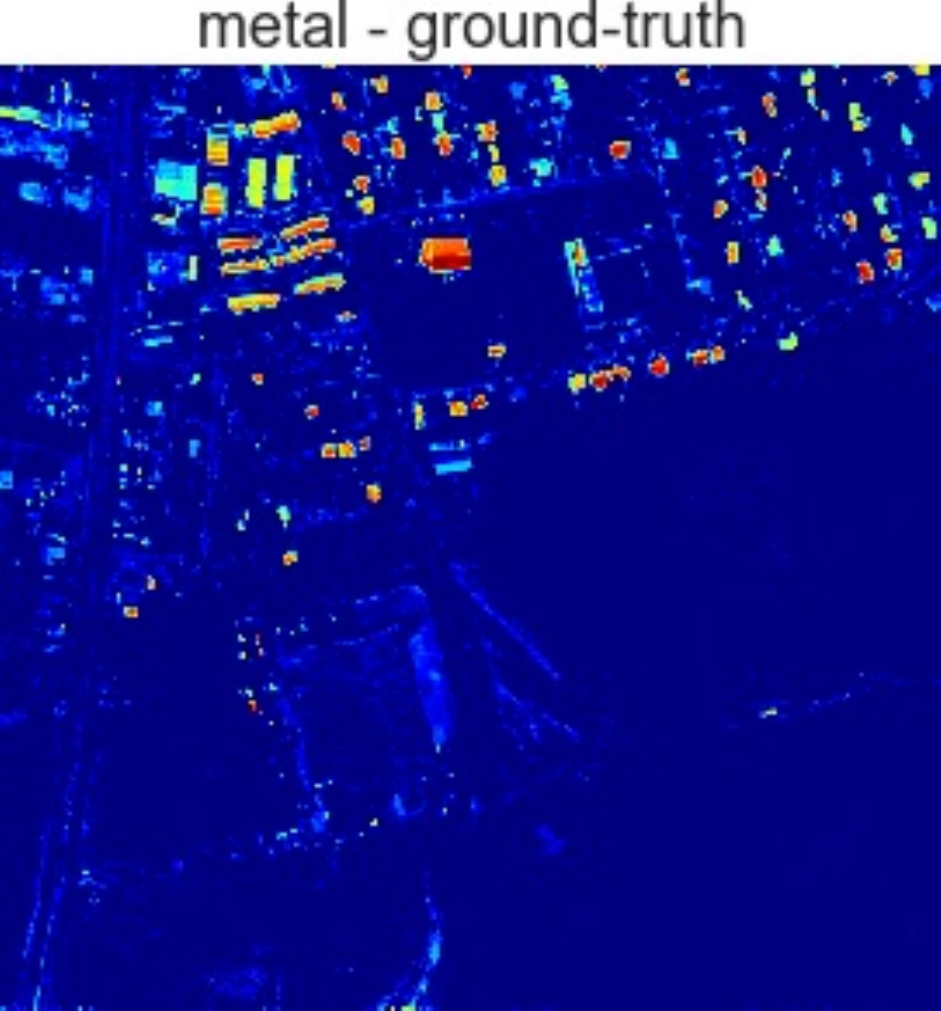}
    \includegraphics[width=0.155\textwidth, trim={0cm 0cm 0cm 1cm},clip]{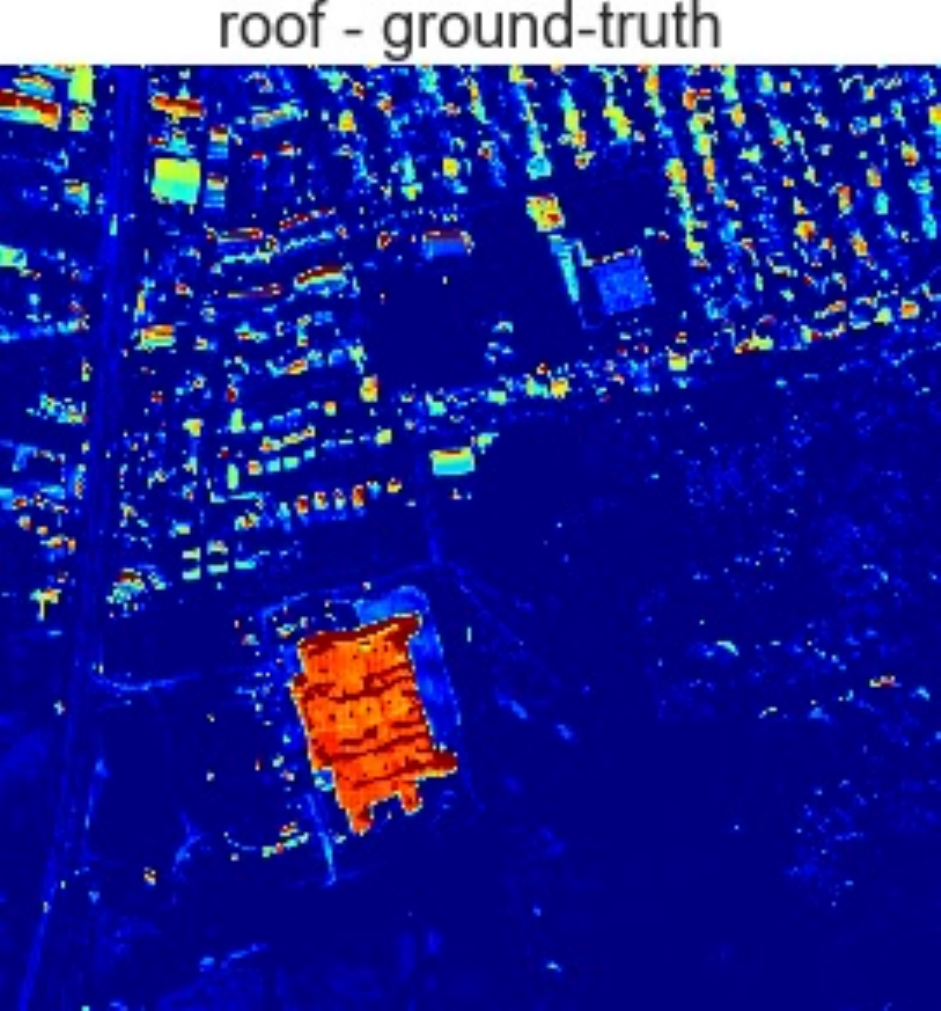}
    \includegraphics[width=0.155\textwidth, trim={0cm 0cm 0cm 1cm},clip]{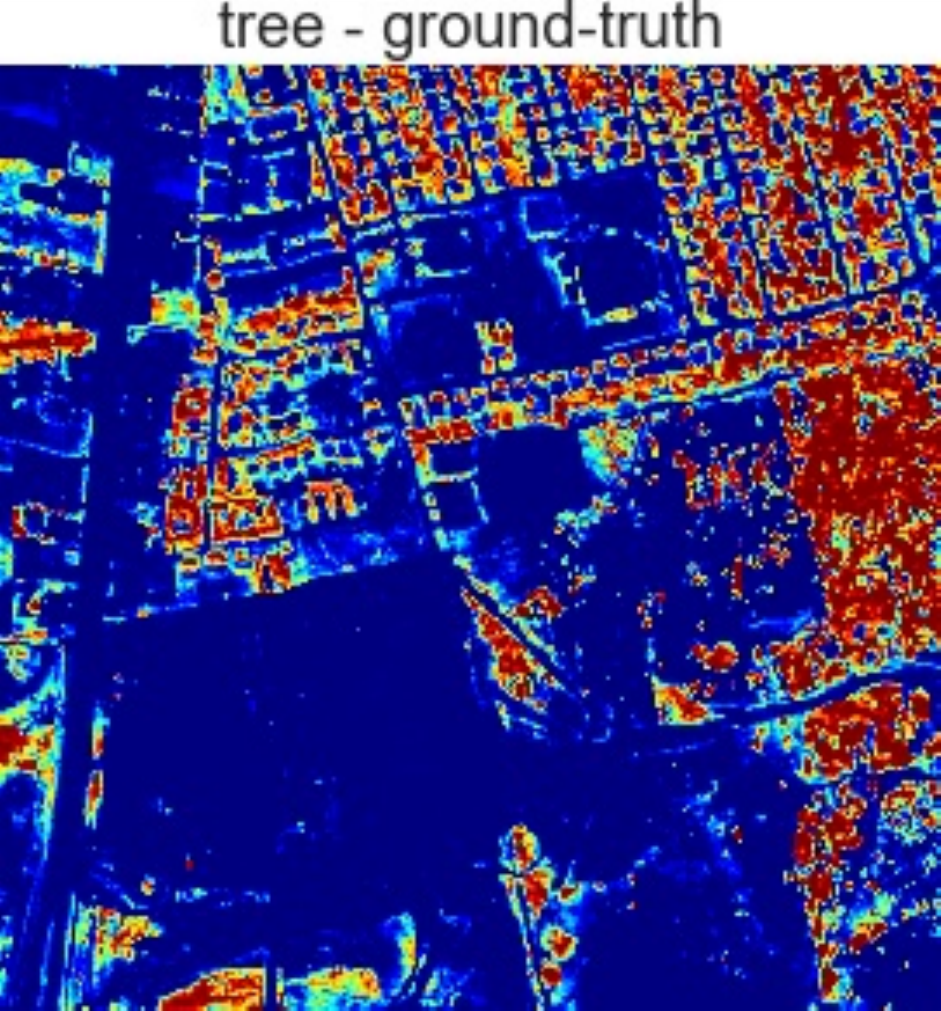}
    \caption{Abundances maps of the HYDICE Urban dataset (top row: LDVAE, bottom row: ground truth). From left to right: asphalt, dirt, grass, metal}
    \label{fig:hydice_urban_abundances}
\end{figure*}
\begin{figure*}
    \centering
    \includegraphics[width=0.36\textwidth,trim={  0cm 1.3cm 0cm 0cm},clip]{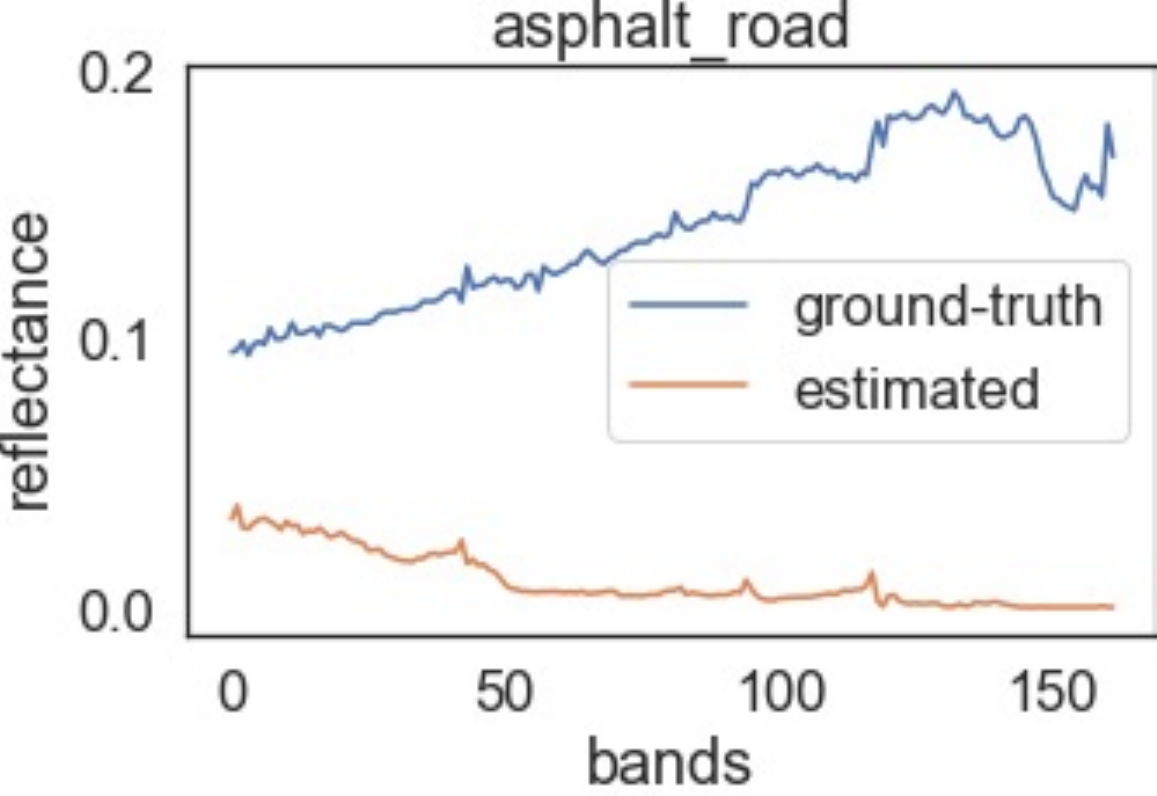}
    \includegraphics[width=0.31\textwidth,trim={1.7cm 1.3cm 0cm 0cm},clip]{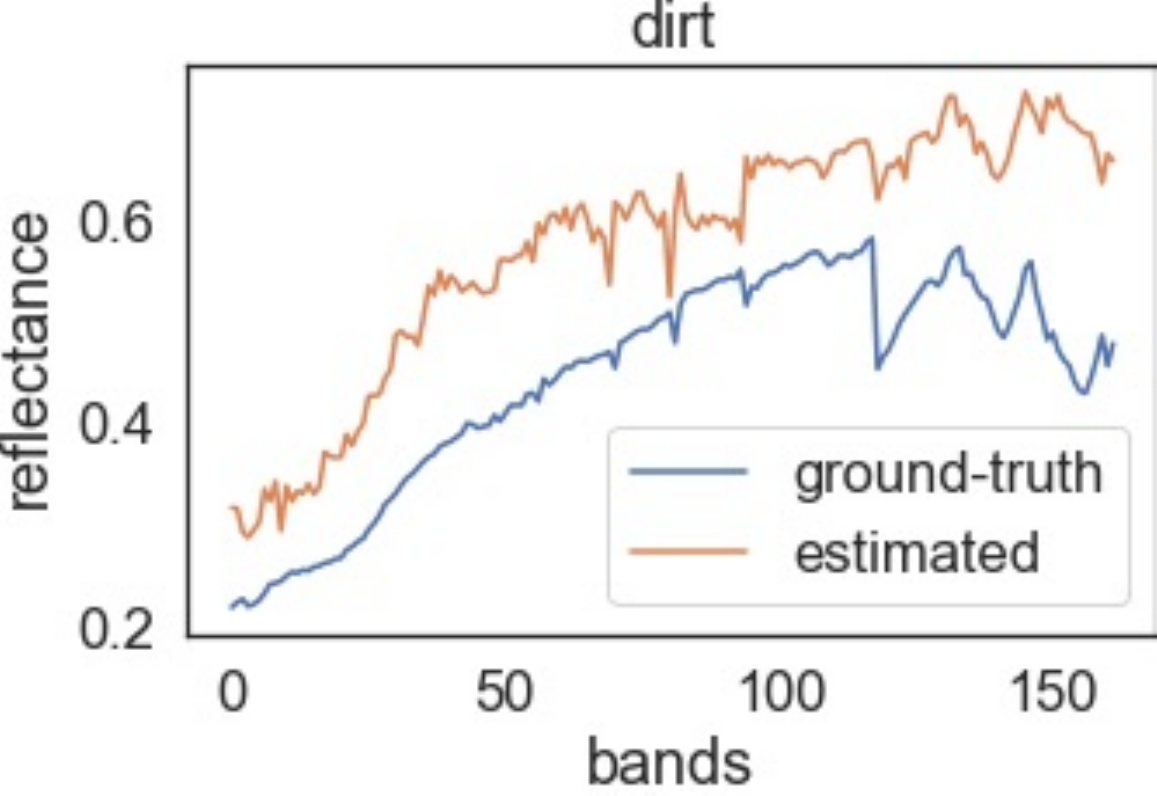}
    \includegraphics[width=0.31\textwidth,trim={1.7cm 1.3cm 0cm 0cm},clip]{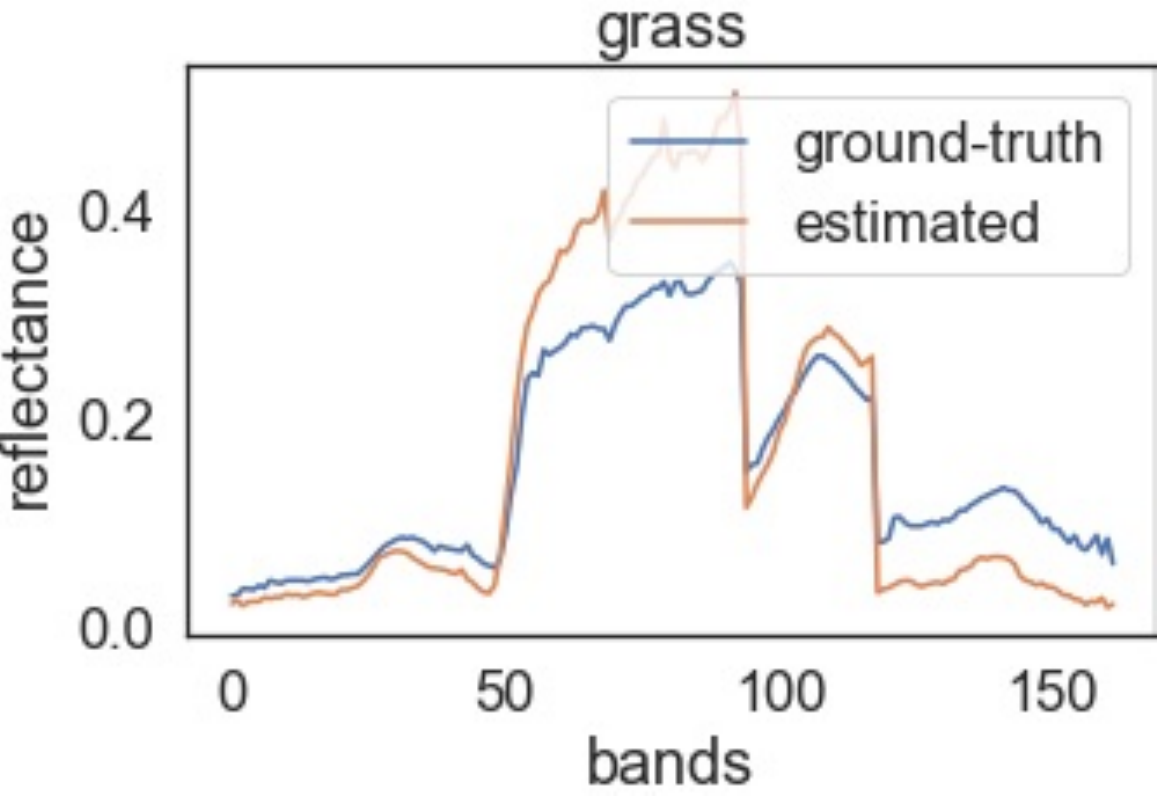}\linebreak
    \includegraphics[width=0.36\textwidth,trim={  0cm   0cm 0cm 0cm},clip]{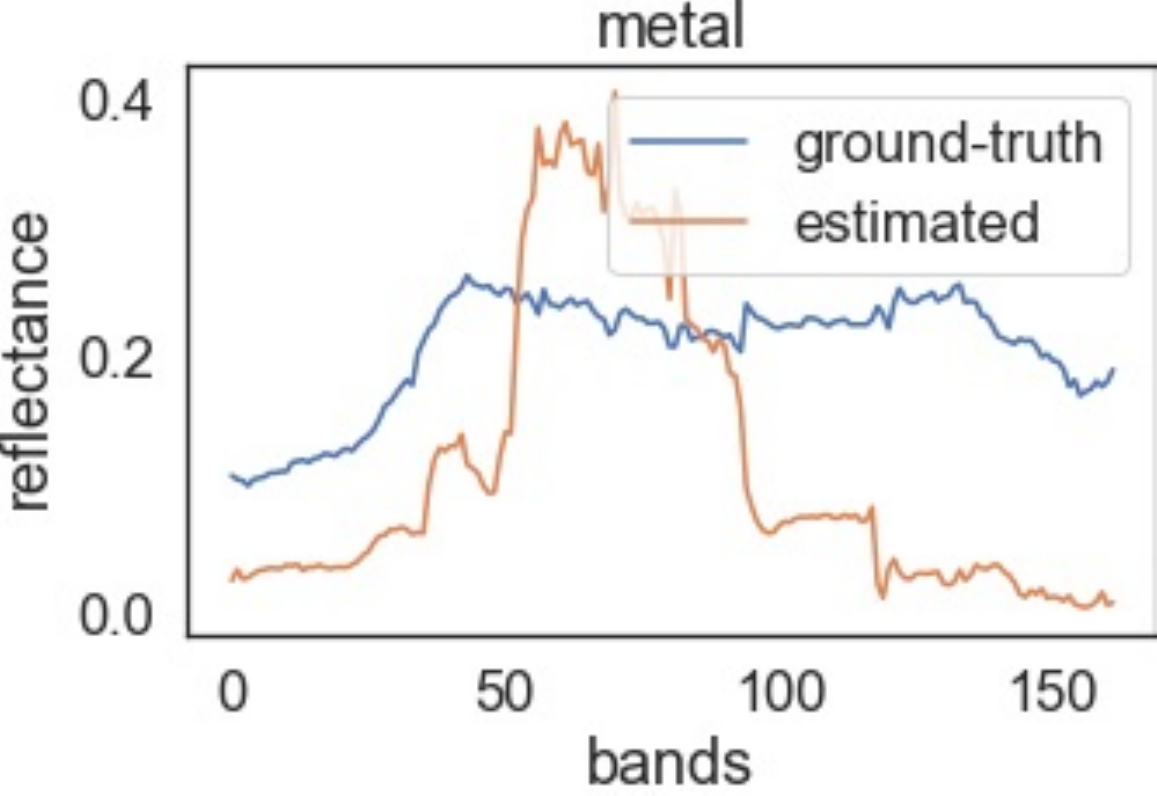}
    \includegraphics[width=0.31\textwidth,trim={1.7cm   0cm 0cm 0cm},clip]{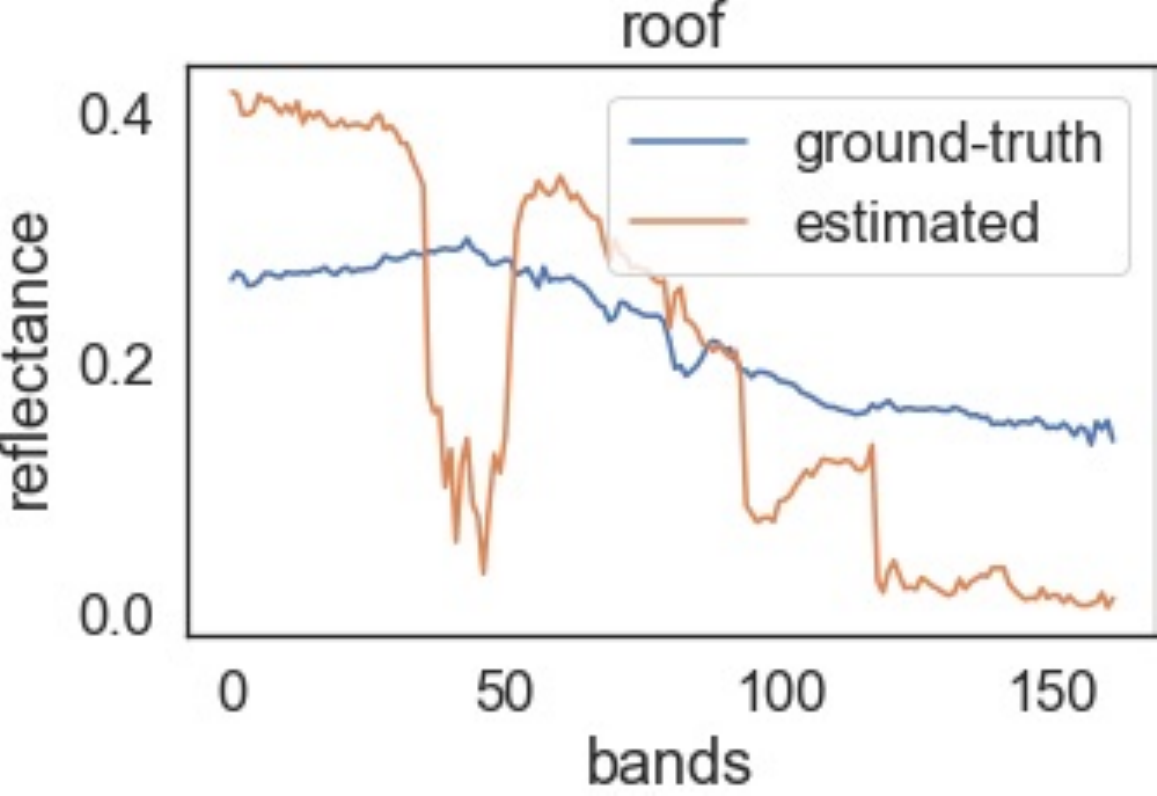}
    \includegraphics[width=0.31\textwidth,trim={1.7cm   0cm 0cm 0cm},clip]{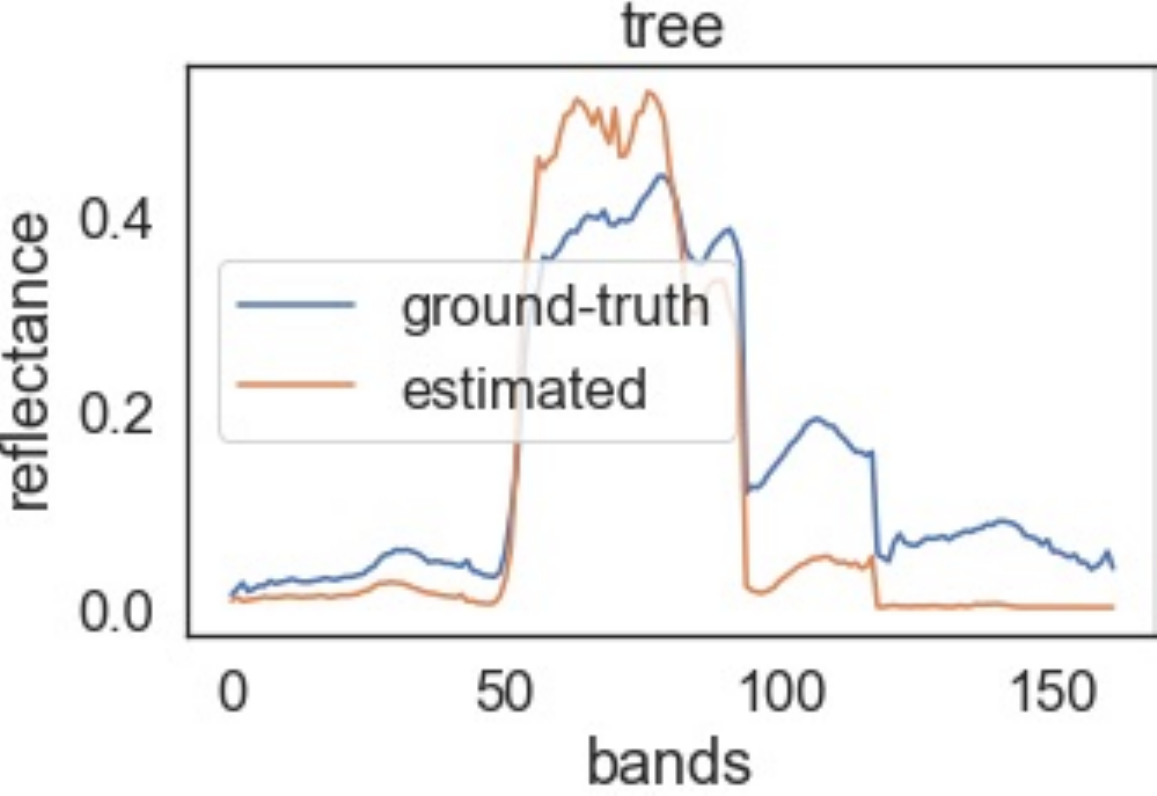}
    \caption{Endmembers of the HYDICE Urban dataset generated by LDVAE: comparison with ground truth.}
    \label{fig:endmembers_hydice_urban}
\end{figure*}

\begin{figure*}
    \centering
    \includegraphics[width=0.150\textwidth,trim={0cm 0cm 0cm 1cm},clip]{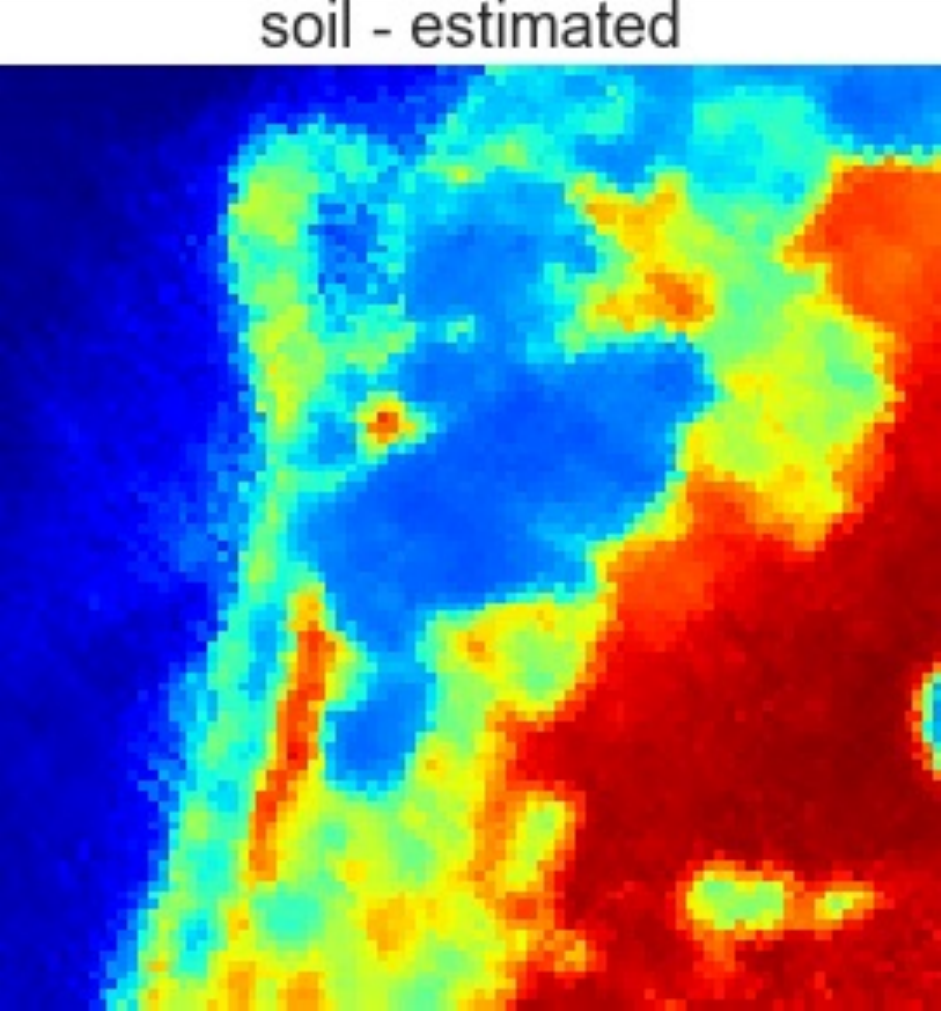}
    \includegraphics[width=0.150\textwidth,trim={0cm 0cm 0cm 1cm},clip]{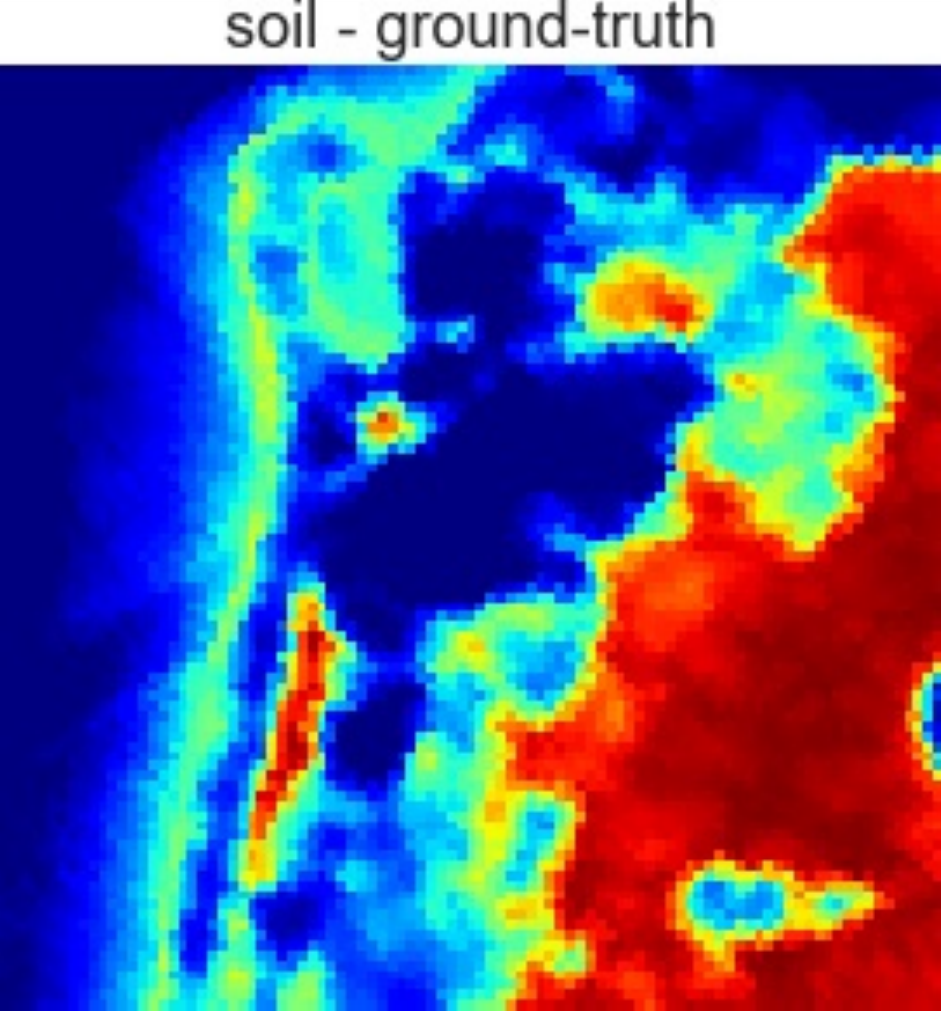} \hspace{5pt}
    \includegraphics[width=0.150\textwidth,trim={0cm 0cm 0cm 1cm},clip]{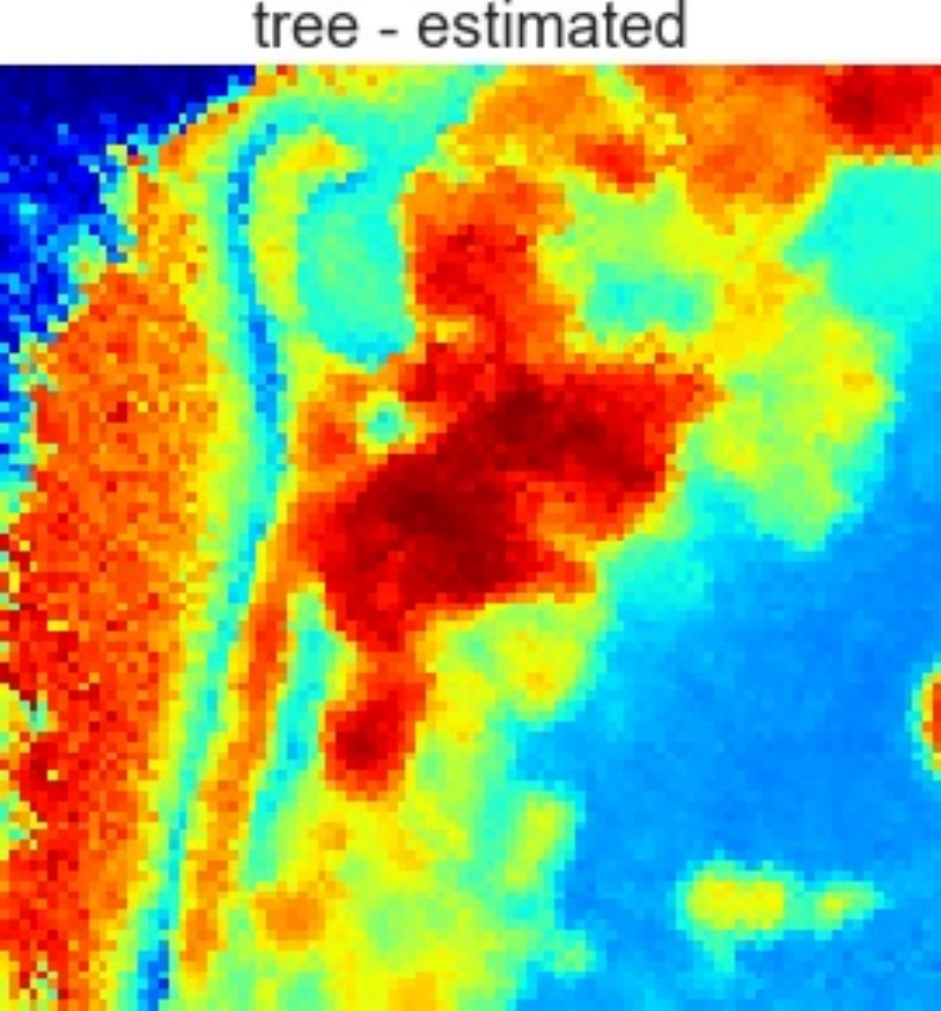}
    \includegraphics[width=0.150\textwidth,trim={0cm 0cm 0cm 1cm},clip]{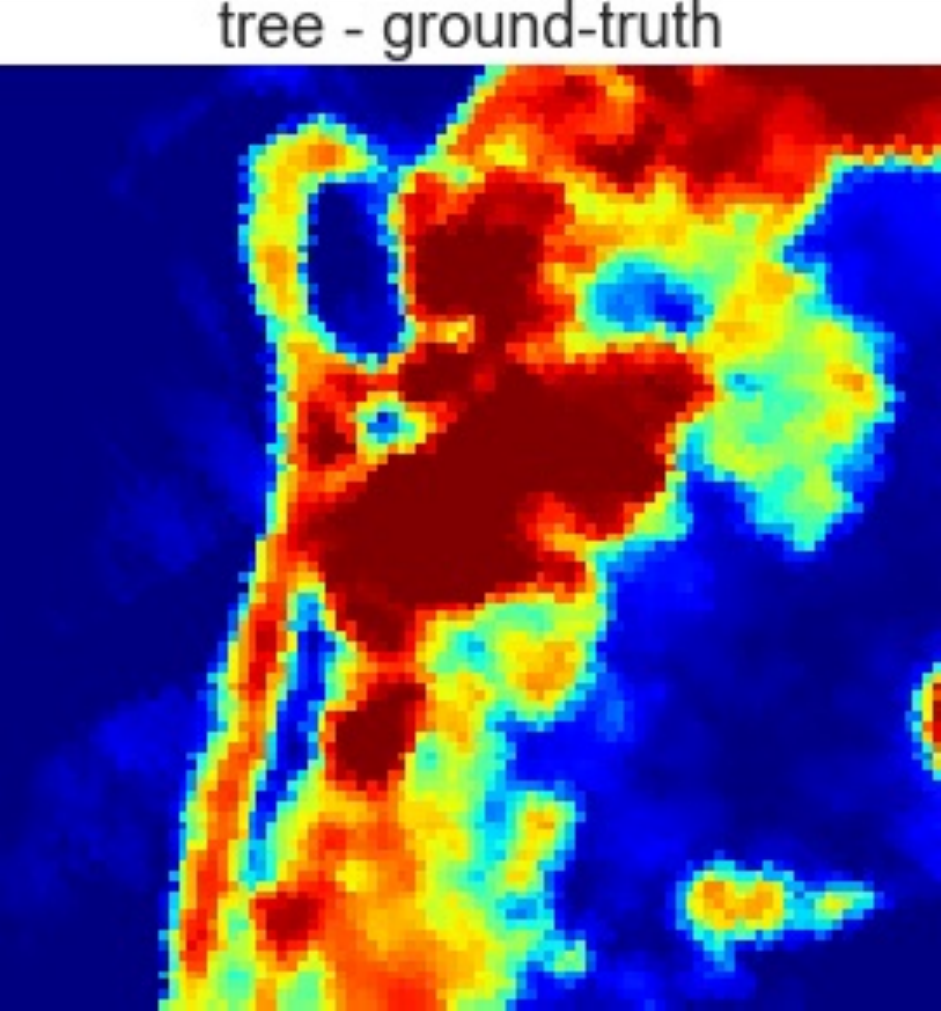} \hspace{5pt}
    \includegraphics[width=0.150\textwidth,trim={0cm 0cm 0cm 1cm},clip]{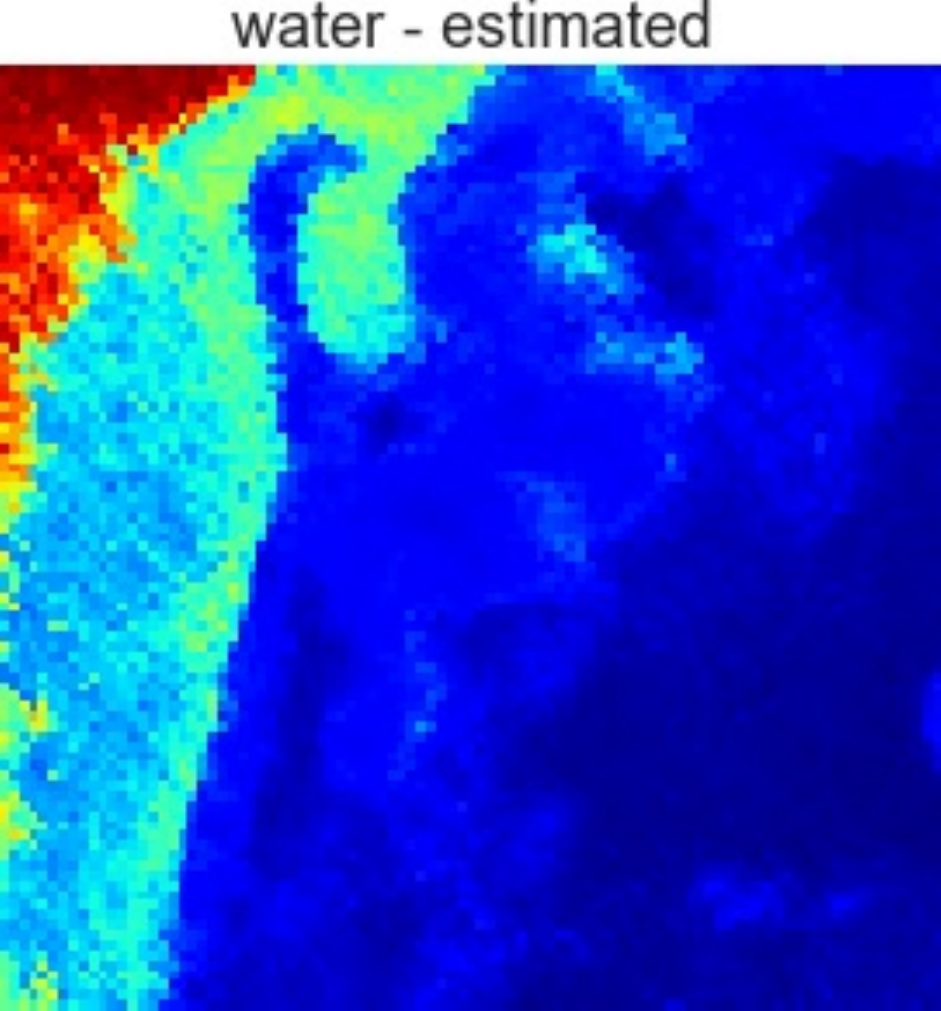}
    \includegraphics[width=0.150\textwidth,trim={0cm 0cm 0cm 1cm},clip]{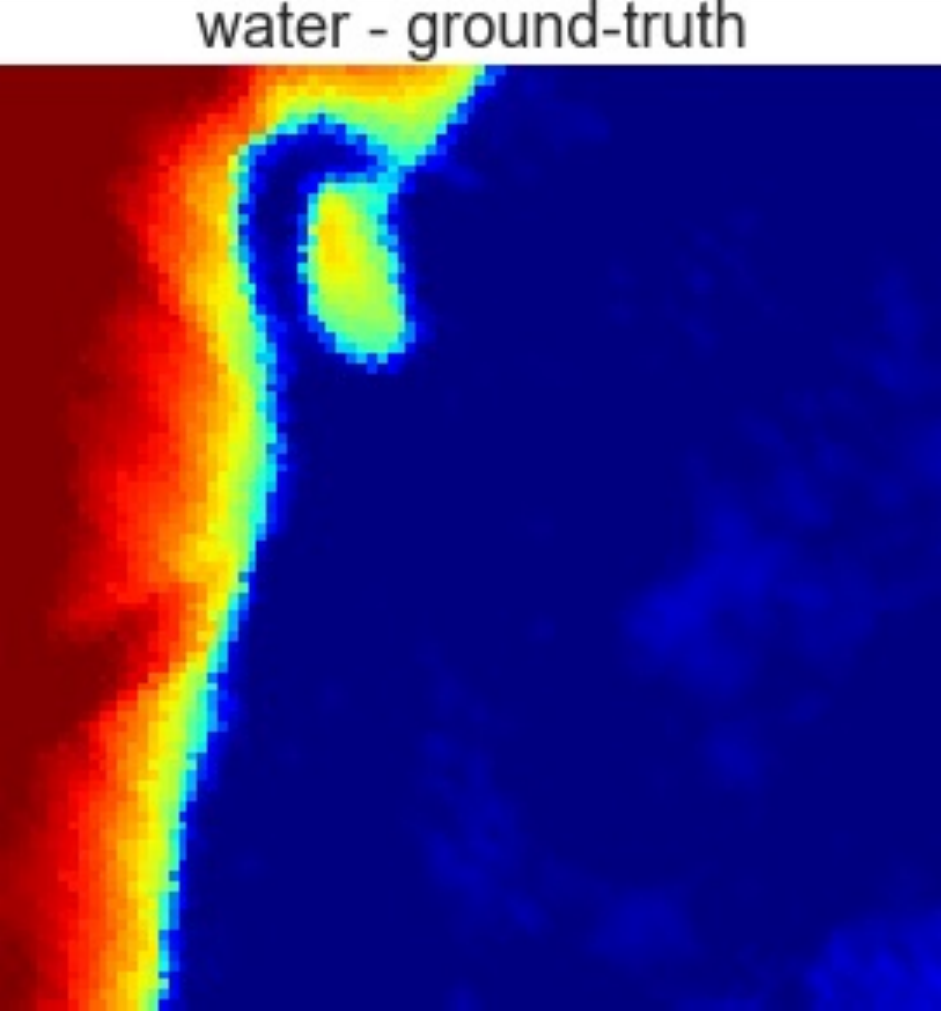}
        \caption{Abundances maps of the Samson dataset. From left to right respectively: Soil (LDVAE), Soil (ground truth), Tree (LDVAE),
        Tree (ground truth), Water (LDVAE), and Water (ground truth).}
    \label{fig:samson_abundances}
\end{figure*}

\begin{figure*}
    \centering
    \includegraphics[width=0.36\textwidth,trim={  0cm 0cm 0cm 0cm},clip]{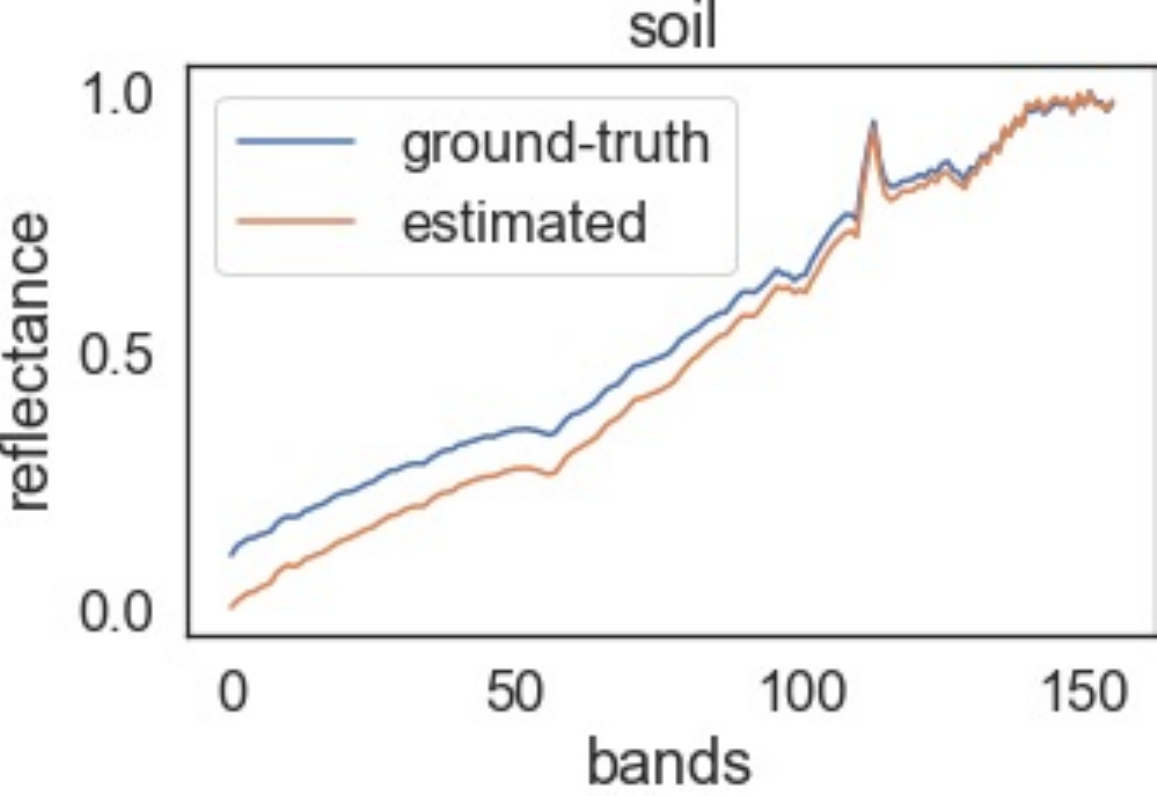}
    \includegraphics[width=0.31\textwidth,trim={1.7cm 0cm 0cm 0cm},clip]{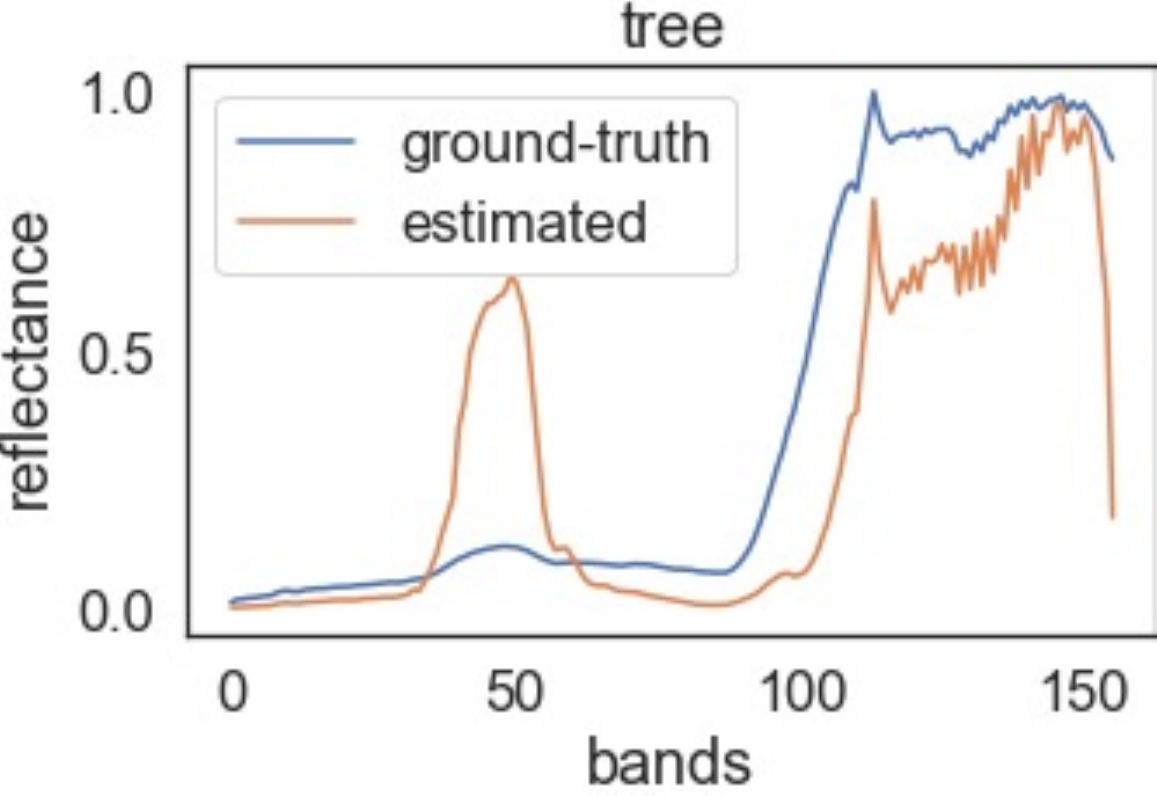}
    \includegraphics[width=0.31\textwidth,trim={1.7cm 0cm 0cm 0cm},clip]{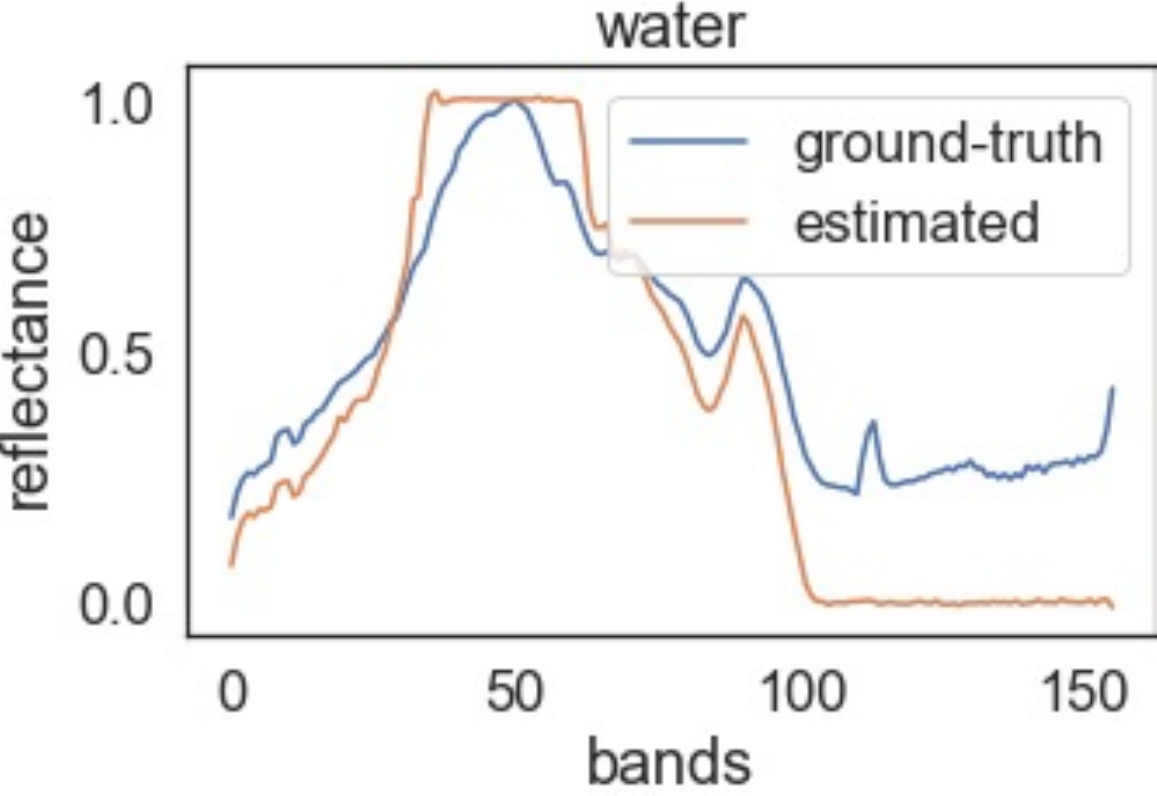}
    \caption{Endmembers of the Samson dataset generated by LDVAE: comparison with ground truth. From left to right: Soil, Tree and Water. The curves show the estimated and the ground truth endmembers.}
    \label{fig:endmembers_samson}
\end{figure*}

The results on Samson dataset were also satisfactory despite the small amount of ground truth data available for training.
We can observe from Figure~\ref{fig:samson_abundances}, that the abundances estimation could detect the prominent classes, but the proportion of each material was not estimated to the highest accuracy.
Figure~\ref{fig:endmembers_samson} shows a moderate performance on endmember extraction, however noisy in some parts of the spectra (see Figure~\ref{fig:endmembers_samson}, material=tree,  bands 100-150).


The RMSE metrics shown in Tables ~\ref{table:results-hydice-urban-rmse} and ~\ref{table:results-samson-rmse} are averages of all pixels in each class. The small variance present in the abundances estimations may stem from the model converging toward a local minima as opposed to a global minima, which also results in higher average RMSE values when compared to other methods. These show 1) the validity of our method and 2) that further investigation is necessary to improve the model training and convergence with respect to model capacity, hyperparameters tuning, and feature engineering.

\section{Conclusions}
\label{sec:conclusion}

We present a Latent Dirichlet Variational Autoencoder (LDVAE) model to solve the problem of hyperspectral pixel unmixing.  Given a pixel spectra, the proposed model is able to infer endmembers and their mixing ratios together. Furthermore, LDVAE's decoder stage is able to synthesize new mixed pixels.  The input to the decoder is an abundance vector, and given this vector the decoder is able  to generate a mixed pixel with known abundances.  These pixel can be use for model training in the absence of labelled data.  We showcase this aspect of LDVAE on the Cuprite dataset, where labelled data is not available for model training.  Rather, the model is trained on synthetic dataset.

We evaluate our model on synthetic data constructed from USGS spectra library and on standard benchmarks for hyperspectral pixel unmixing.  We have compared our approach with commonly used hyperspectral pixel unmixing methods, including a number of recent deep learning based approaches.  The results suggest that our model is able to match, and sometimes exceed, the performance achieved by the current state-of-the-art methods.

A drawback of our approach is that it requires labelled data for training; however, we demonstrate that the proposed model can leverage transfer learning.  Specifically, we show that it is possible to train the model on synthetic data only and use it subsequently on ``real'' data.  This suggests that the proposed model is applicable in real-world scenarios where training data is unavailable.

In the future, we plan to use spatial information in addition to spectral information for the purposes of pixel unmixing.  We also plan to study the feasibility of generating (mixed) pixel spectra for the purposes of hyperspectral image super resolution.  Another line of inquiry would be to study how much training data---synthetic or otherwise---is needed to achieve the desired performance.  We currently do not use SAD loss for model training and an interesting direction for future work is to incorporate SAD loss into the variational autoencoder training setup.   We are currently experimenting with Variational CNN Autoencoder, SAD as regularization term in the loss function, and an extension to LDVAE, called iLDVAE, that eschews training data.  We recently presented our work on iLDVAE on at International Geoscience and Remote Sense Symposium 2023.

\bibliographystyle{IEEEtran}
\bibliography{refs}

\end{document}